\documentclass[aps,showpacs,prc,superscriptaddress,floatfix,nofootinbib]{revtex4}
\usepackage{graphicx}
\usepackage{epsf}
\usepackage{wrapfig}
\usepackage{epsfig}
\usepackage{multirow}
\usepackage{soul}
\usepackage{array}
\usepackage{multirow}
\usepackage{ulem}

\def\bk{{\mbox{\boldmath$k$}}}
\def\bq{{\mbox{\boldmath$q$}}}

\def\br{{\mbox{\boldmath$r$}}}
\def\b0{{\mbox{\boldmath$0$}}}

\def\bk{{\mbox{\boldmath$k$}}}
\def\bq{{\mbox{\boldmath$q$}}}

\def\Vec#1{\mbox{\boldmath $#1$}}

\def\beq{\begin{equation}}
\def\eeq{\end{equation}}

\def\beqy{\begin{eqnarray}}
\def\eeqy{\end{eqnarray}}

\def \b #1{ {\bf #1}}
\newcommand{\be}{\begin{eqnarray}}
\newcommand{\ee}{\end{eqnarray}}

\def \b #1{ {\bf #1}}

\def \b #1{ {\bf #1}}
     \font\tenbifull=cmmib10 scaled 1200 
     \font\tenbimed=cmmib9
     \font\tenbismall=cmmib7
\mathchardef\bbkappa="7114
\mathchardef\bbrho="711A
\mathchardef\bbsigma="711B
\mathchardef\bbtau="711C
\mathchardef\bbvarrho="7125
\mathchardef\bbvarsigma="7126
\mathchardef\bbxi="7118

\begin{document}
\vskip 2mm \date{\today}\vskip 2mm
\title{Nucleon momentum distributions, their spin-isospin dependence and
 short-range correlations. }
\author{M. Alvioli}
\affiliation{ECT$^\star$, European Center for Theoretical Studies in Nuclear Physics
and Related Areas,\\ Strada delle Tabarelle 286, I-38123 Villazzano (TN) Italy}
\altaffiliation{{Present address: CNR-IRPI, Istituto di Ricerca per la Protezione Idrogeologica,
 Via Madonna
Alta 126, I-06128, Perugia, Italy}}
\author{C. Ciofi degli Atti}
\affiliation{
   Istituto Nazionale di Fisica Nucleare, Sezione di Perugia,
   Via A. Pascoli, I-06123, Italy}
\author{L. P. Kaptari}
\affiliation{Department of Physics, University of Perugia and
  Istituto Nazionale di Fisica Nucleare, Sezione di Perugia,
  Via A. Pascoli, I-06123, Italy}
  \affiliation {Bogoliubov Lab. Theor. Phys., 141980, JINR, Dubna,
 Russia} \altaffiliation{ {Supported through the program "Rientro dei
  Cervelli" of the Italian Ministry of University and Research}}
\author{C. B. Mezzetti}
\affiliation{Department of Chemistry and Industrial Chemistry, University of Pisa,
Via Risorgimento 35, I-56126, Italy, and\\
Consorzio Interuniversitario Nazionale per la Scienza e Tecnologia dei Materiali,
 via G. Giusti 9, Pisa, I-50121, Italy}
   \affiliation{Istituto Nazionale di Fisica Nucleare, Sezione di Perugia,
  Via A. Pascoli, I-06123, Italy}
\author{H. Morita}
\affiliation{Sapporo Gakuin University, Bunkyo-dai 11, Ebetsu 069-8555,
  Hokkaido, Japan}
\vskip 2mm

\begin{abstract}
The nucleon momentum distribution $ n_A(k)$ for $A=$2, 3, 4,
 16, and 40
nuclei  is systematically analyzed in terms of  wave functions
resulting from advanced solutions of the nonrelativistic
Schr\"{o}dinger equation, obtained within different many-body
approaches based upon  different realistic bare nucleon-nucleon ($NN$)
interactions featuring similar short-range repulsion and tensor interactions.
 Particular attention is paid to the separation of
the momentum distributions into the mean-field and short-range
correlations (SRC) contributions. It is shown that although at
high values of the momentum $k$  different approaches lead to some
quantitative differences, these do not hinder the general
conclusion that the high-momentum behavior  ($k\gtrsim 1.5-2$
fm$^{-1}$) of all nuclei considered are very similar, exhibiting
the well-known scaling behavior with the mass number $A$,   independently of the
used many-body approach, and the details of the bare  $NN$ interaction.
To analyze and understand the frequently addressed question
concerning the relationships between the nucleus, $n_A(k)$, and
the deuteron, $ n_D(k)$, momentum distributions, the spin
($S$)-isospin (T) structure of few-nucleon systems and complex
nuclei is analyzed in terms of realistic $NN$ interactions and
many-body approaches. To this end the number of $NN$ pairs in a
given ($ST$) state, viz., ($ST$)=(10), (00), (01), and (11), and
the contribution of these states to the
 nucleon momentum
distributions, are calculated. It is shown that, apart from the
(00) state which has very small effects,  all other spin-isospin
states contribute to the momentum distribution in a wide range of
momenta. It is shown that that for all nuclei considered the momentum distributions
in the states  $T=0$ and $T=1$
exhibit at $k\gtrsim 1.5-2$ fm$^{-1}$  very
similar behaviors,
 which represents strong  evidence of the \emph{A}-independent  character of SRCs.
 The ratio $n_A(k)/n_D(k)$ is analyzed in detail stressing that
in the SRC region it always increases with the momentum
 and the origin of such an increase is discussed and elucidated.
 The relationships between the one- and two-body momentum
 distributions, considered in a previous paper, are discussed and clarified, pointing
 out the
 relevant role played by the center-of-mass motion of a correlated pair in the (10)
 state. Eventually, the  values of the
 the probability of high-momentum components in nuclei and the per nucleon probability
 $a_2$ of deuteronlike configurations   in nuclei are  calculated, and
the relationship of the present approach with the many-body methods based upon low-momentum
 effective interactions is briefly discussed.

\end{abstract}
\maketitle

\section{Introduction}\label{sec:intro}

Recent experiments on two-nucleon knockout reactions at  high
values of the four-momentum transfer on carbon using protons,
$A(p,ppN)X$, \cite{Tang:2002ww} and electrons, $A(e,e^\prime
pN)X$ \cite{Shneor:2007tu}, as well as experiments on  inclusive
quasi-elastic (q.e.)  electron scattering $A(e,e^\prime)X$
\cite{Egiyan:2005hs,Frankfurt:1993sp,Fomin:2011ng}, have provided
robust evidence on the long-hunted ground-state nucleon-nucleon
($NN$) short-range correlations (SRC), demonstrating that in both
types of processes the projectile had interacted with a nucleon
belonging to a correlated $NN$ pair \cite{Subedi:2008zz}.  In
 exclusive experiments, where    the knowledge of
  both the three-momentum
transfer $\textbf{q}$, and the momentum of a fast  detected proton ${\bf p}$
allows one  to reconstruct the momentum
${\bf k}_1={\bf p} - {\bf q}\equiv -{\bf p}_{miss}$  that the struck proton had before
 the interaction [provided the final-state interaction (FSI) could be disregarded],
 it has been   found  \cite{Tang:2002ww} that
 in the region $1.4 < |{\bf p}_{miss}| < 2.8$ fm$^{-1}$ the removal of a proton was  almost
always accompanied by the emission of a nucleon $N$ (mostly a neutron) carrying a momentum
roughly equal  to $-{\Vec k}_1$.  At the same time, in the
  q.e. inclusive experiment   $A(e,e^\prime)X$, the ratio of the cross
section off a nucleus $A$ to the cross section  off   the deuteron
or $^3$He in the region of the Bjorken scaling variable $1.5
\lesssim x_{Bj}\lesssim 2$ (the region  of $x_{Bj}$  where q.e.
scattering off a correlated $NN$ pair is expected to occur), exhibits
a  constant behavior,  indicating that, in agreement with
theoretical predictions \cite{Frankfurt:1981mk},  the virtual
photon interacted  with a nucleon of a correlated $NN$ pair. The
exclusive experiment, moreover,  provided evidence not only on SRCs
in general, but also, in particular, of the dominance of
proton-neutron ($pn$) deuteronlike tensor correlations occurring in
states $(ST)=(10)$, where spin is given as ($S$) and isospin as ($T$),
 as predicted by several realistic calculations
\cite{Sargsian:2005ru,Schiavilla:2006xx,Alvioli:2007zz,Alvioli:2012aa};  the  experimental
data
 \cite{Tang:2002ww} have  also provided information on  the  center-of-mass  (c.m.)
  momentum
distribution of the correlated $NN$ pair, finding  agreement with
predictions made long ago \cite{CiofidegliAtti:1995qe}. Recent reviews on
experiments providing information on SRCs and their theoretical interpretations
can be found in
Refs. \cite{Review_1} and \cite{Review_2}.
A detailed
picture of SRC is, however, still limited to the  $^{12}C$ nucleus, so that
extension to other nuclei is necessary to have a general quantitative
picture of SRC through the periodic table. In a systematic study of SRC,
 particular attention should be given to
the experimental and theoretical investigations of:  (i) the
relative and c.m. momentum dependencies of SRC, (ii) their
spin-isospin  structure, (iii) the relative role of two- nucleon
($2N$)  and three-nucleon ($3N$) correlations. We discuss $3N$ SRCs
in a separate paper; here  we  concentrate  on
 $2N$ SRCs.  In configuration space these  can
be defined as those deviations
 from the independent motion  of two nucleons, moving in a mean field,
 when they   approach
 relative distances  $r_{12}=|{\bf r}_1 - {\bf r}_2|\equiv r
\lesssim 1.2-1.5 \,fm$; according to theoretical calculations, in
this region, owing to the very nature of the $NN$ interaction (in particular to  its
 central short-range
repulsion and the  tensor attraction in   (ST)=(10)  state),
the two-body
mean-field density is strongly suppressed and $2N$ correlated motion
dominates. The details of $2N$ SRC  depend upon the spin-isospin
state of the  correlated   $NN$ pair, as well as upon
  the region of the nucleus
one is considering, i.e., upon the c.m. motion  of the pair
 ${\bf R = ({\bf r}_1 +{\bf r}_2)}/2$. To
investigate these  details,  one has to take advantage of the high-momentum components generated by SRCs that lead to  peculiar
configurations of the nuclear wave function in momentum space
 \cite{Frankfurt:1981mk}. As a matter of fact, if nucleons
"1" and "2" become strongly correlated at short distances, the
momentum configurations, in the nucleus c.m. frame, are
characterized by ${\Vec k}_2 \simeq -{\Vec k}_1$, ${\Vec K}_{A-2}=
\sum_{i=3}^A \,{\Vec k}_i \simeq 0$, and not by the mean field
configuration $\sum_{i=2}^A \,{\Vec k}_i \simeq -{\Vec k}_{1}$, i.e. when the high-momentum nucleon is balanced by the entire
 $A-1$ nucleons, each of them carrying  an average momentum of the order
  $\simeq\bk/(A-1)$.  Thus, if a correlated nucleon with
momentum ${\Vec k}_1$ acquires a momentum ${\bq}$ from an external
probe,  and  is removed from the nucleus and detected with
momentum ${\Vec p}= {\Vec k}_1+{\Vec q}$, the partner nucleon
should be emitted with momentum ${\Vec k}_2 \simeq {-\Vec k}_1
={\Vec q}-{\Vec p}={\Vec p}_{miss}$. Such a qualitative
picture is, however, strictly valid only if the c.m. momentum of the
correlated pair was zero before nucleon removal and, moreover, if
the two correlated nucleons leave the nucleus without interacting
between themselves and with the nucleus $(A-2)$. These effects
have to be
 carefully evaluated when attempting to extract the momentum distribution from
 experimental cross sections.   Within a mean-field many-body approach, the main effect of SRCs is to deplete the occupancy
of single-particle shell-model states and to make  the occupation
of levels above the Fermi sea  different from zero; this  leads to
a decrease of the momentum distribution  at values of $|\bk|$
roughly less than  the Fermi momentum $k_F$, and to an increase of
it, by orders of magnitude, with respect to the mean-field
distribution \cite{Bohigas:1979kk}. In this context, it has been
pointed out    that   even a low resolution measurement of the
one-body momentum distribution at $|\bk|\gtrsim 2-3 \,fm^{-1}$,
where mean-field effects are negligible,  may provide precious
information on SRCs \cite{CiofidegliAtti:1990rw}. Though the most
direct way to investigate SRCs would be via experiments  that
detect a pair of back-to-back nucleons in the final state, also
experiments which are sensitive to the one-body momentum
distributions could be very useful. We have analyzed two-nucleon
momentum distributions in two previous papers
\cite{Alvioli:2007zz,Alvioli:2012aa};  here
 we concentrate on the one-nucleon momentum distribution $n_A(k)$,  with
the aim of clarifying some points concerning, particularly, its
SRC and spin-isospin structures. We quantitatively clarify to
what extent the high-momentum part of $n_A(k)$ can be associated
to the deuteron momentum distribution $n_D(k)$. To this end we show that  such an association,
which is only qualitatively correct, has been motivated   either from
the results of approximate many-body approaches
\cite{Zabolitzky:1978cx,VanOrden:1979mt,BenharCiofi,Ji:1989nr},
 or from just assuming it as an input in pioneering Monte Carlo many-body
 calculations of  $n_A(k)$   \cite{Schiavilla_old}.
Recently,  the momentum distributions of  few-nucleon  systems and
complex nuclei have
 been calculated
 within sophisticated
 many-body approaches
\cite{Alvioli:2007zz,Alvioli:2012aa,Pieper,Feldmeier:2011qy,Akaishi,Hiko,Kievsky:1992um,
Gloeckle:1995jg,Arias de Saavedra:2007qg,
Wiringa_review,Roth:2010bm,Alvioli:2005cz, Varga:1995dm,
Suzuki:2008fg, Suzuki_1},
 using
  modern realistic interactions \cite{RSC,Paris,AV8,AV14,AV18}. For this reason it seems to us appropriate
  to update
 the situation concerning the relationship between the momentum distributions of a
  nucleus $A$, where all $2N$ spin-isospin states may contribute,
 and the momentum distribution of the deuteron, where only the state $(ST)=(10)$  is present.
 Our paper is organized as follows.
 Our formalism, based upon  proper spin-isospin dependent
 one- and two-body density matrices which allows one to calculate the various
 spin-isospin components of the nucleon momentum distribution $n_A (k)$, is presented
  in Sec. \ref{sec:sec2}.
 In Sec. \ref{sec:sec3} we (i) provide some general definitions
 of the one-  and two-body momentum distributions, (ii) illustrate the  way SRCs
 influence the momentum distribution, (iii) critically analyze the way the probability of
 SRCs can be defined and (iv) present  a systematic
 comparison of the  momentum distributions of $A=2$, $3$, $4$, $16$, and $40$, nuclei  resulting from
 different many-body calculations and $NN$ interactions. The values of the calculated
 number of pairs  in different spin-isospin states  in a nucleus $A$,
 and
 the momentum distributions in these  states  are given in Sec.
 \ref{sec:sec4}. The comparison between the momentum distributions of complex nuclei
 and the
 deuteron momentum distributions is illustrated in
 Sec.\ref{sec:sec5}. In this section the result of calculation
 of the probability of $2N$ correlations in nuclei is also
 presented.
  Finally,
 the Summary and Conclusions are given in Sec. \ref{sec:sec6}.

 \section{Spin-isospin dependent density matrices, momentum
 distributions and short-range correlations}\label{sec:sec2}

\subsection{Nuclear ground-state wave function and spin-isospin dependent
 density matrices}
In this paper we consider  the nuclear wave function of a nucleus
with $Z$ protons and $N$ neutrons ($Z+N=A$), resulting from the
nonrelativistic Schr\"{o}dinger equation containing two- and
three-body interactions, {\it viz} \beqy {\left[
-\frac{\hbar^2}{2\,m_N}\, \sum_{i=1}\,\hat{\nabla}^2_i\,
+\,\sum_{i<j} \,\hat{v}_{2}(i,j) +\,\sum_{ i<j<k  }
      \,\hat{v}_3(i,j,k)\right]\,\psi_f^A(\{ {\bf x}_i \}_A)\,
      =\,E_f\,\psi_f^A(\{ {\bf x}_i \}_A)}.
\label{standard}
\eeqy
 In Eq. (\ref{standard}) $m_N$ is the nucleon mass,
  and $f$ and $\{ {\bf x}_i \}_A$
stand, respectively,  for the set of quantum numbers of the state
$f$, and  the set of $A$ generalized coordinates ${\bf x}_i\equiv\{
{\bf r}_i, {\bf s}_i, {\bf t}_i\}$, with ${\bf s}_i$ and  ${\bf
t}_i$ denoting the nucleon spin and isospin and ${\bf r}_i$ denoting the
position coordinates measured from the  c.m. of the nucleus
$(\sum\limits_{i=1}^A \textbf{r}_i=0)$.
 Once
 $\psi_f^A(\{ {\bf x}_i \}_A)$ is known, various density matrices pertaining to
 the nuclear ground  state $\psi_{f=0}^A\equiv \psi_{ JM_J}^A$ can be calculated.
For ease of presentation we consider in what follows  complex
nuclei with zero total momentum
 $J=0$ in the ground-state and use the notation
  $\psi_{00}^A\equiv \psi_{0}^A$.
 In this paper we investigate the
 number of pairs in various spin-isospin states and the
  spin-isospin dependent two-body and one-body densities and  momentum distributions.
This requires
the knowledge of  two-body and one-body spin-isospin
dependent density matrices,  which  can  be obtained
by introducing the spin-isospin projector operators   ${\hat P}_{ij}^{T=1}=
(3+\Vec{\tau}_i\cdot\Vec{\tau}_j)/4$  and
and ${\hat P}_{ij}^{T=0}=(1-\Vec\tau_i\cdot \Vec\tau_j)/4$,
 (with the same form for the spin operators).
 A list of the density matrices
  that we
 need in our calculations is given below:

1. {\it The non diagonal  spin-isospin dependent two-body density matrix, viz}
\beqy
\rho_{(ST)}^{N_1N_2}({\bf r}_1, {\bf r}_2;{\bf r}^{\prime}_1, {\bf
r}^{\prime}_2) = \int\psi_{0}^{A*}(\widetilde {{\bf
x}}_1,\widetilde {{\bf x}}_2 \dots,\widetilde {{\bf
x}}_A)\,\sum_{i<j}\widehat{\rho}_{ij}^{(ST)} ({\bf r}_1,{\bf
r}_2;{\bf r}^{\prime}_1,{\bf r}_2^{\prime})
\psi_{0}^A(\widetilde{{\bf x}}_{1}^{\prime},\widetilde{{\bf
x}}_{2}^{\prime},\cdots,\widetilde{{\bf x}}_{A}^{\prime})
\prod\displaylimits_{i=1}^A d\widetilde{{\bf x}}_i
d\widetilde{{\bf x}}_i^{\prime}, \label{2BNDDMatrix} \eeqy where
the non diagonal two-body spin-isospin dependent density matrix
operator is \beqy
\widehat{{\rho}}_{ij}^{(ST)}
({\bf r}_1,{\bf r}_2;{\bf r}^{\prime}_1,{\bf r}_2^{\prime})
={\hat P}_{ij}^{S}\,{\hat P}_{ij}^{T}\,\delta(\widetilde{{\bf r}}_i- {\bf r}_1)
\delta(\widetilde{{\bf r}}_j- {\bf r}_2)\delta({\bf r}_i^{\prime}-
\widetilde{{\bf r}}_1^{\prime})
\delta(\widetilde{{\bf r}}_j^{\prime}- {\bf r}_2^{\prime})
\prod\displaylimits_{k\neq \{i,j\}}^A \delta(\widetilde{{\bf r}}_k - \widetilde{{\bf r}}_k^{\prime})
\prod\displaylimits_{n=1}^A
\delta_{{s}_{3_n}\, {s}_{3_n}^{\prime}}
 \delta_{{t}_{3_n}\, {t}_{3_n}^{\prime}},
\label{2BNDDOperator}
\eeqy
2. the {\it half-diagonal two-body  spin-isospin dependent density matrix, viz}

\beqy
\rho_{(ST)}^{N_1N_2}({\bf r}_1, {\bf
r}_2;{\bf r}^{\prime}_1) =
\int\psi_{0}^{A*}(\widetilde {{\bf x}}_1,\widetilde {{\bf x}}_2
\dots,\widetilde {{\bf
x}}_A)\,\sum_{i<j}\widehat{\rho}_{ij}^{(ST)} ({\bf r}_1,{\bf
r}_2;{\bf r}^{\prime}_1) \psi_{0}^A(\widetilde{{\bf
x}}_{1}^{\prime},\widetilde{{\bf
x}}_{2}^{\prime},\cdots,\widetilde{{\bf x}}_{A}^{\prime})
\prod\displaylimits_{i=1}^A d\widetilde{{\bf x}}_i\,
d\widetilde{{\bf x}}_i^{\prime}, \label{2BHalfDDMatrix} \eeqy
where
\beqy
\widehat{\rho}_{ij}^{(ST)}
({\bf r}_1,{\bf r}_2;{\bf r}_1^{\prime})
={\hat P}_{ij}^{S}\,{\hat P}_{ij}^{T}\,\delta(\widetilde{{\bf r}}_i- {\bf r}_1)
\delta(\widetilde{{\bf r}}_j- {\bf r}_2)\delta(\widetilde{{\bf r}}_i^{\prime}-
\widetilde{\bf r}_1^{\prime})
\prod\displaylimits_{k\neq {\it i}}^A
 \delta(\widetilde{{\bf r}}_k - \widetilde{{\bf r}}_k^{\prime})
 \prod\displaylimits_{n=1}^A
 \delta_{{s}_{3_n}\, {s}_{3_n}^{\prime}}
 \delta_{{t}_{3_n}\, {t}_{3_n}^{\prime}},
\label{2BHalfDOperator}
\eeqy

3. the {\it diagonal two-body  spin-isospin dependent density matrix, viz}
\beqy
\rho_{(ST)}^{N_1N_2}({\bf r}_1, {\bf r}_2)=
\int\psi_{0}^{A*}(\widetilde {{\bf x}}_1,\widetilde {{\bf x}}_2
\dots,\widetilde {{\bf
x}}_A)\,\sum_{i<j}\widehat{\rho}_{ij}^{(ST)} ({\bf r}_1,{\bf r}_2)
\psi_{0}^A(\widetilde{{\bf x}}_{1}^{\prime},\widetilde{{\bf
x}}_{2}^{\prime},\cdots,\widetilde{{\bf x}}_{A}^{\prime})
\prod\displaylimits_{k=1}^A d\widetilde{{\bf
x}}_k\,d\widetilde{{\bf x}}_k^{\prime}, \label{2BDDMatrix} \eeqy
 where
 \beqy
\widehat{\rho}_{ij}^{(ST)}
({\bf r}_1,{\bf r}_2)
={\hat P}_{ij}^{S}\,{\hat P}_{ij}^{T}\,\delta(\widetilde{{\bf r}}_i- {\bf r}_1)
\delta(\widetilde{{\bf r}}_j- {\bf r}_2)
\prod\displaylimits_{k=1}^A \delta(\widetilde{{\bf r}}_k - \widetilde{{\bf r}}_k^{\prime})
\delta_{{s}_{3_k}\, {s}_{3_k}'}
 \delta_{{ t}_{3_k}\, {t}_{3_k}'}.
\label{2BDOperator}
\eeqy

The  following relations between the various density matrices and their
normalization should be stressed:
\beqy
\int  \rho_{(ST)}^{N_1N_2} ({\bf r}_1, {\bf r}_2;{\bf r}^{\prime}_1, {\bf r}^{\prime}_2)
\delta({\bf r}_1-{\bf r}_1') \delta({\bf r}_2-{\bf r}_2') \, d{\bf
r}^{\prime}_1, d {\bf r}^{\prime}_2= \rho_{(ST)}^{N_1N_2}({\bf r}_1, {\bf r}_2,)
\label{Rel1}
 \eeqy
\beqy
\int \rho_{(ST)}^{N_1N_2}({\bf r}_1, {\bf r}_2)\,d{\bf r}_1
\,d {\bf r}_2=N_{(ST)}^A,
\label{Norm1new}
 \eeqy

where $N_{(ST)}$ is the number of nucleon pairs in state $(ST)$, so that
\beqy
 \sum\displaylimits_{(ST)} N_{(ST)}^{A}=
\frac{A(A-1)}{2}\equiv N_A.
\label{Norm2new}
 \eeqy

As for the spin-isospin independent  density matrices, they are
normalized in the usual way, namely $\int \rho_{A}({\bf r}_1, {\bf
r}_2)\,d{\bf r}_1\,d {\bf r}_2=A(A-1)/2$, $\int \rho_{A}({\bf
r}_1, {\bf r}_2;{\bf r}^{\prime}_1) \,d {\bf r}_2 =
[(A-1)/2)]\,\rho_A({\bf r}_1,{\bf r}^{\prime}_1)$, and $\int
\rho_A({\bf r}_1) d\,{\bf r}_1 = A$.
Note that because  the two-body state has to be antisymmetric, the possible
  $(ST)$ states are: $(ST)=(10), (01),\,L=even$ and
$(ST)=(11), (00),\,L=odd$, where $L$ is the relative orbital
momentum of the pair.

 \subsection{The spin-isospin independent and spin-isospin dependent two- and one-nucleon momentum distributions}
Having defined the spin-isospin dependent density matrices,
 we can introduce  the two-body spin-isospin dependent momentum distribution of a pair
of nucleons  in state  $(ST)$, namely
\beqy
n^{N_1N_2}_{(ST)}(\Vec{k}_{1},\Vec{k}_{2})=
\,\frac{1}{(2\pi)^6}\int
d\Vec{r_1}\,d\Vec{r_2}\,d\Vec{r_1}^\prime\,d\Vec{r_2}^\prime\,
e^{i\,\Vec{k}_{1}\cdot\left(\Vec{r}_1-\Vec{r}_1^\prime\right)}\,
e^{i\,\Vec{k}_{2}\cdot\left(\Vec{r}_2-\Vec{r}_2^\prime\right)}\,
\rho^{N_1N_2}_{(ST)}(\Vec{r}_1,\Vec{r}_2;\Vec{r}_1^\prime,\Vec{r}_2^\prime).
\label{2Nmomdis}
\eeqy
By summing Eq. (\ref{2Nmomdis})
over $T$ and $S$, the  spin-isospin averaged two-nucleon momentum distribution is obtained \footnote{In case of nonisoscalar nuclei interference
between different spin-isospin states may occur. Such a contribution in case of the three nucleon systems is negligible and is omitted
in the presentation.}
\beqy
n_A(\Vec{k}_{1},\Vec{k}_{2})=\sum\displaylimits_{(ST)}n_{(ST)}^{N_1N_2}(\Vec{k}_{1},\Vec{k}_{2})=\,\frac{1}{(2\pi)^6}\int
d\Vec{r}_1\,d\Vec{r}_1^\prime\, d\Vec{r}_2\, d\Vec{r}_2^{\prime}
e^{i\,\Vec{k}_{1}\cdot\left(\Vec{r}_1-\Vec{r}_1^\prime\right)}\,
e^{i\,\Vec{k}_{2}\cdot\left(\Vec{r}_2-\Vec{r}_2^\prime\right)}\,
\,\rho_A (\Vec{r}_1,
\Vec{r_2};\Vec{r}_1^\prime,\Vec{r}_2^{\prime}),
\label{2NMaverage} \eeqy
where
\beqy
\rho_A (\Vec{r}_1, \Vec{r_2};\Vec{r}_1^\prime,\Vec{r}_2^{\prime})=
\sum_{(ST)}\, \rho^{N_1N_2}_{(ST)} (\Vec{r}_1, \Vec{r_2};\Vec{r}_1^\prime,\Vec{r}_2^{\prime}).
\label{2NDaverage}
\eeqy
The two-body momentum distributions obey the following normalization
 \beqy \int
n_{(ST)}^{N_1N_2}(\Vec{k}_{1},
\Vec{k}_{2})\,d\Vec{k}_1\,d\Vec{k}_2=\int
\rho_{(ST)}^{N_1N_2}(\Vec{r}_{1},
\Vec{r}_{2})\,d\Vec{r}_1\,d\Vec{r}_2=N_{(ST)}^A \label{IntenneST}
\eeqy
and
\beqy
\int n_A(\Vec{k}_{1}, \Vec{k}_{2})\,d\Vec{k}_1\,d\Vec{k}_2=\int
\rho_{A}(\Vec{r}_{1},
\Vec{r}_{2})\,d\Vec{r}_1\,d\Vec{r}_2=\frac{A(A-1)}{2}.
\label{IntenneA} \eeqy

In this paper we are interested in the various spin-isospin
components $n_{(ST)}^{{N}_1}(\Vec{k}_{1})$ of the one-body
momentum distribution of nucleon $N_1$
 \beqy
 n_A^{\it{N_1}}(\Vec{k}_{1})=
\,\frac{1}{(2\pi)^3}\frac{1}{A} \int
d\Vec{r}_1\,d\Vec{r}_1^\prime
e^{i\,\Vec{k}_{1}\cdot\left(\Vec{r}_1-\Vec{r}_1^\prime\right)}\,
\rho_{A}(\Vec{r}_1,\Vec{r}_1^\prime), \label{Usual1NMDST}
\eeqy
with normalization
 \beqy \int
n_A^{N_1}(\Vec{k}_{1})\,d\Vec{k}_1=\int \rho_{A}^{N_1}(\Vec{r}_{1}) \,d\Vec{r}_1=1, \label{Normennerho}
\eeqy
where $\rho_{A}^{N_1}=\rho_{A}/A$. More specifically,  we have to find the  spin-isospin dependent momentum distribution
 of a nucleon
$N_1$ that belongs to all possible $N_1N_2$ pairs with given value
of S and T. To this end, we need  the  two-nucleon momentum
distribution of all pairs which contains nucleon N$_1$. Because the
isotopic spin, unlike the spin, which is mixed by the tensor
force, is a conserved
  quantity, we first consider the two-body momentum distribution corresponding to a
  fixed value of T, i.e. the spin-isospin  two-body momentum distribution,
Eq. (\ref{2Nmomdis}),  summed over the spin $S=0, 1$; this
quantity is denoted by
 $n_{T}^{{N}_1N_2}(\Vec{k}_{1},\Vec{k}_{1})$, and, according to Pauli principle,
we have
\beqy
   n_{T=0}^{{N}_1N_2}(\Vec{k}_{1}, \Vec{k}_{2}) =
   \left[n_{(00)}^{{N}_1N_2}(\Vec{k}_{1}, \Vec{k}_{2})
+n_{(10)}^{{N}_1N_2}(\Vec{k}_{1}, \Vec{k}_{2})\right]
\label{enne2T0}
\eeqy
and
  \beqy
   n_{T=1}^{{N}_1N_2}(\Vec{k}_{1}, \Vec{k}_{2}) =
   \left[n_{(01)}^{{N}_1N_2}(\Vec{k}_{1}, \Vec{k}_{2}) +n_{(11)}^{{N}_1N_2}
(\Vec{k}_{1}, \Vec{k}_{2})\right], \label{enne2T1} \eeqy where
each of  the four quantities $n_{(ST)}^{{N}_1N_2}(\Vec{k}_{1},
\Vec{k}_{2})$ is defined by   Eq. (\ref{2Nmomdis}).

   By integrating the two-body momentum distribution in   isospin state   T,
   we  find the one-body momentum distribution of a nucleon $N_1$
belonging to a pair with isospin T
 \beqy
 n_{T}^{(N_1N_2)}(\Vec{k}_{1})= \frac{1}{N_{T}^{A}}
\int n_{T}^{N_1N_2}(\Vec{k}_{1}, \Vec{k}_{2})\,d\Vec{k}_2 = \frac{1}{N_{T}^{A}}
\frac{1}{(2\pi)^3}\int
e^{i\,\Vec{k}_{1}\cdot\left(\Vec{r}_1-\Vec{r}_1^\prime\right)}\,
\left[\int \rho_{T}^{N_1,N_2}(\Vec{r}_1,\Vec{r}_2;
\Vec{r}^\prime)d\,{\br}_2 \right]d\Vec{r}_1\,d\Vec{r}_1^\prime
\label{EnneAST12}
\eeqy
with normalization
\beqy
\int n_{T}^{(N_1N_2)}(\Vec{k}_{1})\,d\Vec{k}_1= \frac{1}{N_{T}^{A}}
\int n_{T}^{N_1N_2}(\Vec{k}_{1}, \Vec{k}_{2})\,d\Vec{k}_1\,\,d\Vec{k}_2 =1.
\label{EnneAST13}
\eeqy
where $N_T^A$ is the number of pairs $N_1N_2$ with isospin T. In Eq. (\ref{EnneAST12})
 $\rho_{T}^{N_1,N_2}(\Vec{r}_1,\Vec{r}_1^\prime;
\Vec{r}_2)$ is
the half-diagonal two-body density matrix, Eq. (\ref{2BHalfDDMatrix}),
summed over the
spin; it is the central quantity necessary to calculate one-body momentum distributions.
The analogs of Eqs. (\ref{enne2T0}) and (\ref{enne2T1}) for the T-dependent one-body momentum
distribution readily follow, namely
\beqy
n_{T=0}^{(N_1N_2)}(\Vec{k}_{1})=
\left[n_{(00)}^{{N}_1N_2}(\Vec{k}_{1}) +n_{(10)}^{{N}_1N_2} (\Vec{k}_{1})\right].
  \label{enne1T0}
\eeqy
\beqy
  n_{T=1}^{(N_1N_2)}(\Vec{k}_{1})=
 \left[n_{(01)}^{{N}_1N_2}(\Vec{k}_{1}) +n_{(11)}^{{N}_1N_2} (\Vec{k}_{1})\right].
\label{enne1T1}
 \eeqy

To obtain an explicit equation
for the momentum distribution we have to know the weights
of a given isospin state in nucleus A. Let us
 consider the proton distribution, which gets contributions from
$pn$ and $pp$ pairs; the former can be in $T=0$ and $T=1$ states,  whereas the latter can
only be in T=1 state. We have therefore to find the weight of $T=0$ and $T=1$ $pn$ pairs
 in
nucleus A, because the weight of $pp$ pairs in $T=1$ state is one.
 The total number of $pn$ pair in T=0 state in nucleus $A$ with isospin  $T_A$ is (see Ref. \cite{Forest:1996kp})
\beqy
N_{T=0}^A=N_{00}^A+N_{10}^A=\frac{1}{8}\left[A(A+2)-4T_A(T_A+1) \right].
\label{Totalennezero}
\eeqy
Dividing Eq. (\ref{Totalennezero}) by the total number of $pn$ pairs, NZ, we
find the weight $w_{T=0}^{pn}$  of a $pn$ pair in nucleus A, namely
   \beqy
 w_{T=0}^{pn}=\frac{1}{8ZN}\left[A(A+2)-4T_A(T_A+1) \right] ,
\label{probability}
\eeqy
with the weight of a $pn$ pair in $T=1$ given by
$w_{T=1}^{pn}=1-w_{T=0}^{pn}$.
Thus, we obtain the momentum distribution of nucleon $N_1$
 in terms of the explicit $T=0,1$
contributions
\beqy
n_{A}^{N_1}(\Vec{k}_{1})=
\frac{1}{A-1}
\{ Z \left[ w_{T=0}^{pn}\, n_{T=0}^{(pn)}(\Vec{k}_{1})+ w_{T=1}^{pn}\,
n_{T=1}^{(pn)}(\Vec{k}_{1})\right]
+(Z-1) n_{T=1}^{(pp)}(\Vec{k}_{1})\} ,
\label{Momdisfin}
\eeqy
which is correctly normalized to one because
 $w_{T=0}^{pn}+w_{T=1}^{pn}=1$  and all $n_{T}^{(N_1N_2)}(\Vec{k}_{1})$
are normalized to one
\beqy
\int n_{T=0}^{(pn)}(\Vec{k}_{1})\,d\Vec{k}_1 =\int n_{T=1}^{(pn)}(\Vec{k}_{1})
\,d\Vec{k}_1=\int n_{T=1}^{(pp)}(\Vec{k}_{1})\,d\Vec{k}_1=1.
\label{NormeST}
\eeqy
Note that using Eq. (\ref{probability}) in Eq. (\ref{Momdisfin}) an even simpler equation is obtained
for isoscalar nuclei, namely
\beqy
n_{A}^{N_1}(\Vec{k}_{1})=\frac{1}{A-1}\left[ \frac{A+2}{4}n_{T=0}^{(pn)}(\Vec{k}_{1})+
3\frac{A-2}{4}n_{T=1}^{(pn)}(\Vec{k}_{1}) \right].
\label{Momdisfin1}
\eeqy

Using Eqs. (\ref{enne1T1}) and (\ref{enne1T0}) we can write
\beqy
n_A(\Vec{k}_1)^{N_1}= n_{A}^{(10)}(\Vec{k}_1)+n_{A}^{(00)}
( \Vec{k}_1)+n_{A}^{(01)}(\Vec{k}_1 )
+n_{A}^{(11)}( \Vec{k}_1),
\label{Separation}\eeqy
where all A-dependent coefficients are incorporated in the proper   $n_{A}^{(ST)} $.

The calculation of the quantities $n_A^{(ST)}(\Vec{k}_{1})$ are presented in
Sec. \ref{sec:sec4}.
\section{$NN$ interactions, many-body approaches,  nucleon momentum distributions
and SRCs} \label{sec:sec3}
We now address the question concerning  the content  of SRC in the nuclear wave
function,
in particular the question concerning the definition of the probability of two-nucleon
SRCs, for,  here,   a  certain degree of ambiguity may easily arise.
The ground-state wave function $\psi_{J\,M_J}^A$ is the solution of the many-body
Schr\"odinger equation and  it describes
both  mean- field and correlated motions. The latter includes  both
long- and short-range correlations; long-range correlations manifest themselves mostly
in open-shell nuclei,  making partially occupied
states which are empty in a simple independent particle model,
 with small effects on high-momentum components;
SRCs, however,  generate high virtual particle-hole
excitations even in closed-shell nuclei and strongly affect
the high-momentum content of the wave function.
Therefore, assuming
 that the momentum
distributions could be extracted
from some experimental data,  we have to figure out a clear-cut  way  to disentangle
the momentum content generated by the mean-field from the one
 arising from SRCs.
 To this end let
us use the following procedure \cite{CPS_momdistr}. If we denote
by $\{|\psi_f^{A-1}>\}$ the complete set of plane waves and eigenfunctions of
the $(A-1)$ Hamiltonian of the $(A-1)$ nucleus, containing the same interaction as the
Hamiltonian which generated the ground-state wave function $\psi_0^A$,  and
use the completeness relation
 \beqy
 \sum_{f=0}^{\infty}
|\psi_f^{A-1}><\psi_f^{A-1}|=1
 \label{completness}
\eeqy
 in Eq.(\ref{Usual1NMDST}),
it is easy to see that the one-nucleon momentum distribution becomes
 \cite{CPS_momdistr}
\beqy
n_A^{N_1}(\Vec{k}_{1})= n_{gr}^{N_1}(\Vec{k}_{1})
+n_{ex}^{N_1}(\Vec{k}_{1}),
 \label{nsplit}
\eeqy
where
 \beqy
 n_{gr}^{N_1}(\Vec{k}_{1})=\frac{1}{(2\pi)^3}  \sum_{f=0,\sigma_1}\Big |
 \int e^{i\,\Vec{k}_{1}\cdot\Vec{r}_1} d\,\Vec{r}_1 \int \,\chi_{\frac{1}{2}\sigma_1}^
 {\dagger}\,\psi_{f=0}^{(A-1)\,*}
 ( {\Vec r}_2 \dots {\Vec r}_A)\psi_{0}^A( {\Vec r}_1,{\Vec r}_2 \dots {\Vec r}_A)
 \prod_{i=2}^A d\,\Vec{r}_i \Big |^2
 \label{ennegr}
  \eeqy
and
 \beqy
n_{ex}^{N_1}(\Vec{k}_{1})=\frac{1}{(2\pi)^3}
 \sum_{f \neq 0,\sigma_1} \Big |\int e^{i\,\Vec{k}_{1}\cdot\Vec{r}_1}
  d  \Vec{r}_1 \int \,\chi_{\frac{1}{2}\sigma_1}^
 {\dagger}\,\psi_{f\neq 0}^{(A-1)\,*}
 ( {\Vec r}_2 \dots {\Vec r}_A)\psi_{0}^A( {\Vec r}_1,{\Vec r}_2
 \dots {\Vec r}_A) \prod_{i=2}^A  d  \Vec{r}_i \Big |^2 .
 \label{enneex}
  \eeqy
In the last equation the sum over $"f"$ stands also for
 an integral over the continuum energy states,  which are present in Eq.
 (\ref{completness}).
We see that the momentum distribution can be expressed through the
overlap integrals between the ground-state wave function $\psi_{0}^A$ of nucleus $A$ and the wave function  $\psi_{f}^{(A-1)}$
of the state $f$ of nucleus $(A-1)$. The squared modulus  of the overlap
integral represents the weight of the ground and excited states of $(A-1)$ in the
ground-state of $A$, so that the quantities
 \beqy
 \mathcal{P}_{gr}^{N_1}=
 \int  n_{gr}^{N_1}(\Vec{k}_{1})\,d\,\Vec{k}_1
 \label{essegr}
  \eeqy
and
 \beqy
 \mathcal{P}_{ex}^{N_1}=\int  n_{ex}^{N_1}(\Vec{k}_{1})\,d\,\Vec{k}_1,
  \label{esseex}
  \eeqy
  with
  \beqy
\mathcal{P}_{gr}^{N_1}+\mathcal{P}_{ex}^{N_1}=1,
\label{sumesse}
  \eeqy
   can be associated to the lack of ground-state
  correlations ($\mathcal{P}_{gr}^{N_1}$) and  to the presence of them ($\mathcal{P}_{ex}^{N_1}$). The separation of the
  momentum distributions in $n_{gr}^{N_1}$ and $n_{ex}^{N_1}$ is particularly useful in
  the case of $A=3,\, 4$ systems, i.e. when the excited states of $(A-1)$ are in the continuum.
  In the case of a complex nucleus, where many discrete hole excited states are present,
  it is more convenient to use another representation
  where the
  particle-hole structure of the  realistic solutions
of Eq. (\ref{standard}) is explicitly exhibited, namely
\beqy
\psi_{0}^A( {\Vec r}_1,{\Vec r}_2 \dots {\Vec r}_A)= c_0
\Phi_{0p\,0h}^A( {\Vec r}_1,{\Vec r}_2 \dots {\Vec r}_A)+
c_2\Phi_{2p\,2h}^A( {\Vec r}_1,{\Vec r}_2 \dots {\Vec r}_A)+ \dots
\label{wfexpansion}
\eeqy
In Eq. (\ref{wfexpansion}) , $\Phi_{0p\,0h}^A$ is a Slater determinant describing the mean-field motion
of $A$ nucleons occupying all states below the Fermi level, $\Phi_{2p\,2h}^A$
describes 2p-2h excitations owing to SRCs, and the dots include
higher order p-h excitations. The modulus squared of the various expansion
 coefficients $c_i$ is nothing but the probability to have $np-nh$ excitations
 in the ground-state
 wave function. In particular $|c_2|^2 \equiv a_2$ will determine the amount
 of ground-state SRCs. Within such a representation, one can write  \cite{CiofiLiuti}
\beqy
n_A^{N_1}(\Vec{k}_{1})=
n_{0}^{N_1}(\Vec{k}_{1}) +n_{1}^{N_1}(\Vec{k}_{1}),
 \label{nsplit1}
\eeqy
 where
 \beqy
 n_{0}^{N_1}(\Vec{k}_{1})=\frac{1}{(2\pi)^3}
  \sum_{f\leq F \,, \sigma_1} \Big |\int e^{i\,\Vec{k}_{1}\cdot\Vec{r}_1} d \Vec{r}_1
 \int \,\chi_{\frac{1}{2}\sigma_1}^
 {\dagger}\, \psi_{f}^{(A-1)\,*}
 ( {\Vec r}_2 \dots {\Vec r}_A)\psi_{ 0 }^A( {\Vec r}_1,{\Vec r}_2
 \dots {\Vec r}_A) \prod_{i=2}^A d\Vec{r}_i \Big |^2
 \label{ennezero}
  \eeqy
  and
 \beqy
 n_{1}^{N_1}(\Vec{k}_{1})=\frac{1}{(2\pi)^3}
 \sum_{f > F, \sigma_1}\Big |\int e^{i\,\Vec{k}_{1}\cdot\Vec{r}_1}
  d\Vec{r}_1 \int \,\chi_{\frac{1}{2}\sigma_1}^
 {\dagger}\,\psi_{f}^{(A-1)\,*}
 ( {\Vec r}_2 \dots {\Vec r}_A)\psi_{ 0  }^A( {\Vec r}_1,{\Vec r}_2
 \dots {\Vec r}_A) \prod_{i=2}^A d\Vec{r}_i \Big |^2
 \label{enneuno}
  \eeqy
where the summation over $f$ in Eq. (\ref{ennezero}) includes all the discrete
 shell-model levels below the Fermi level ($F$) in $(A-1)$
 ("hole states" of $A$),  and in
Eq. (\ref{enneuno})  it includes all the discrete and continuum states above the Fermi sea created by SRCs.
In a fully uncorrelated mean-field approach, we have

\beqy
n_A^{N_1}(\Vec{k}_{1})= n_{0}^{N_1}(\Vec{k}_{1})=\sum_{\alpha \leq F}
\left |\phi_{\alpha}(\Vec{k}_{1})
\right |^2;  \qquad
n_{1}^{N_1}({k}_{1})= 0,
\label{nmeanfield}
\eeqy
and the  the analogs of Eqs. (\ref{essegr}) and (\ref{esseex}) are

\beqy
\mathcal{P}_{0}^{N_1}=\int
n_0^{N_1}(\Vec{k}_{1})\, d\,\Vec{k}_{1}
\label{essezero}
\eeqy
\beqy
\mathcal{P}_{1}^{N_1}=\int
n_1^{N_1}(\Vec{k}_{1})\, d\,\Vec{k}_{1}
\label{esseuno}
\eeqy
with
\beqy
\mathcal{P}_{0}^{N_1}+\mathcal{P}_{1}^{N_1}=1.
\label{normtotal}
\eeqy
The quantities $\mathcal{P}_0^{N_1}$ and $\mathcal{P}_0^{N_1}$ yield, respectively,   the probability to find
a mean-field and a correlated nucleon in the range
$0 \leq k_1 \leq \infty$. Therefore they can be assumed as the mean-field and
SRC total probabilities.
It is clear that both low- and high-momentum components contribute to mean-field and correlated
momentum distributions, but we shall see, as expected, that $n_0^{N_1}$ ($n_1^{N_1}$) gets contribution mainly
from low- (high-)
momentum components. Assuming that   $n_0^{N_1}$ and $n_1^{N_1}$ could experimentally be obtained, it might well be that only a limited range of momenta
is available experimentally, in which case it is useful to define the partial probabilities

\beqy
{\mathcal P}_{0(1)}^{N_1}(k^{\pm}_{1})=4\,\pi
\int_{{k^{-}_{1}}}^{{k^{+}_{1}}}
n_{0(1)}^{N_1}(\Vec{k}_{1}){k}_{1}^2\,d\,{k}_{1}
\label{partialprob1}
\eeqy
i.e.  the probability to observe a mean-field (correlated) nucleon with momentum
 in the range $k^{-}_{1} \leq k_1 \leq k^{+}_{1}$.
Although we do not  discuss in this paper how
the momentum distribution could, in principle, be extracted  from the experimental data,
 we would like nevertheless
to briefly comment on  this point.  It is clear from the very definition of the momentum
distribution  that to obtain information on it one
has to figure out an experiment in which  a nucleon is struck
from a nucleus $A$ and the nucleus $(A-1)$ is left in a well-defined  energy state.
To fully understand the point, it is useful  to introduce  the nucleon spectral function, i.e., the following
quantity:
\beqy
 &&S_A^{N_1}({\Vec k}_1,E)=<\psi_{0}^A|a_{{\bf k}_1,\sigma_1}^{\dag}
 \delta(E-\hat H+E_A)
 a_{{\bf k}_1,\sigma_1} |\psi_{0}^A>=\\
 \label{1BSF_1}
 &&=\sum_{f,\sigma_1} \Big|\int e^{i{\bk}_1\cdot{\br}_1} d\,{\br}_1 \int \,\chi_{\frac{1}{2}\sigma_1}^
 {\dagger}\,\psi_{f}^{(A-1)\,*}
 ( {\Vec r}_2 \dots {\Vec r}_A)\psi_{0}^A( {\Vec r}_1,{\Vec r}_2
 \dots {\Vec r}_A) \prod_{i=2}^A d\,{\br}_i \Big|^2\,
 \delta(E-E_{A-1}^f -E_A))=\\
 &&=S_0^{N_1}({\bk}_1,E_0) + S_1^{N_1}({\bk}_1,E),
 \label{1BSF_2}
 \eeqy
 where Eq. (\ref{1BSF_2})  has been obtained from Eq. (\ref{1BSF_1})
 using the completeness relation (Eq. (\ref{completness})), $a_{{\bk}_1}^{\dag}
 (a_{{\bk}_1})$ is a creation (annihilation) operator,
  $E_A=M_A-M_{A-1}-m_N$, $E=E_A+
 E_{A-1}^f$ is the
 nucleon removal energy, and
\beqy
S_0^{N_1}({\Vec k}_1,E)= \sum_{f \leq F} \Big|\int e^{i{\bk}_1\cdot{\br}_1} d\,{\br}_1
\int \,\chi_{\frac{1}{2}\sigma_1}^
 {\dagger}\,\psi_{ f}^{(A-1)\,*}
 ( {\Vec r}_2 \dots {\Vec r}_A)\psi_{0}^A( {\Vec r}_1,{\Vec r}_2 \dots {\Vec r}_A)
  \prod_{i=2}^A d\,{\br}_i \Big|^2\,
 \delta(E-E_{A-1}^f -E_A),
 \label{Spec_0}
 \eeqy
 \beqy
S_1^{N_1}({\Vec k}_1,E)= \sum_{f>0} \Big|\int e^{i{\bk}_1\cdot{\br}_1} d\,{\br}_1
\int \,\chi_{\frac{1}{2}\sigma_1}^
 {\dagger}\,\psi_{f}^{(A-1)\,*}
 ( {\Vec r}_2 \dots {\Vec r}_A)\psi_{0}^A( {\Vec r}_1,{\Vec r}_2 \dots {\Vec r}_A)
 \prod_{i=2}^A d\,{\br}_i \Big|^2 \,
 \delta(E-E_{A-1}^f -E_A).
 \label{Spec_1}
 \eeqy
The spectral function   represents the
 probability that, after particle "1" is adiabatically
 removed from the bound state and placed in the continuum,
 the nucleus $(A-1)$ remains in the state $E_{A-1}^f$. The relation between the
 spectral function and the momentum distribution is given by the momentum sum rule
\beqy
 \int S_A^{N_1}({\Vec k}_1,E)\,d\,E= n_A^{N_1}({\Vec k}_1).
 \label{1BSFint}
 \eeqy
The partial and full momentum distributions can therefore
be obtained  in principle by detecting the final nuclear system $(A-1)$  in correspondence
of $f<F$ and $f>F$.
The exclusive processes $A(e,e^\prime N)(A-1)_f$ in plane wave impulse approximation
(PWIA)
depends directly upon  $S({\bf k}_1,E)$. Thus by performing these types of experiments in a
wide range of excitation
energies of the final $(A-1)$ nucleus and by performing the integration over
 $E$  the momentum distributions can be obtained.
FSIs
 make the cross section deviate from the PWIA, and, moreover,
 for a complex nucleus,
the  sum  over the entire continuum spectrum of $(A-1)$ is difficult, if not impossible,
to perform. In the case of
few-body systems this difficulty can be overcome, because the number of possible final states
is strongly reduced and, as a matter of fact, experimental information
of $n_{gr}$ and $n_{ex}$ for $^3$He and $^4$He
is already available \cite{eeprime_exp1,eeprime_exp2}.

  We reiterate that the aim of this paper is the theoretical investigation
  of  some general properties of momentum distributions, concerning in particular their
   SRC and  spin-isospin structures. To this end
  for A=3 and 4  "exact" wave functions obtained, either by a direct
  solution of the
Schr\"odinger equation or by variational  procedures,  are used, whereas for complex nuclei
 momentum distributions obtained from various methods,
 ranging from the
 Brueckner-Bethe-Goldstone approach  to the
  cluster expansion techniques are adopted. In the next section the momentum
  distributions of several nuclei are presented and the values  of the quantity
  ${\mathcal P}_{0(1)}^{N_1}(k^{\pm}_{1})$ (Eq. (\ref{partialprob1})) are given.
\subsection{The momentum distributions of few- nucleon systems and complex nuclei}
\begin{figure}[!htp]
  \centerline{\includegraphics[height=8.0cm]{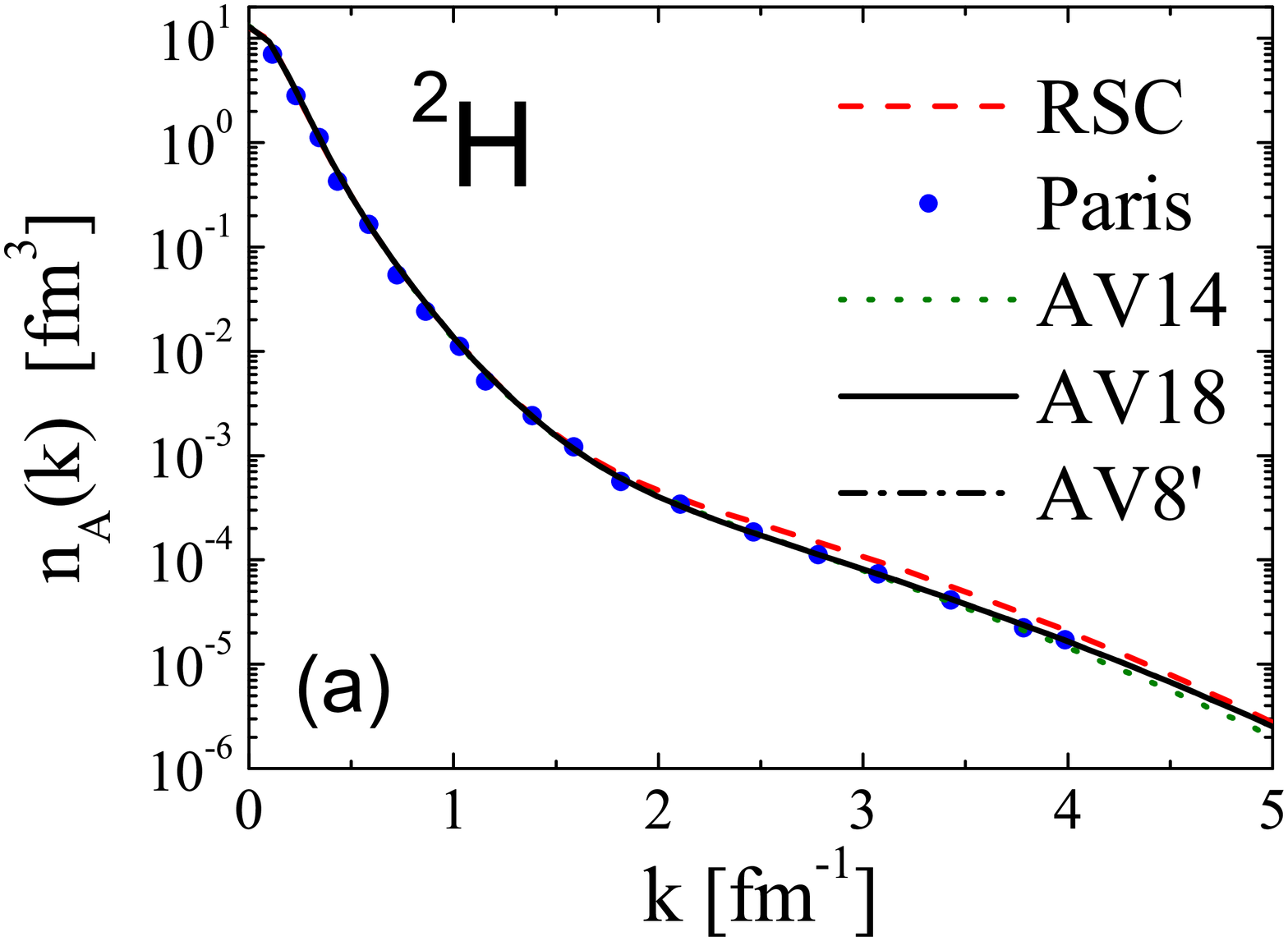}
    \hspace{-1.0cm}
    \includegraphics[height=8.0cm]{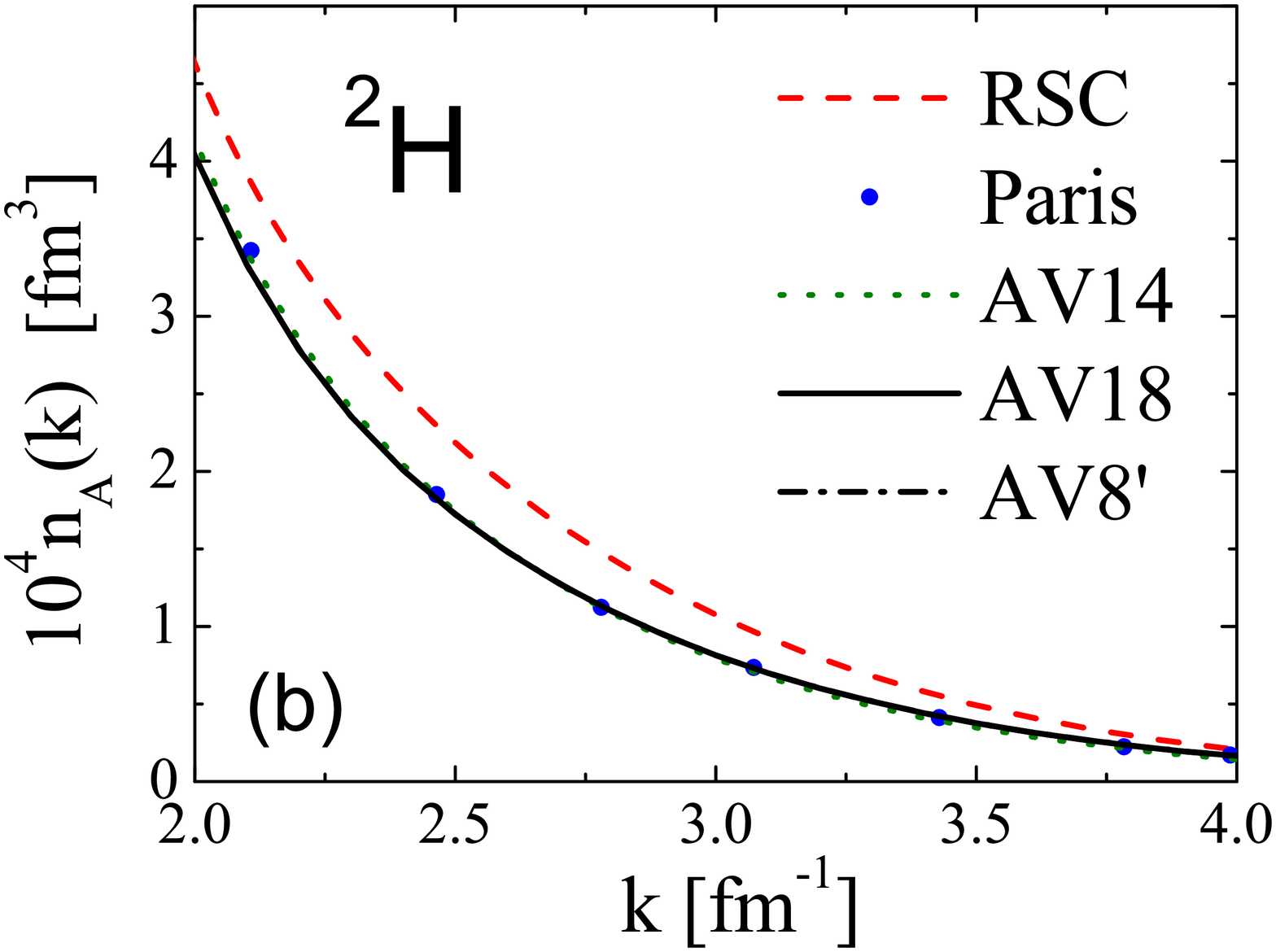}}
  \caption{(Color online) Deuteron momentum distributions in logarithmic (a) and linear (b) scales corresponding
    to various  $NN$ interactions: RSC \cite{RSC}, Paris \cite{Paris}, AV8$^\prime$ \cite{AV8},
    AV14 \cite{AV14} and AV18 \cite{AV18}. Unless otherwise stated, here, and in the other
      figures, the normalization is $4\pi\,\int
      k^2\,d\,k\,n_A(k)=1$. In this and the following
      figures $|{\bf k}_1| \equiv k$ and $n_A(k)\equiv n_A^{(N_1)}(k) [Eq. (\ref{Usual1NMDST})]$.}
  \label{Fig1}
\end{figure}

\begin{figure}[!htp]
  \centerline{\includegraphics[height=8.0cm]{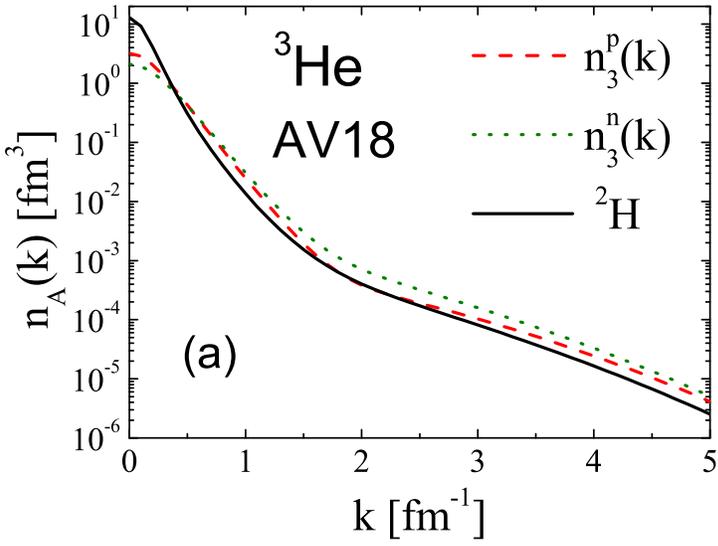}
    \hspace{-1.0cm}
    \includegraphics[height=8.0cm]{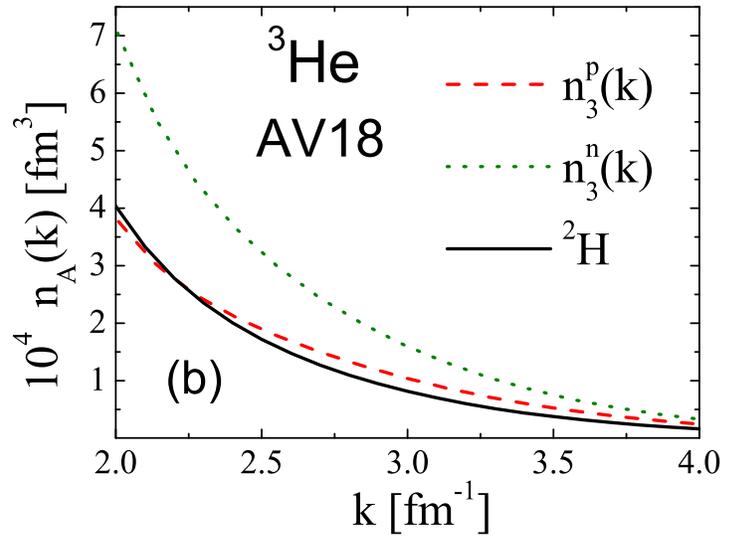}}
  \caption{(Color online) The proton and neutron momentum distributions of $^3$He
    in logarithmic  (a) and linear  (b) scales. Three-nucleon wave functions
     from Ref. \cite{Kievsky:1992um}. The full curve represents   the deuteron momentum
      distribution. Both $^3$He and deuteron wave
    functions correspond to the  AV18 interaction \cite{AV18}.}
  \label{Fig2}
\end{figure}

\begin{figure}[!htp]
  \centerline{\includegraphics[height=8.0cm]{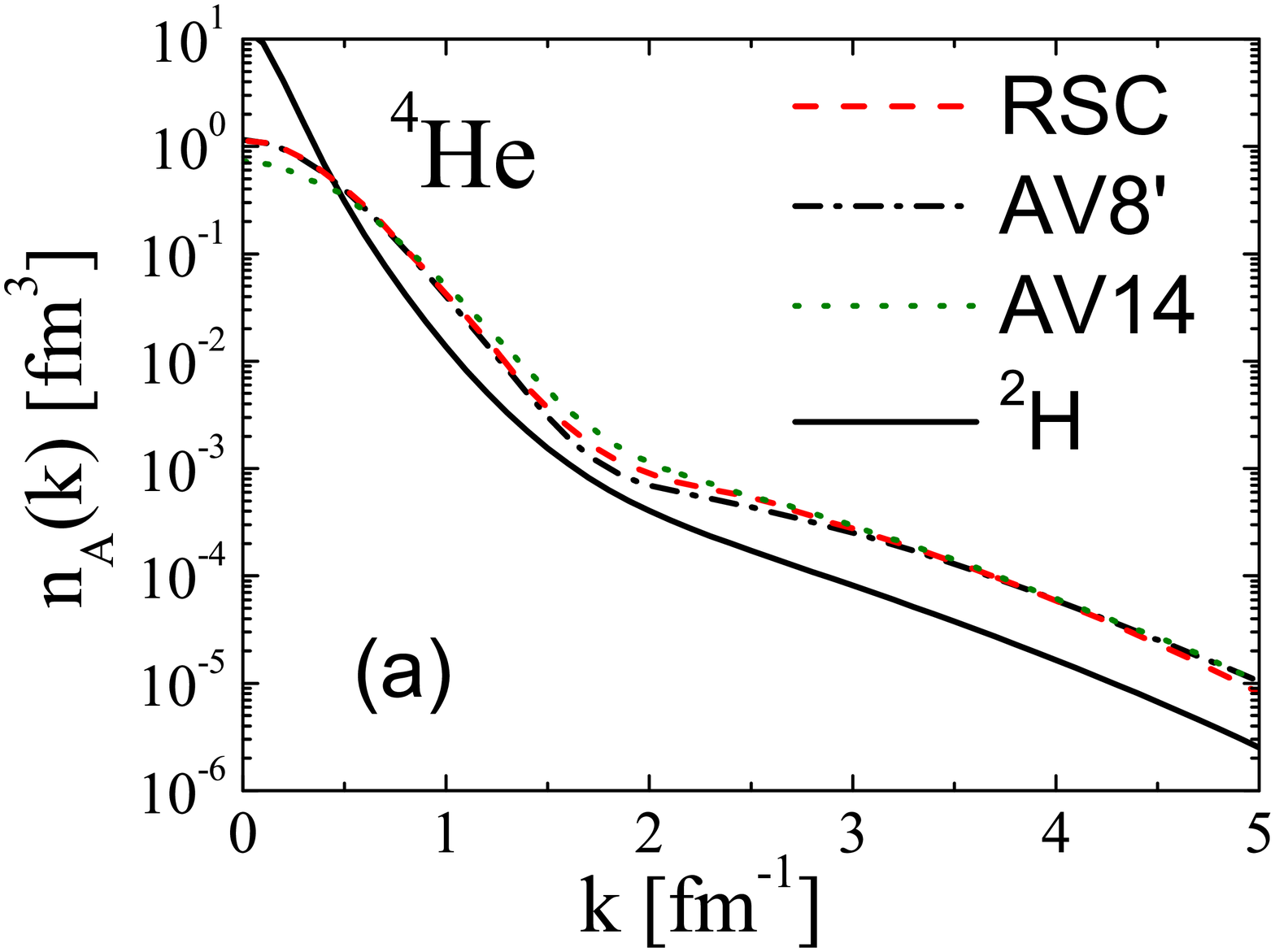}
    \hspace{-1.0cm}
    \includegraphics[height=8.0cm]{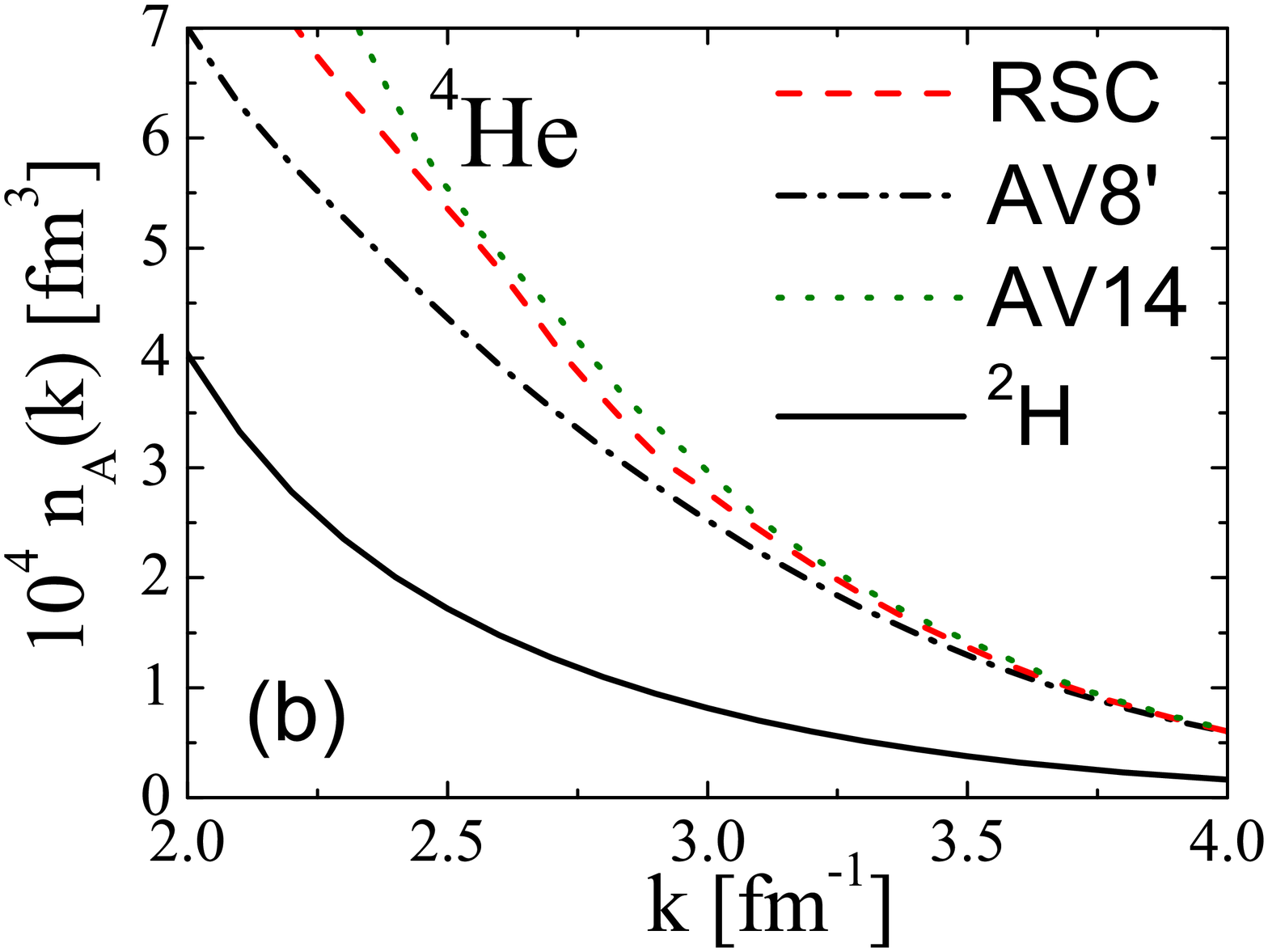}}
  \caption{(Color online) The nucleon  momentum distributions of $^4$He in   logarithmic  (a) and linear  (b) scales corresponding
    to different four-body wave functions and $NN$ interactions. Dashed curve,
    Ref. \cite{Akaishi}; dot-dashed curve, Ref.
    \cite{Hiko}; dotterd curve, Ref. \cite{Pieper}. The full curve represents the deuteron momentum
    distribution
    corresponding to the AV18 interaction.}
  \label{Fig3}
\end{figure}

\begin{figure}[!htp]
  \centerline{
    \includegraphics[height=8.0cm]{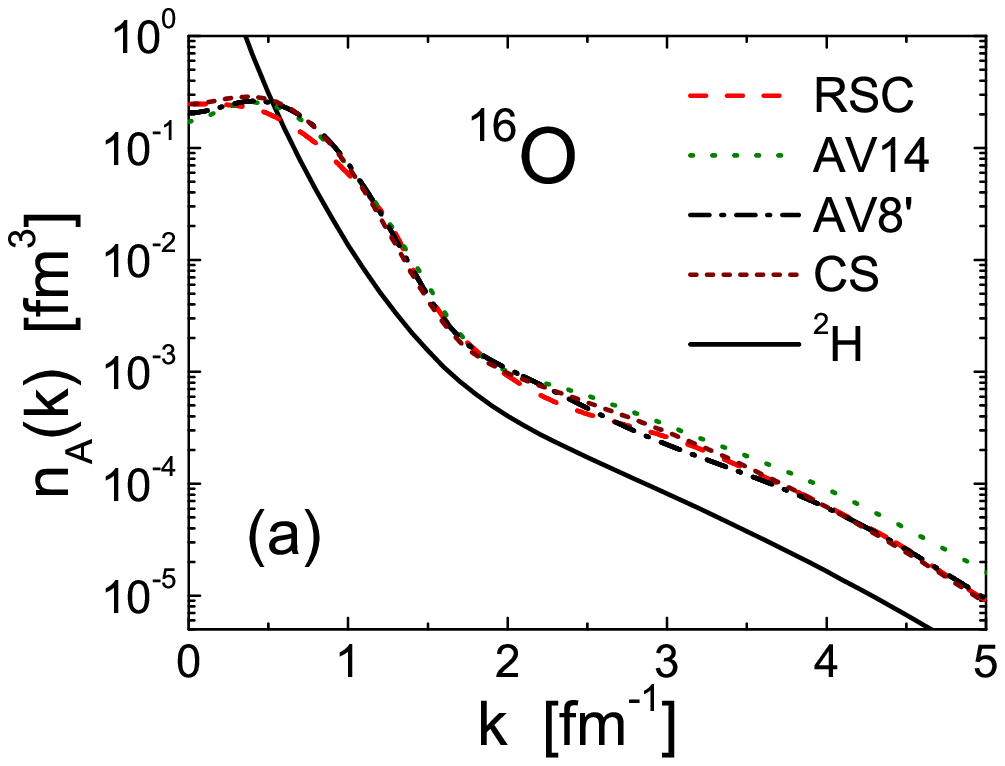}
    \hspace{-1.0cm}
    \includegraphics[height=8.0cm]{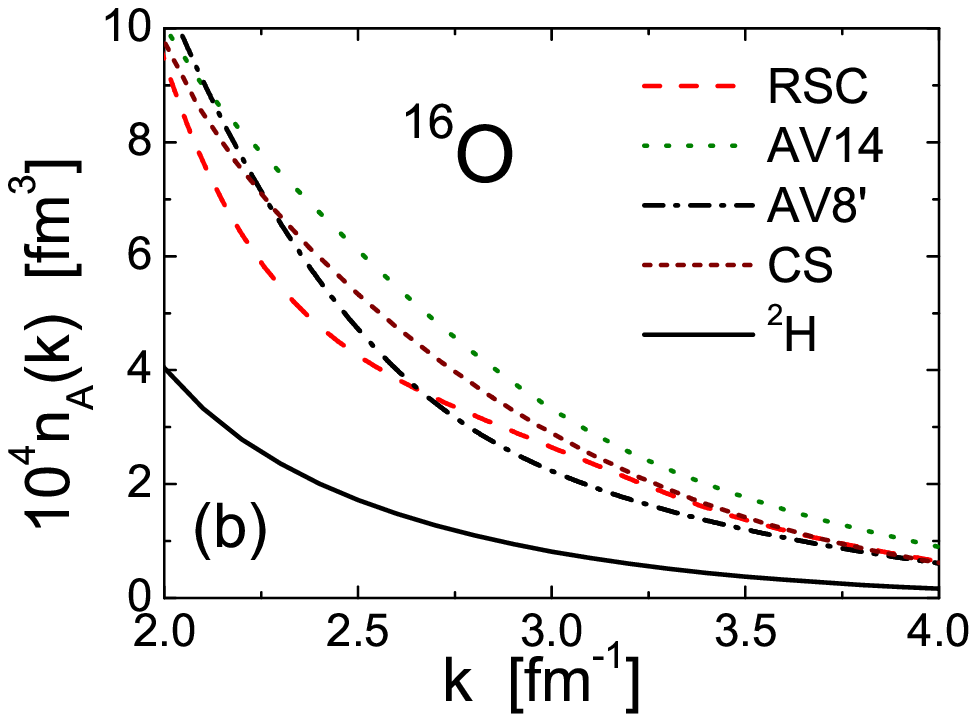}}
  \caption{(Color online) The  momentum distribution of $^{16}$O in logarithmic (a) and linear (b) scales,
    corresponding to different wave functions and $NN$ interactions: Dashed curve, Ref.
    \cite{Zabolitzky:1978cx};  Dotted curve,
     Ref. \cite{Pieper}; dot-dashed curve, Ref. \cite{Alvioli:2005cz}.
    The parametrization of Ref. \cite{CiofidegliAtti:1995qe} is also shown by the
    short-dashed curve (CS).
     The full curve represents the deuteron momentum
    distribution
    corresponding to the AV18 interaction.}
  \label{Fig4}
\end{figure}

\begin{figure}[!htp]
  \centerline{
    \includegraphics[height=8.0cm]{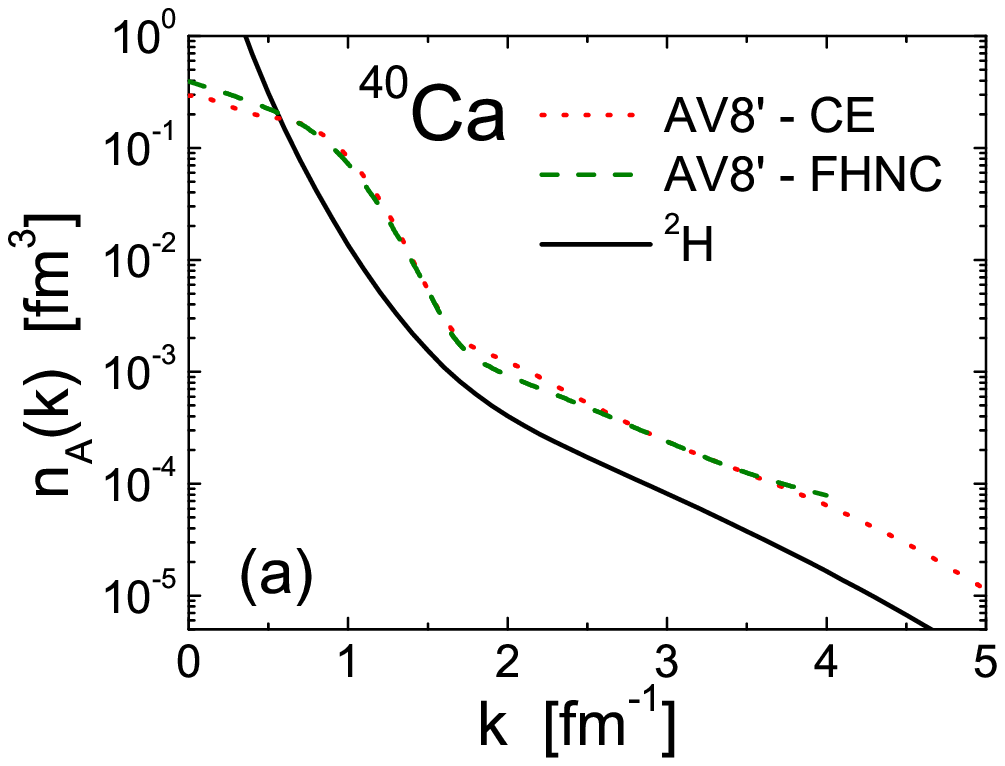}
    \hspace{-1.0cm}
    \includegraphics[height=8.0cm]{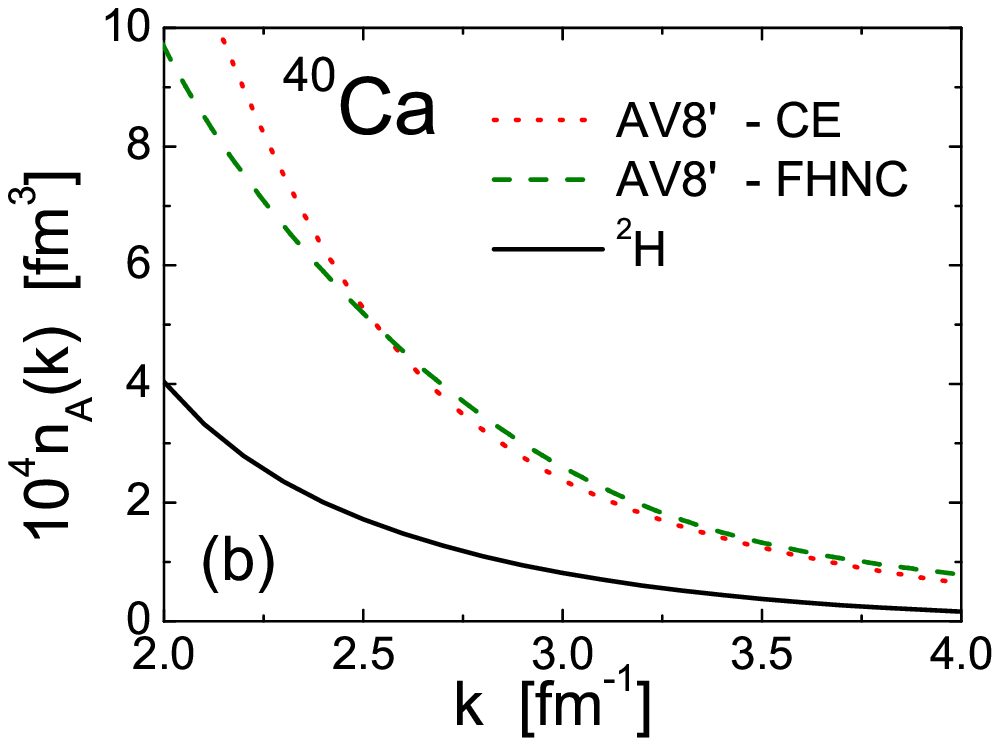}}
  \caption{(Color online) The  momentum distribution of $^{40}$Ca
  in logarithmic (a) and linear (b) scales,
    corresponding to the AV8$^\prime$ $NN$ interaction  calculated within two different
    many-body approaches.  Dashed curve, cluster expansion (FHNC) up to FHNC/SOC order \cite{Arias de Saavedra:2007qg};
    dotted curve, cluster expansion (CE) at second order \cite{Alvioli:2005cz}. The full curve represents the deuteron momentum
    distribution
    corresponding to the AV18 interaction.
    }
  \label{Fig5}
\end{figure}

\begin{figure}[!htp]
  \centerline{
    \includegraphics[scale=0.7]{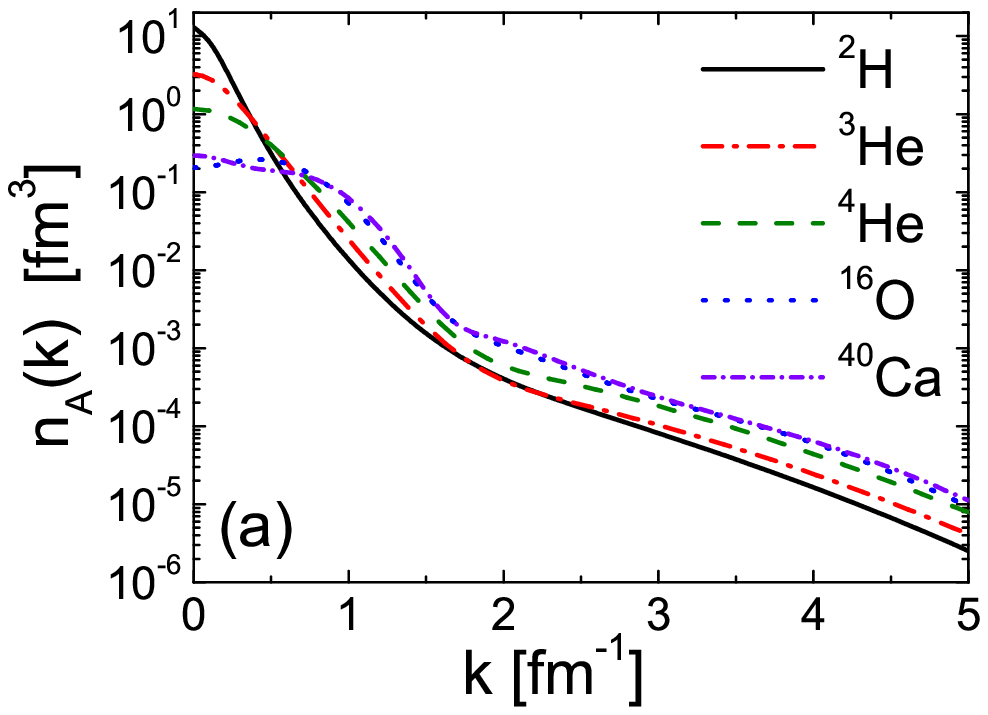}
    \hspace{-0.2cm}
    \includegraphics[scale=0.7]{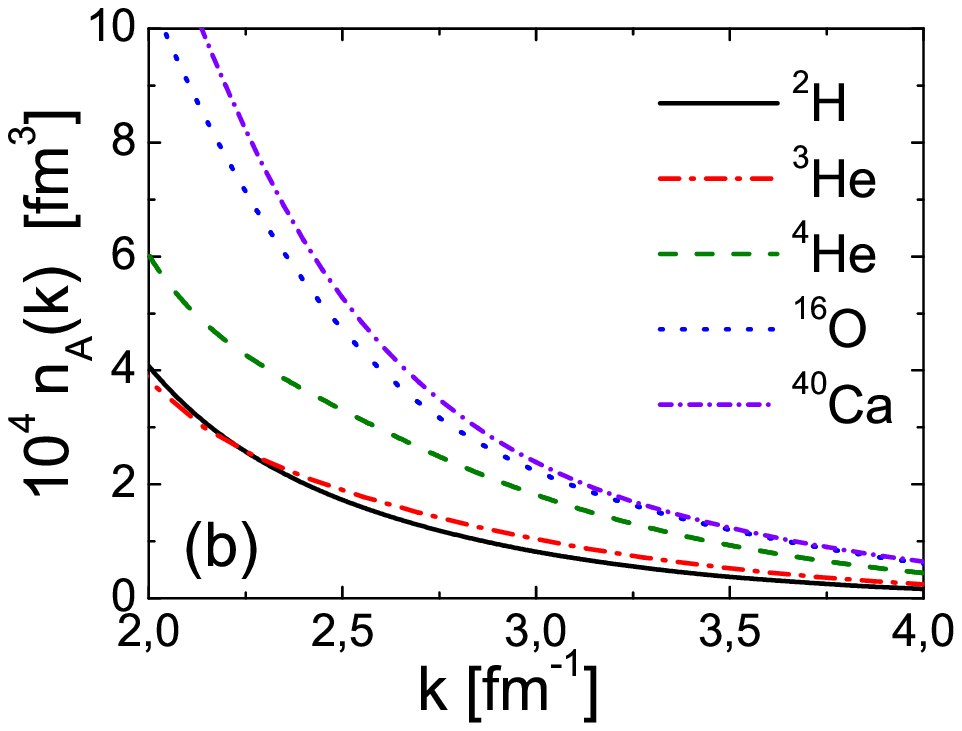}}
  \vskip 0.5cm
  \caption{(Color online) The  proton momentum distribution of   nuclei
    considered in this work in logarithmic  (a) and  linear (b) scales, calculated within different
    many-body approaches with
    equivalent $NN$  interactions, namely the AV18 one, in the case of
    $^2$H and $^3$He, and the AV8$^\prime$ one, in the
    case of $^4$He, $^{16}$O, and $^{40}$Ca.
    }
  \label{Fig6}
\end{figure}

\begin{figure}[!htp]
  \centerline{\includegraphics[height=12.0cm]{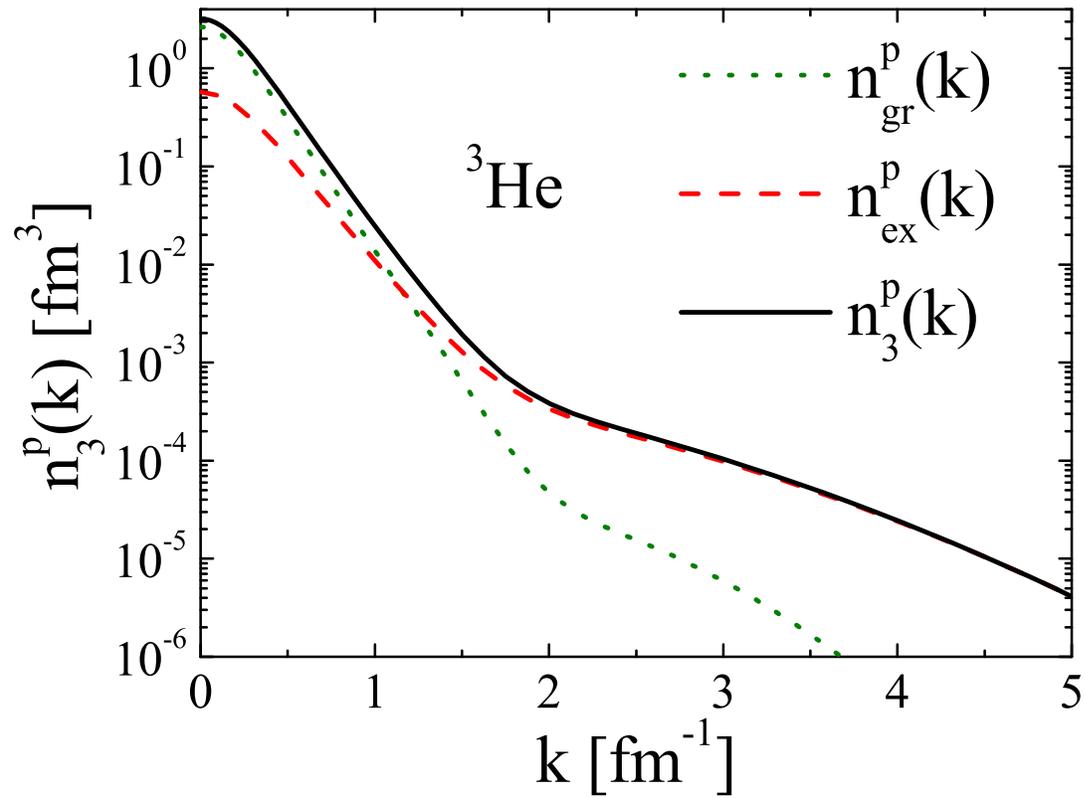}}
  \caption{(Color online)
    The separate contributions $n_{gr}$ and $n_{ex}$ to the proton momentum distributions of
    $^3$He. Wave function from Ref. \cite{Kievsky:1992um}, AV18 interaction. The
    values of
    $\mathcal{P}_{gr}^p=4 \pi\int k^2\,dk \,n_{gr}^p(k)$ and
     $\mathcal{P}_{ex}^p=4 \pi\int k^2\,dk \,n_{ex}^p$ are listed in Table
    \ref{Table1} and the values of Eq. (\ref{partialprob1}) in Table
    \ref{Table2}.}
  \label{Fig7}
\end{figure}

\begin{figure}[!htp]
  \centerline{\includegraphics[height=12.0cm]{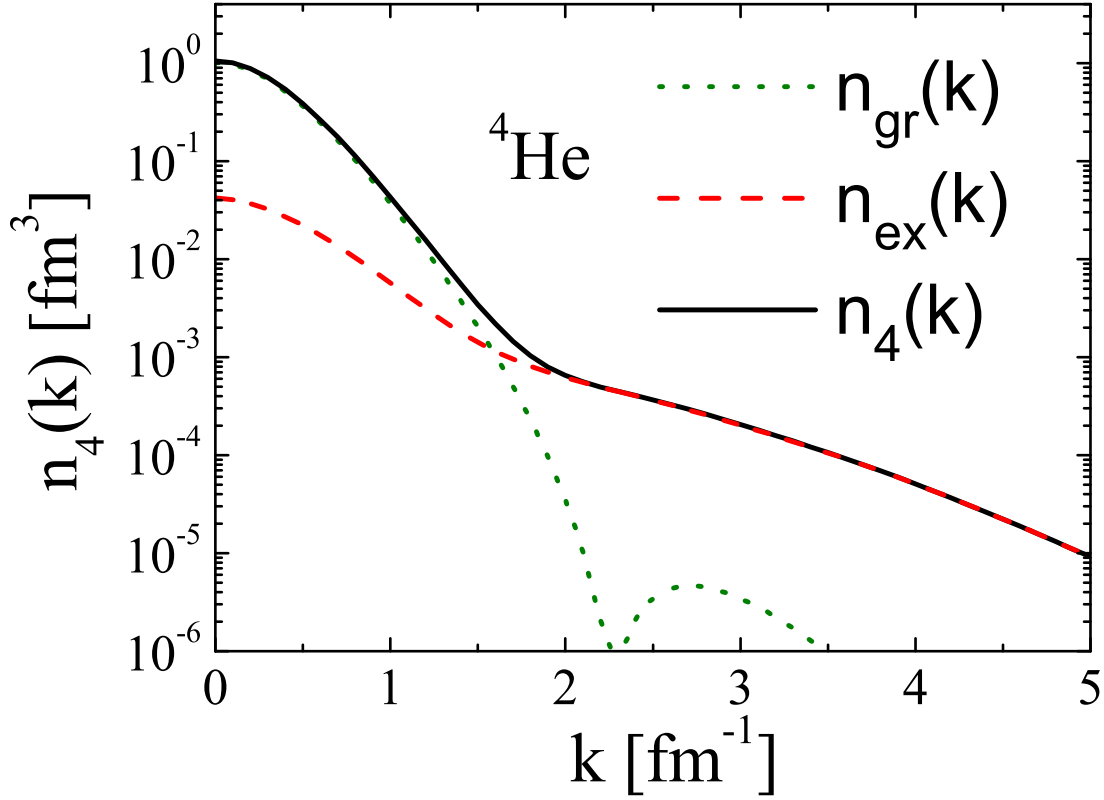}}
  \caption{The same as in Fig. \ref{Fig7}
    but for $^4$He. Wave function from Ref. \cite{Hiko}, AV8$^\prime$ interaction.
    The values of $\mathcal{P}_{gr}=4 \pi \int k^2\,dk \,n_{gr}(k)$ and
     $\mathcal{P}_{ex}=4 \pi\int k^2\,dk \,n_{ex}(k)$
    are listed in Table \ref{Table1} and the values of Eq. (\ref{partialprob1}) in Table
    \ref{Table2}. }
  \label{Fig8}
\end{figure}

\begin{figure}[!htp]
  \centerline{\includegraphics[height=12.0cm]{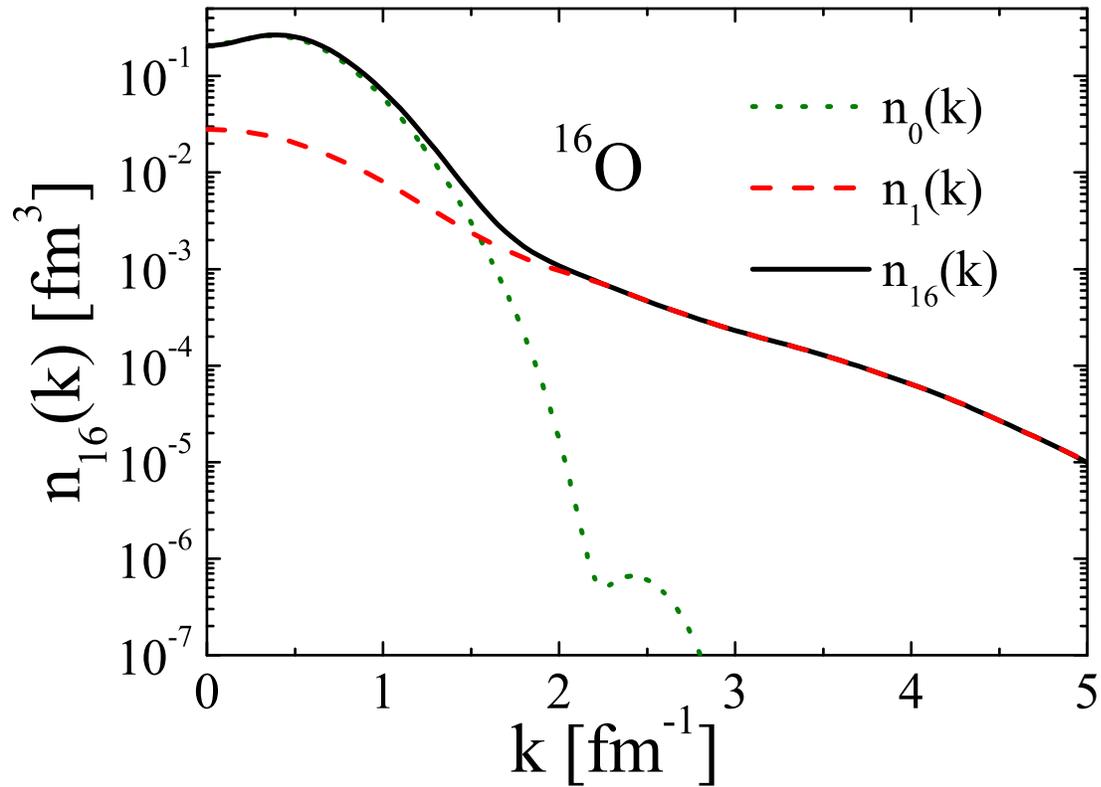}}
  \caption{(Color online) The same as in Fig. \ref{Fig7}
    but for  $^{16}$O. Wave functions from Ref. \cite{Alvioli:2005cz}, AV8$^\prime$ interaction.
    The values of $\mathcal{P}_{0}=4 \pi \int k^2\,dk \,n_{0}(k)$ and
     $\mathcal{P}_{1}=4 \pi \int k^2\,dk \,n_{1}$
    are listed in Table \ref{Table1} and the values of Eq. (\ref{partialprob1}) in Table
    \ref{Table2}.}
  \label{Fig9}
\end{figure}

\begin{figure}[!htp]
  \centerline{\includegraphics[height=12.0cm]{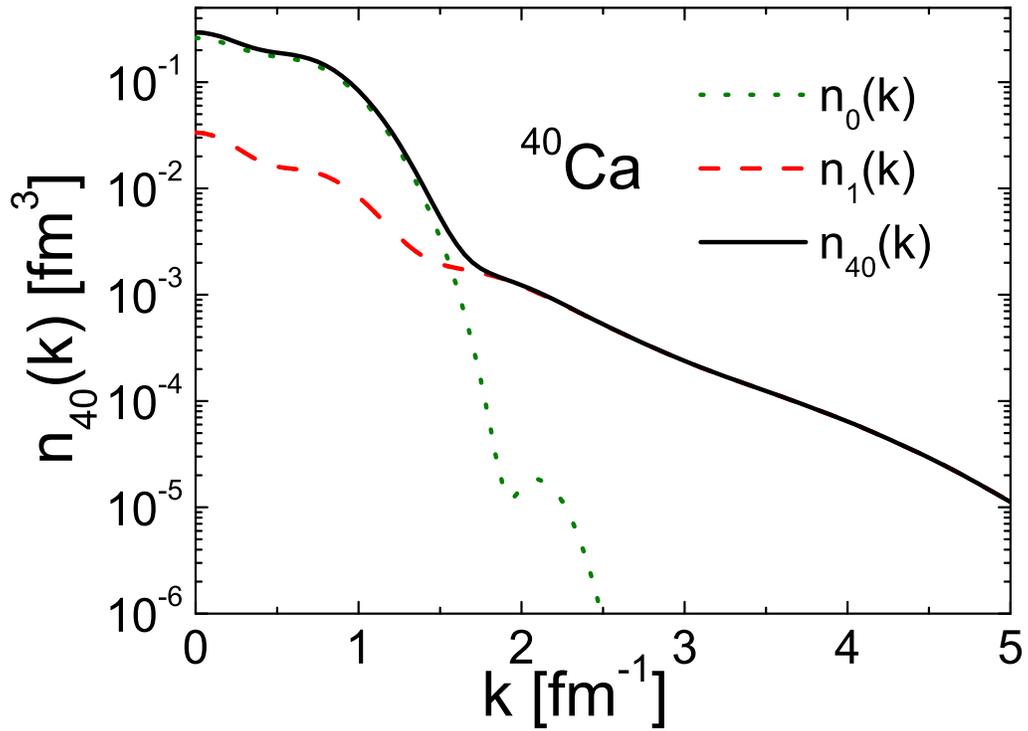}}
  \caption{(Color online) The same as in Fig. \ref{Fig7}
    but for  $^{40}$Ca. Wave function from Ref. \cite{Alvioli:2005cz}, AV8$^\prime$ interaction.
    The values of $S_{0}=4 \pi \int k^2\,dk \,n_{0}(k)$ and
     $S_{1}=4 \pi \int k^2\,dk \,n_{1}(k)$
    are listed in Table \ref{Table2}.}
  \label{Fig10}
\end{figure}

 In this section the momentum distributions of $^2$H, $^3$H, $^3$He, $^4$He,  $^{16}$O
 and $^{40}$Ca,
 calculated within different approaches and using various two-nucleon interactions,  will be presented.
 The full momentum distributions are shown
 in Figs. \ref{Fig1}-\ref{Fig6},  whereas  their separation into
 the mean-field and correlation contributions,
  according to Eqs. (\ref{ennegr}),
  (\ref{enneex}), (\ref{ennezero}) and (\ref{enneuno}), are presented in Figs.
  \ref{Fig7}-\ref{Fig10}. Note that from now on  the
notation ${\bk}\equiv {\bk}_1$ and   $k\equiv |{\bk}_1|$ is used.
\subsubsection{The momentum distributions of $^2$H}
The momentum distributions of  $^2$H
obtained by solving exactly the
Schr\"{o}dinger equation
is crucial for our analysis. It is presented in   Fig. \ref{Fig1},
where it  can be seen that, apart from
 the RSC  interaction  \cite{RSC},
the Paris   interaction  \cite{Paris} and the family of Argonne interactions
 AV8$^\prime$ \cite{AV8}, AV14 \cite{AV14}, and
AV18 \cite{AV18}  provide essentially the same result.
All these potentials exhibit a strong short-range repulsion which gives
 rise to a strong suppression of the deuteron wave function at internucleon
separation $r=|{\bf r}_1-{\bf r}_2| \lesssim 1.5$  fm . This, together with the effects from
the tensor force, generate
 high-momentum components in the  momentum distribution.

\subsubsection{The momentum distributions of $^3$H and  $^3$He}
As already stated in Sec. \ref{sec:sec3},   the three- and four-nucleon
systems $^3$H,
$^3$He and $^4$He are
very important, in that  $n_{gr}$ and $n_{ex}$ have been
explicitly calculated within accurate few-body techniques. Moreover, being $^3$He
and  $^3$H
 non-isoscalar nuclei, their  proton and neutron distributions
 are different. As a matter of fact, in $^3$He the
 proton momentum distribution is given by
\beqy n_{3}^p(k)= n_{gr}^p(k)+ n_{ex}^p(k) \label{3proton}
\eeqy
and the neutron distribution, owing to the absence of
 a two-body bound state  in the  final state (cf. Eq. (\ref{enneex}) ),   is given by
\beqy
n_{3}^n(k)= n_{ex}^n(k). \label{3neutron}
\eeqy
In the above
equations,  $n_{gr}^p(k)$ is the Fourier transform of the overlap
between the ground-state wave functions of $^3$He and $^2$H (cf.
Eq. [\ref{ennegr})] and $n_{ex}^p(k)$ is the Fourier transform of
the overlap between the ground-state wave function of $^3$He and
the continuum state of $pn$ pair (cf. Eq. (\ref{enneex})). Thanks to
isospin invariance, Eqs.(\ref{3proton}) and (\ref{3neutron})
 represent, respectively, the proton and neutron momentum distributions in
$^3$H.
 The proton and neutron momentum distributions in
$^3$He resulting from Faddeev and variational calculations in correspondence
 of  the AV18 interaction are shown in Fig.~\ref{Fig2}.
It can indeed be seen  that they are different, with the former strongly
differing from the deuteron momentum distributions. The origin
of such a difference is discussed in detail
in Sec. \ref{sec:sec4}.
\subsubsection{The momentum distributions of $^4$He}
The nucleus $^4$He is the lightest  isoscalar nucleus, with identical
 proton and neutron momentum distributions. These have been calculated in Ref. \cite{Hiko}
within  the approach of Ref. \cite{Akaishi}
using  the AV8$^\prime$ interaction. They are compared in Fig. \ref{Fig3}
 with the results of the variational Monte Carlo method performed with the AV14 interaction
   \cite{Pieper}.
\subsubsection{The momentum distributions of $^{16}$O and $^{40}$Ca}

The momentum distributions of complex nuclei is  by far more complicated to
calculate with the same accuracy attained  in the case of three- and
four-nucleon  systems.
Nonetheless, several calculations for $^{16}$O
have been performed  within different approaches and  using various $NN$ interactions,
 namely with the RSC potential \cite{RSC},  in
 Ref. \cite{Zabolitzky:1978cx,BenharCiofi,VanOrden:1979mt,Ji:1989nr},
 with the AV8$^\prime$ potential \cite{AV8},
in Ref. \cite{Arias de Saavedra:2007qg} and
  Ref. \cite{Alvioli:2005cz},  and with  the AV14 potential \cite{AV14} in Ref.
 \cite{Pieper} ;
 the various methods that have been used are the unitary operator approach
  \cite{Zabolitzky:1978cx}, the  Brueckner-Bethe-Goldstone  approach
  \cite{VanOrden:1979mt,Ji:1989nr}, the cluster expansion
  approach  truncated at different orders \cite{BenharCiofi,Alvioli:2005cz},
  the fermion-hypernetted-chain method \cite{Arias de Saavedra:2007qg},
  and the  variational Monte Carlo correlated approach  \cite{Pieper}.
   The various results are compared  in Fig. \ref{Fig4}. As
  for $^{40}$Ca,  two available results obtained with the
 V8$^{\prime}$ interactions are shown in Fig. \ref{Fig5}.

\subsubsection{The A-dependence of the momentum distributions}
The momentum distributions of the   considered nuclei obtained with
 the V8$^{\prime}$ interaction  (the AV18 in the  $^2H$ and  $^3$He cases),
 are compared in Fig. \ref{Fig6}.
The general features  that emerge from such a comparison
 can be summarized as follows: i) at low values of the momentum
 $k=|{\bf k}_1|$
the shape of $n_A(k)$ is determined by the asymptotic
 behavior of the wave function of the least bound nucleon, and therefore it
is very different for different
nuclei, ii) in the high-momentum region ($k \gtrsim 1.5-2\, fm^{-1}$)  a qualitative similarity between the
momentum distributions of deuteron and heavier
nuclei can be  observed. In what follows we show
 that in this region  $n_A(k)$ is dominated by  the correlated
 part of the distributions, namely
$n_{ex}$ and $n_1(k)$,  and that the  similarity between
deuteron and complex nuclei  is only a qualitative one, with the
high-momentum behavior of $n_A(k)$ being governed by the
the various spin-isospin components contributing to $n_A(k)$, and not only by the deuteronlike
state $(ST)=(10)$.

\subsubsection{The mean-field and SRC  contributions to the momentum distributions}
The separation of the momentum distribution according
 to Eqs. (\ref{nsplit}) and (\ref{nsplit1}) is shown in Figs. \ref{Fig7}-\ref{Fig10}.
 It can be seen that:
 (i) in the region $k\lesssim 1.5-2.0\,fm^{-1}$ SRC reduce the mean-field distribution
 without practically changing its shape, the effect being
 attributable to the decrease of the occupation probability of the shell-model states below
 the Fermi level; (ii) in the region $k\gtrsim 2.0\,fm^{-1}$ the
 momentum distribution  are entirely exhausted by SRCs. Having at disposal
  both $n_{gr}(k)\ (n_{0}(k))$ and
 $n_{ex}(k)\ (n_{1}(k))$  the probabilities given by
 Eqs. (\ref{essegr}), (\ref{esseex}), (\ref{essezero}) and (\ref{esseuno})
 can be calculated. These
 are listed in Table \ref{Table1}, whereas
 the partial probabilities defined by Eq.
(\ref{partialprob1}) are listed in Table \ref{Table2}.

\subsection{Summary of Section \ref{sec:sec2}}
From what is exhibited in the present section, some general features of the momentum
distributions can be identified, which are, to a large extent,
 independent of the many-body approach and the two-nucleon interaction
 used in the calculations, namely:
i) at $k \gtrsim 2 \,fm^{-1}$ the momentum distributions of both few-nucleon and complex nuclei qualitatively
resemble the deuteron momentum distributions; (ii) in the region of high momenta,
 the realistic momentum distributions of complex nuclei overwhelm
 the mean-field distributions
  by
 several orders of magnitude; (iii) whereas for few-nucleon systems
 the method of calculations
 is very well established, for complex nuclei different methods and potentials
 provide at high momenta values of the distributions
  which can  differ up
 to a factor of two, and it is not yet clear to which extent such a difference should be ascribed to the different potentials or to the different
 methods. It should be mentioned that the momentum distributions extracted from $A(e,e^\prime p)X$ and
 from the $y$-scaling analysis of inclusive $A(e,e')X$ scattering
 \cite{CiofidegliAtti:1990rw} agree with many-body calculations; although
 the errors of the extracted momentum distributions are very large
 at high momenta, they are much smaller than the difference between correlated
 and mean-field distributions, with the latter being totally inadequate
 to predict high-momentum components.
 In what follows our analysis of the momentum distributions continues
  using
 the most advanced available calculation methods and two-nucleon interactions.
 To understand the microscopic origin of the correlated part of $n_A(k)$, we analyze in the next section its spin-isospin structure.

\section{The spin-isospin structure of the nucleon momentum distribution and SRC}
\label{sec:sec4}
The spin-isospin structure of SRCs is a fundamental quantity because it reflects
the details of the $NN$ interaction in the medium. It is therefore
important
to investigate how such a structure can affect various  quantities which
are  related to SRCs, such as, e.g.,  the nucleon momentum distributions.
 In Ref. \cite{Feldmeier:2011qy}
a detailed analysis pertaining to few-nucleon systems has been presented
of the various $(ST)$ channel contributions to
 the relative
momentum distribution  $n_{(ST)}^{N_1N_2}(\Vec{k}_{1},\Vec{k}_{2})$,
 Eq. (\ref{2Nmomdis}), integrated over the c.m.
momentum, namely \beqy
 n_{(ST)}^{N_1N_2}(\Vec{k}_{rel})=\int n_{(ST)}^{N_1N_2}(\Vec{k}_{1},
 \Vec{k}_{2})\,d {\bf K}_{c.m.}=
\int n_{(ST)}^{N_1N_2}(\Vec{k}_{rel},\Vec{K}_{c.m.})\,d {\bf K}_{c.m.},
\label{2BMDrelative}
 \eeqy
whereas in Ref. \cite{Alvioli:2012aa} the dependence of the
two-body momentum distribution
$n_{(ST)}^{N_1N_2}(\Vec{k}_{1},\Vec{k}_{2})=n_{(ST)}^{N_1N_2}({k}_{rel},{K}_{c.m.},\theta)$,
upon the values of ${k}_{rel}$, ${K}_{c.m.}$ and $\theta$ has been
investigated in the case of $A=3$ and $4$.

In this paper we proceed further on into this direction by
analyzing the contribution of various $(ST)$ channels
to the one-body momentum distribution
of a nucleon $N_1$ belonging to a $N_1N_2$ pair in a  spin-isospin state $(ST)$.
Our aim is to understand the quantitative relevance and the momentum dependence
of these contributions, in particular  as far as  the
 deuteronlike state $(10)$ is concerned.
In this respect, it should be stressed that in Ref. \cite{Alvioli:2012aa},
 it has been shown that in $^3$He and $^4$He and for
  back-to-back nucleons  ($K_{\text{c.m.}}=0$, the deuteronlike momentum
 configuration) the quantity
\beqy
 R_{(10)}^{(pn)}(k_{rel},
K_{\text{c.m.}}=0)=n^{pn}_{(10)}(k_{\text{rel}},K_{\text{c.m.}}=0)/n_D(k_{\text{rel}}),
\label{Ratiorel}
\eeqy
i.e. the ratio of the  relative momentum  distribution of a
  $pn$ pair in state   $(ST)=(10)$   to the
  deuteron momentum distribution,
   exhibits a constant  behavior starting from
$k_{rel} \gtrsim 1.5-2\,fm^{-1}$; this  means that at short relative distances,
the motion of a back-to-back $(pn)$ pair in a nucleus behaves at short
distances like in a deuteron. However, a constant behavior is not expected to be observed
in the ratio of the $(ST)$  one-nucleon momentum
  distribution to the deuteron distribution, because the  former, being the integral of the two-body
  momentum distribution, besides the deuteronlike configuration,
  includes many other $NN$  configurations.
  The separation of various $(ST)$
contributions to the one-body momentum distribution is an
involved  task. The problem can be solved by considering the
half-diagonal spin-isospin dependent two-body density matrix and its
integral over $\Vec{r}_2$. To this
end, it is useful first of all  to calculate the number of nucleon pairs in a given
spin-isospin state.
\subsection{The number of $NN$ pairs in various spin-isospin states }
The two-body interaction acts differently in states with different
spin, isospin    and relative orbital momentum  $L$, whose values
are fixed by the Pauli principle, namely $ S+T+L= odd $.
To investigate the spin-isospin
 dependence of the momentum distributions,
it is useful to start counting the number   of pairs $N^{N_1N2}_{(ST)}$ in various $(ST)$ states
in a nucleus with $Z$ protons and $N$ neutrons, with $Z+N=A$. This quantity is
given by
 Eq. (\ref{Norm1new}) and satisfies the sum rule Eq. (\ref{Norm2new}).
   The value of $N_{(ST)}^{N_1N_2}$
has been calculated in various papers, e. g.  in
Refs. \cite{Wiringa:2006ih,Forest:1996kp} for $A \leq 16$, in Ref. \cite{Feldmeier:2011qy}
 for $A \leq 4$ and in Refs. \cite{Vanhalst:2011es,Vanhalst:2012} by considering only pairs with $L=0$.
Here  our approach to this topic and the results for  $A=3,\,4,\, 16,$ and  $40$ \,
and $L=even$ and
$odd$ will be presented.

To start with, let us consider  a  full independent-particle (IP)
shell-model. In the case of s-shell nuclei the number
of pairs in $(ST)$ states can  readily be obtained. As a matter of
fact in A=3 and 4  nuclei the  relative orbital momentum of all
pairs is zero, so that only two $(ST)$ states survive, namely
$(10)$ and $(01)$. A $pn$ pair can be either in $(10)$ state, with
probability $3/4$, or
 in $(01)$ state, with probability $1/4$, whereas  a $pp (nn)$ pair can only be
  in  $(01)$ state, with probability
 $1$. Multiplying these probabilities by the number of $pn$, $pp$ and $nn$ pairs
 ($NZ$,  $Z(Z-1)/2$, $N(N-1)/2$, respectively), the total number of pairs
 is obtained

\be
N_{3(4)}=NZ\left (\frac{3}{4}(10)_{pn} +\frac{1}{4}(01)_{pn}\right)+
\frac{Z(Z-1)}{2}(01)_{pp} +\frac{N(N-1)}{2}(01)_{nn}
\label{totalnumber}
\ee
with the total number of pairs in a given $(ST)$  given by
 \beqy
 N_{(10)}^3= \frac{3}{2} \quad  N_{(01)}^3= \frac{3}{2}
 \eeqy
in $^3$He, and
 \beqy
 N_{(10)}^4= 3 \quad  N_{(01)}^4= 3
  \label{N4He}
 \eeqy
 in $^4$He (note that the  state ($10$) refers to $pn$ pairs only,
 whereas the state $(01)$ includes $pp$,
 $nn$ and $pn$ pairs and it is for this reason that no nucleon labels appear in $N_{(ST)}$).
 In $A>4$ nuclei also the states $(11)$ and $(00)$ contribute.
  In Ref. \cite{Wiringa:2006ih} a general approach to calculate, within the IP model,
  the number of pairs in various
 $(ST)$ states, based upon  counting  even and odd pairs in spatial
 configurations corresponding to a given Young tableaux, has been given,
 and explicit formulas can
 be found there.
In our approach   the values of $N_{(ST)}^A$,
for the three- and four-nucleon
systems have been
 obtained  using the wave functions of
 Ref. \cite{Kievsky:1992um} and \cite{Akaishi,Hiko} corresponding to the AV18
 and AV8$^{\prime}$
 interaction, respectively,  whereas for complex nuclei the cluster expansion
 of Ref. \cite{Alvioli:2005cz} which includes two-, three-, and four-body cluster
 contributions has been used to calculate   the integral of the diagonal
  spin-isospin dependent
 two-body density matrix (Eq. (\ref{2BDDMatrix})) yielding
 \beqy
N_{(ST)}^{A}=\int\rho_{(ST)}^{N_1N_2}(\Vec {r}_1,\Vec {r}_2)d\,\Vec {r}_1\,d\,\Vec {r}_2=
\int n_{(ST)}^{N_1N_2}(\Vec {k}_1,\Vec {k}_2)d\,\Vec {k}_1\,d\,\Vec {k}_2=
\int n_{(ST)}^{N_1N_2}(\Vec {k}_{rel},\Vec {k}_{c.m.})d\,\Vec {k}_{rel}\,d\,\Vec {K}_{c.m.}
\label{NAST_1}
 \eeqy
 If IP wave functions are used in Eq. (\ref{NAST_1}),
 the IP values of $N_{(ST)}^A$ have to coincide with the values
 provided by the formulas of Ref. \cite{Wiringa:2006ih}, as indeed
 it
 is the case.
When  the IP model picture is released and a full many-body
approach with interacting nucleons is considered, odd values of
the relative orbital momentum appear also in  A=3 and 4
 nuclei so that (i) the  states $(00)$ and $(11)$ are
generated in $^3$H, $^3$He, and $^{4}$He; (ii) the amount of
various $(ST)$ states in complex nuclei is changed. Thanks  to
isospin conservation, the number
 of states $(01)$ is decreased in
favor of states $(11)$ and the number  of deuteronlike states
$(10)$ is also decreased in favor of the state $(00)$.  In Ref.
\cite{Forest:1996kp} $N_{(ST)}^{N_1N_2}$ has been calculated for
$^3$He, $^4$He, $^6$Li, $^7$Li, and  $^{16}$O using variational
Green's Functions Monte Carlo wave functions   and various Argonne
interactions; in Ref. \cite{Feldmeier:2011qy} $N_{(ST)}^{A}$
has been obtained for nuclei $^3$He, $^3$H and $^4$He
 using wave functions resulting   from the
correlated Gaussian basis approach \cite{Varga:1995dm} and the
V8$^\prime$ interaction, finally, in Refs.
\cite{Vanhalst:2011es,Vanhalst:2012} the number of pairs in the $L=0$ state has been
evaluated through the periodic table using phenomenological
correlated wave functions. We reiterate that in the present paper we have
calculated $N_{(ST)}^A$ for $A=3, \,4,\, 16,\, 40$ using wave functions
obtained within the hyperspherical
 harmonic variational method
 \cite{Kievsky:1992um} and
the $AV18$ interaction, for $A=3$,  the ATMS  method of Refs.
\cite{Akaishi,Hiko} and the AV8$^\prime$ interaction, for $A=4$,
the linked-cluster expansion of Ref.  \cite{Alvioli:2005cz} and
the AV8$^\prime$, for $A=16$ and $A=40$. The results of our
calculations, which   are presented in Table \ref{Table3},  clearly show
that (i) there is satisfactory general agreement between our
results and the ones of Ref.
\cite{Forest:1996kp,Feldmeier:2011qy}; (ii) as previously found
in those papers,   when the IP model picture is
released and $NN$ correlations are taken into account, the  value of
$N_{(10)}$ is practically unchanged,
 whereas the number of pairs in the $(01)$  state  is decreased in favor of the state
 $(11)$. The reason for that was  nicely explained in
Ref. \cite{Forest:1996kp,Feldmeier:2011qy}: it is attributable to some kind of
 many-body effects induced by
tensor  correlations between particles "2" and "3" , generating
 a spin flip of particle "2", and giving rise to the state $(11)$ between particles "2"
 and "1". These effects are automatically
 included in our calculations,
 because "exact" wave
 functions are used in case of few-nucleon systems and a cluster expansion embodying
 many-body clusters is adopted  in our approach for complex nuclei.

\subsection{The spin-isospin contributions to the momentum distributions}

\begin{figure}[!htp]
\centerline{
  \includegraphics[height=12.0cm]{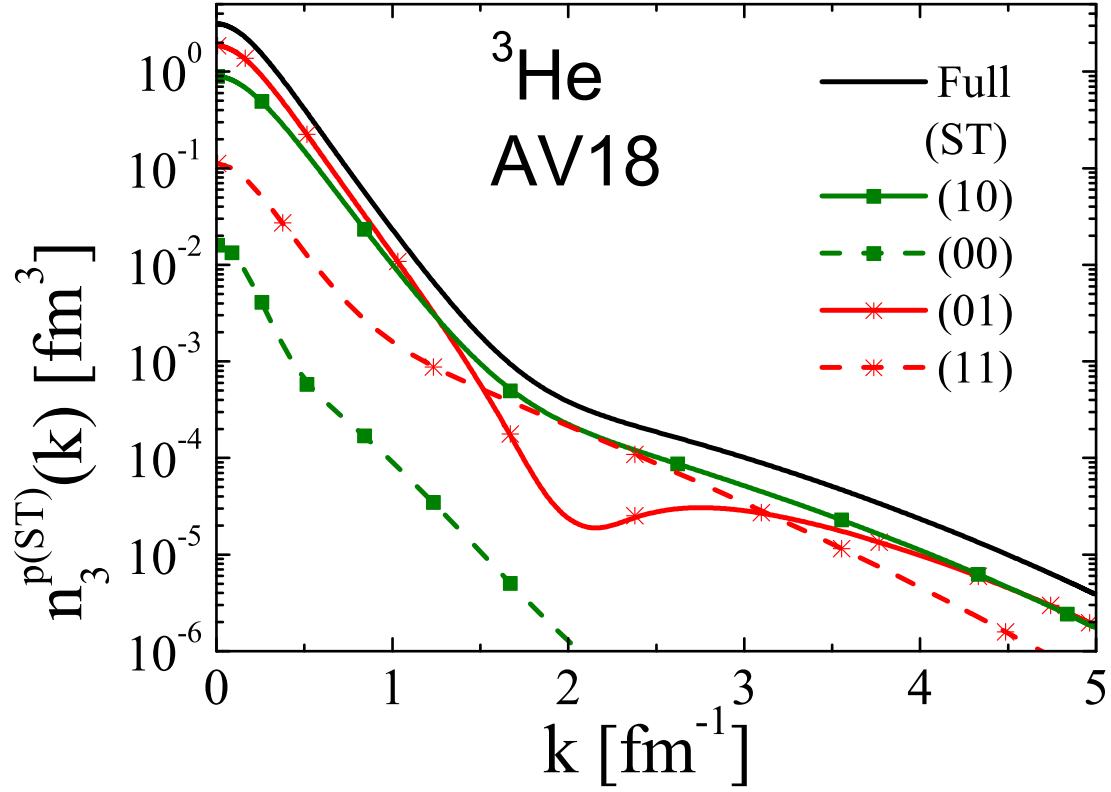}}
  \caption{(Color online) The various spin-isospin contributions, $(ST)$,  to the proton momentum distribution
    of $^3$He. Wave function from Ref. \cite{Kievsky:1992um}, AV18 interaction. The continous line without
    symbols is the sum of the four contributions (cf. Eq. (\ref{n3p1})).}
\label{Fig11}
\end{figure}

\begin{figure}[!htp]
  \centerline{
    \includegraphics[height=12.0cm]{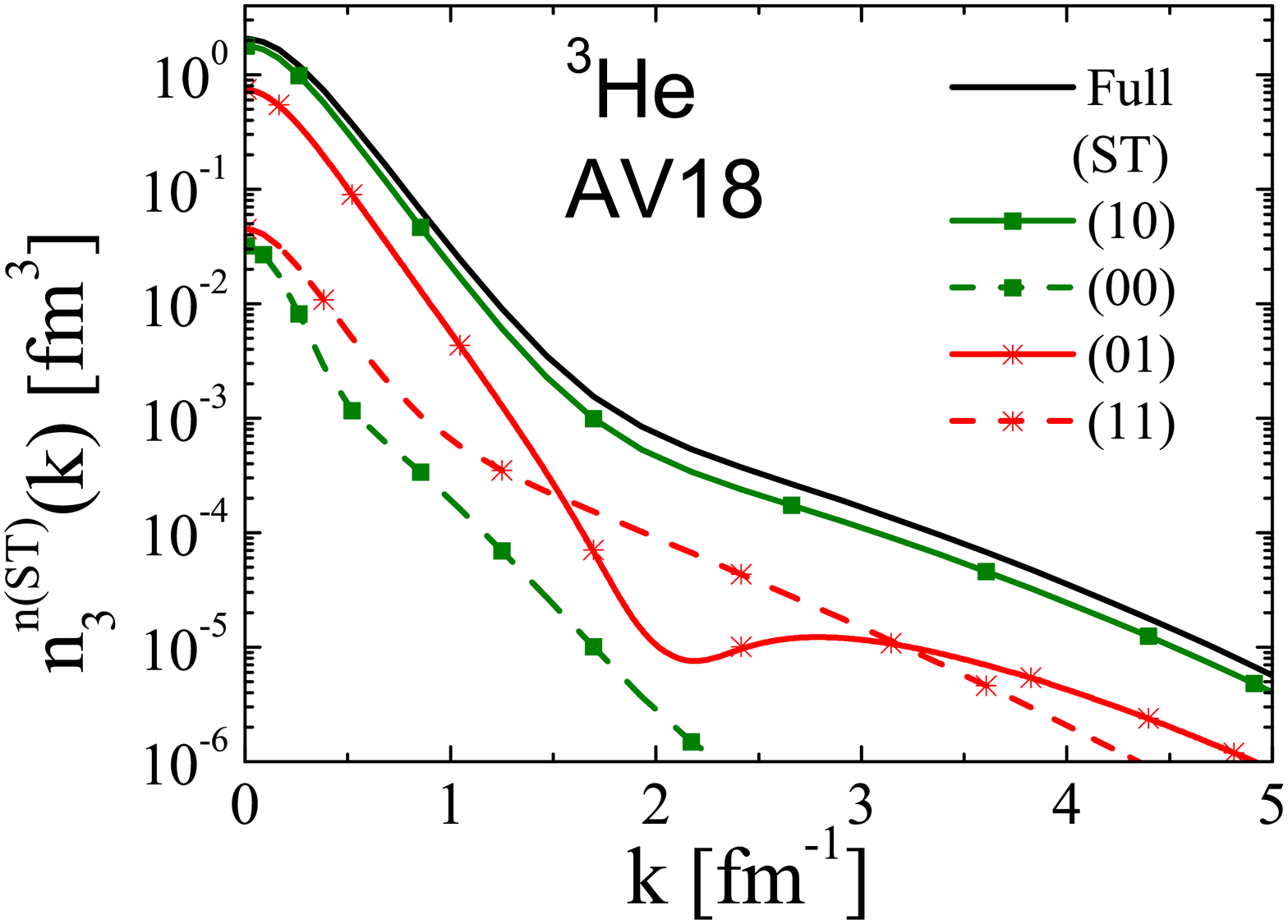}}
  \hspace{0.0cm}
  \caption{(Color online) The same as in Fig. \ref{Fig11} but for the neutron distribution (cf. Eq.
  (\ref{n3n1})).}
  \label{Fig12}
\end{figure}

\begin{figure}[!htp]
  \centerline{
    \includegraphics[height=12.0cm]{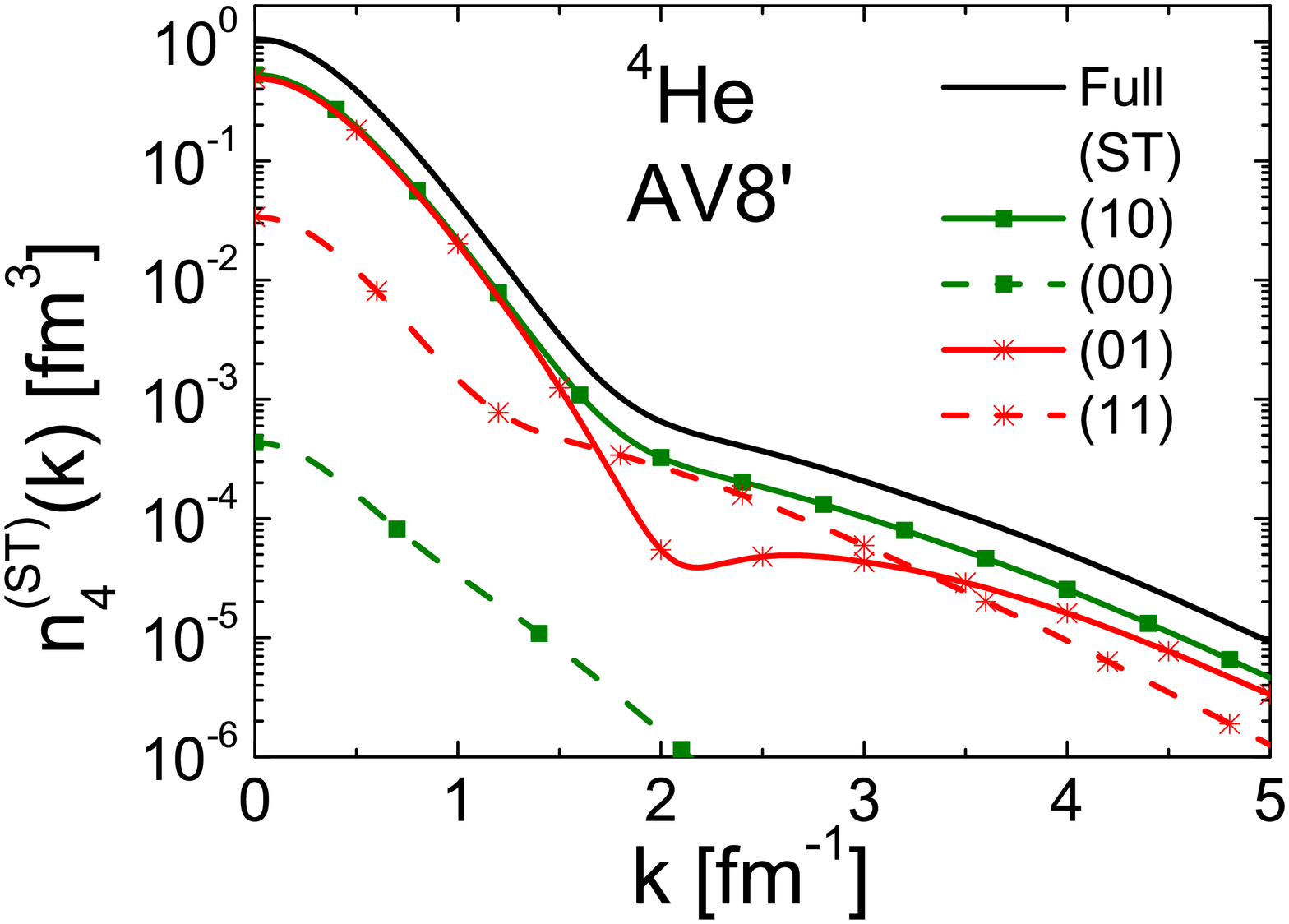}}
  \vskip 1cm \caption{(Color online)
The various spin-isospin contributions to the proton momentum distribution
    of $^4$He (cf. Eq. (\ref{n4_2})). Wave function from Ref. \cite{Hiko}, AV8$^\prime$ interaction.}
  \label{Fig13}
\end{figure}

\begin{figure}[!htp]
  \centerline{
    \includegraphics[height=12.0cm]{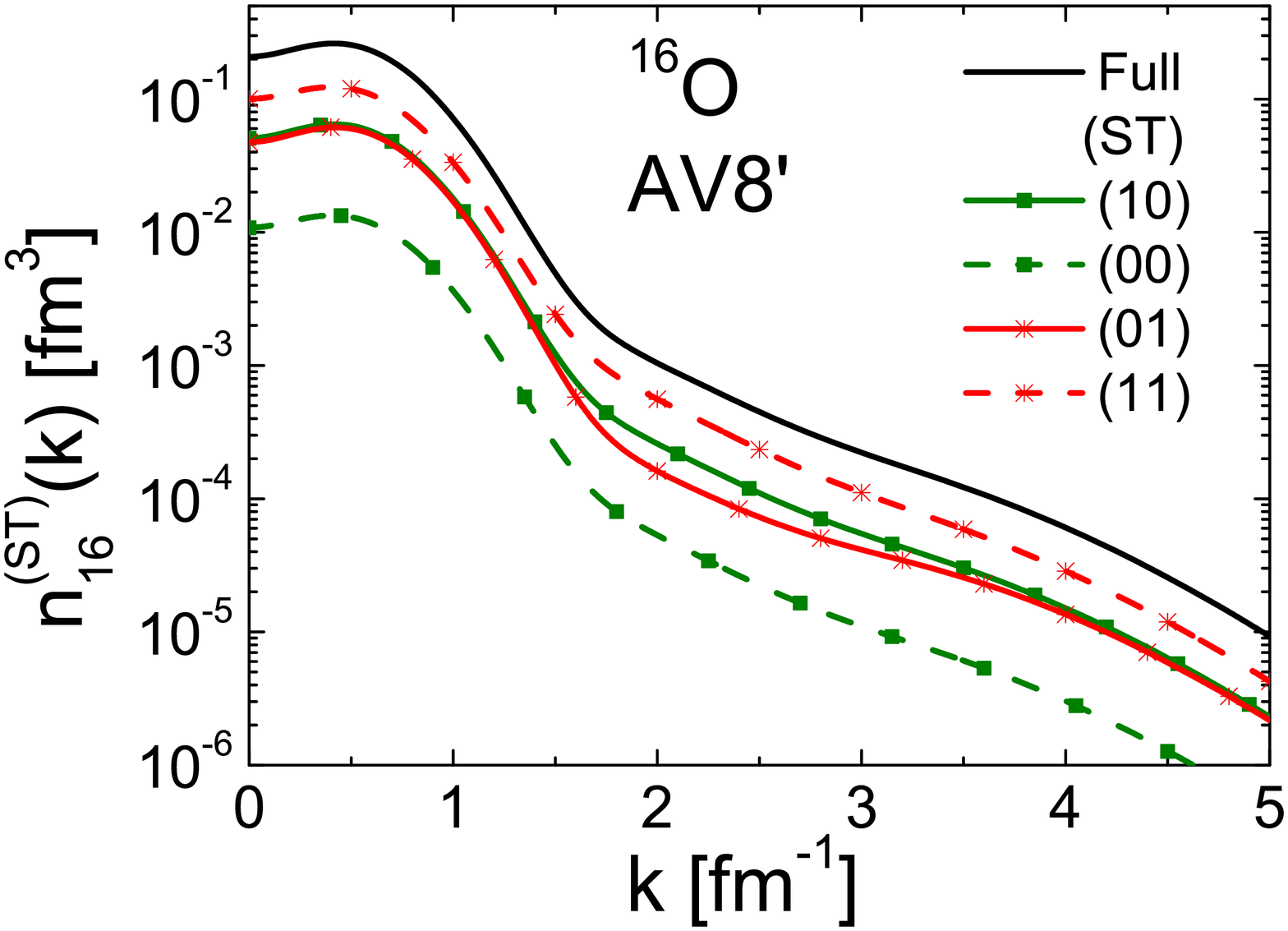}}
  \hspace{0.0cm}
  \caption{(Color online) The various spin-isospin contributions to  the momentum distribution of $^{16}$O
  (cf. Eq. (\ref{n16_2})).
    Wave function from Ref. \cite{Alvioli:2005cz}, AV8$^\prime$ interaction.}
  \label{Fig14}
\end{figure}

\begin{figure}[!htp]
  \centerline{
    \includegraphics[height=12.0cm]{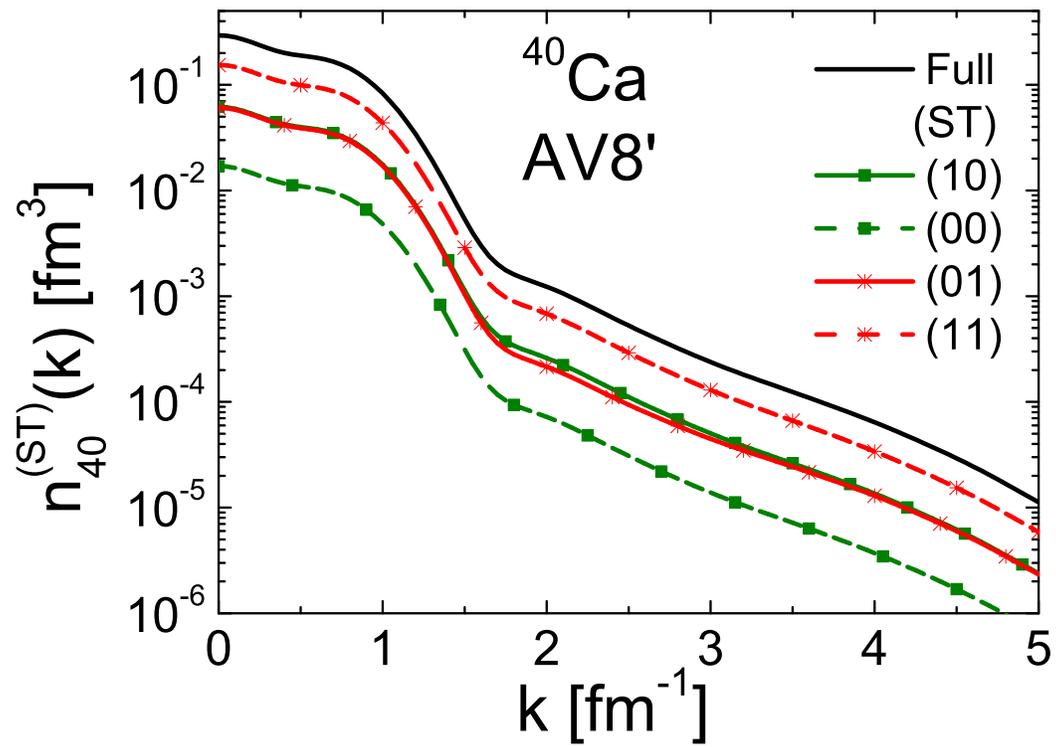}}
  \hspace{0.0cm}
  \caption{(Color online) The same as  in Fig. \ref{Fig14}, but for $^{40}$Ca (cf. Eq. (\ref{n40_2}).}
  \label{Fig15}
\end{figure}

We apply here Eq. (\ref{Momdisfin}) (with $\Vec{k}_1 \equiv \Vec{k}$), obtaining for
the proton  momentum distributions in $^3$He
\be
n_{3}^{p}(\Vec{k})&=&
\frac{3}{8}n_{T=0}^{pn}(\Vec{k})+
\frac{5}{8}n_{T=1}(\Vec{k})=
\label{n3p}\\
&=&
n_{3}^{p(10)}(\Vec{k})+n_{3}^{p(00)}(\Vec{k})+n_{3}^{p(01)}(\Vec{k})
+n_{3}^{p(11)}(\Vec{k}) \label{n3p1},
\ee because there is only
one $pp$ and one $pn$ pair containing proton $"1"$, whereas the
neutron distribution is given by
\be n_{3}^{n}(\Vec{k})&=&
\frac{3}{4}n_{T=0}^{pn}(\Vec{k}) +
\frac{1}{4}n_{T=1}(\Vec{k})=
\label{n3n}\\
&=&
n_{3}^{n(10)}(\Vec{k})+n_{3}^{n(00)}(\Vec{k})+n_{3}^{n(01)}(\Vec{k})
+n_{3}^{n(11)}(\Vec{k}) \label{n3n1},
\ee because there are two
$pn$ pairs containing neutron "1" and no $pp$ pairs.  The momentum distributions
of $^4He$, $^{16}O$ and $^{40}Ca$ are given, respectively, by
\be
n_{4}(\Vec{k})&=& \frac{1}{2}n_{T=0}^{(pn)}(\Vec{k}) + \frac{1}{2}n_{T=1}(\Vec{k})=
\label{n4_1}\\
&=&
n_{4}^{(10)}(\Vec{k})+n_{4}^{(00)}(\Vec{k})
 +n_{4}^{(01)}( \Vec{k})
+n_{4}^{(11)}(\Vec{k})
\label{n4_2},
\ee
\be
n_{16}(\Vec{k}_1)&=& \frac{3}{10}n_{T=0}^{(pn)}(\Vec{k}) + \frac{7}{10}n_{T=1}(\Vec{k})=
\label{n16_1} \\
&=&
n_{16}^{(10)}(\Vec{k})+n_{16}^{(00)}(\Vec{k})+n_{16}^{(01)}(\Vec{k})
+n_{16}^{(11)}(\Vec{k})
\label{n16_2},
  \ee
 \be
n_{40}(\Vec{k})&=& \frac{7}{26}n_{T=0}^{(pn)}(\Vec{k}) + \frac{19}{26}n_{T=1}(\Vec{k})=
 \label{n40_1} \\
 &=&
 n_{40}^{(10)}(\Vec{k})+n_{40}^{(00)}(\Vec{k})+n_{40}^{(01)}(\Vec{k})
 +n_{40}^{(11)}(\Vec{k})
  \label{n40_2},
  \ee
where Eqs. (\ref{enne1T1}) and (\ref{enne1T0}) have been used,
\beq
n_{T=1}^{(pn)}( {\bf k})=n_{T=1}^{(pp)}( {\bf k}
)\equiv n_{T=1}( {\bf k}),
\label{NT_1}
 \eeq
and
\beq
\int n_A(\Vec{k})\,d \Vec{k}=\int n_{T=0}^{pn}(\Vec{k})\,d \Vec{k}=
\int n_{T=1}(\Vec{k})\,d \Vec{k}=1.
\label{norme}
\eeq
The results of calculations of the spin-isospin
contributions to the
 momentum distribution of $^3$He, $^4$He,  $^{16}$O and
$^{40}$Ca,  are presented in Figs. \ref{Fig11}-\ref{Fig15}. The following
remarks are in order:
 (i) the contribution from the $(00)$ state is negligible
in both few-nucleon systems and complex nuclei; (ii) the $(11)$
state in $^3$He and $^4$He is small,
 both at low
and large values of $k$,  but it plays a relevant role in the region
$1.5 \lesssim k \lesssim 3\,fm^{-1}$; (iii) in the proton distribution of $^3$He
 (Fig. \ref{Fig11}) the $(01)$ contribution  is important  everywhere except in the
 region $1.5 \lesssim k \lesssim
3\,fm^{-1}$, whereas in the neutron distributions (Fig. \ref{Fig12}), thanks to the
different weight of the $(01)$ state ($1/4$ instead of $5/8$; cf. Eqs. (\ref{n3p})
 and
(\ref{n3n})), the contribution from this  state is much  smaller;
(iv) in complex nuclei  the
 $(11)$ state (odd relative orbital momenta) plays  a dominant role, both  in the
 independent particle  model and in the many-body approach
   (cf. Table \ref{Table3} and Figs. \ref{Fig14} and \ref{Fig15}).
Thus, in summary, we found that
all spin-isospin components, except the $(00)$ one, contribute to
the high-momentum content of the  momentum distributions and only
in the case of the neutron distribution in the non isoscalar
nucleus   $^3$He,  the deuteronlike state $(10)$ is the
dominant contribution.

\begin{figure}[!htp]
  \centerline{
    \includegraphics[height=12.0cm]{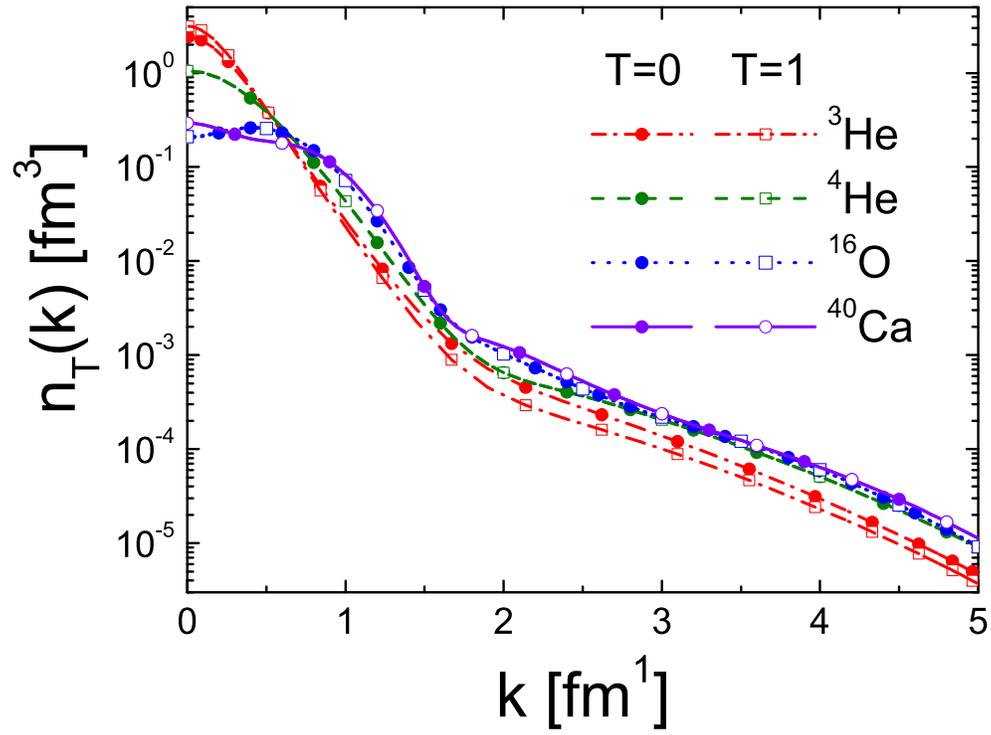}}
  \hspace{0.0cm}
  \caption{(Color online) The isospin $T=0$ and $T=1$ contributions to the proton
momentum distributions
  (Eqs. (\ref{enne1T0}) and (\ref{enne1T1})).}
  \label{Fig15a}
\end{figure}

To provide  further evidence of  the A independence of SRC,  we show in Fig.
\ref{Fig15a} the "elementary" quantities $n_{T=0}^{(pn)}(\Vec{k}_1)$
and $n_{T=1}(\Vec{k}_1)$ for different nuclei, and
it can be seen that, starting from
 $k \equiv|\Vec{k}_1| \simeq 2 fm^{-1}$ , they follow the same pattern.
\section{The momentum distributions of nuclei {\it vs}
the deuteron momentum distribution}\label{sec:sec5}
\begin{figure}[!htp]
\centerline{
\includegraphics[height=12.0cm]{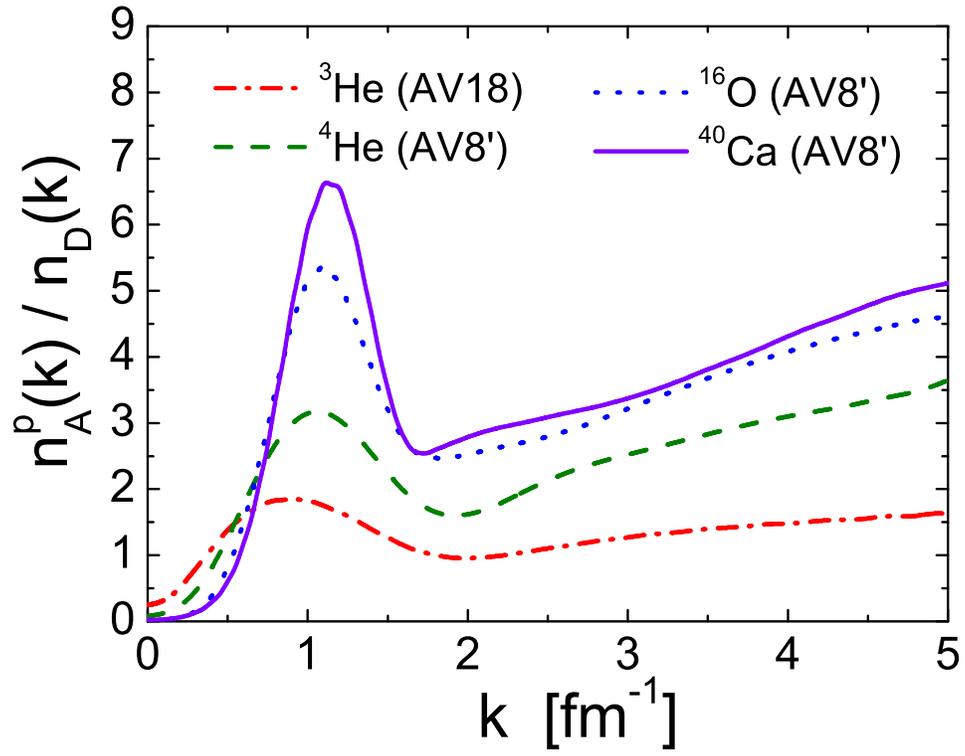}}
  \hspace{0.0cm}
  \caption{(Color online)
  The ratio of the proton momentum distribution of nucleus $A$ shown in the previous figures to the deuteron
  momentum distributions.}
  \label{Fig16}
\end{figure}

\begin{figure}[!htp]
\centerline{
\includegraphics[height=12.0cm]{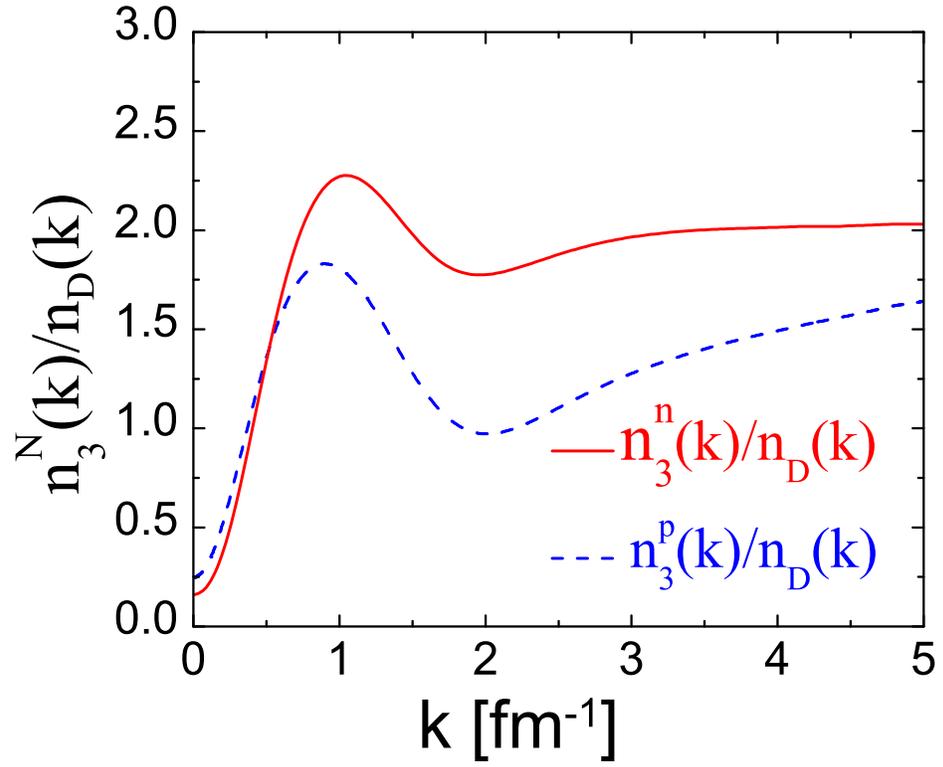}}
  \caption{
  (Color online) The ratio of the neutron, $n_3^n(k)$,  and proton,  $n_3^p(k)$, distributions in $^3He$ to
  the deuteron
  momentum distributions,  $n_D(k)$.}
  \label{Fig17}
\end{figure}

\begin{figure}[!htp]
\centerline{
\includegraphics[height=12.0cm]{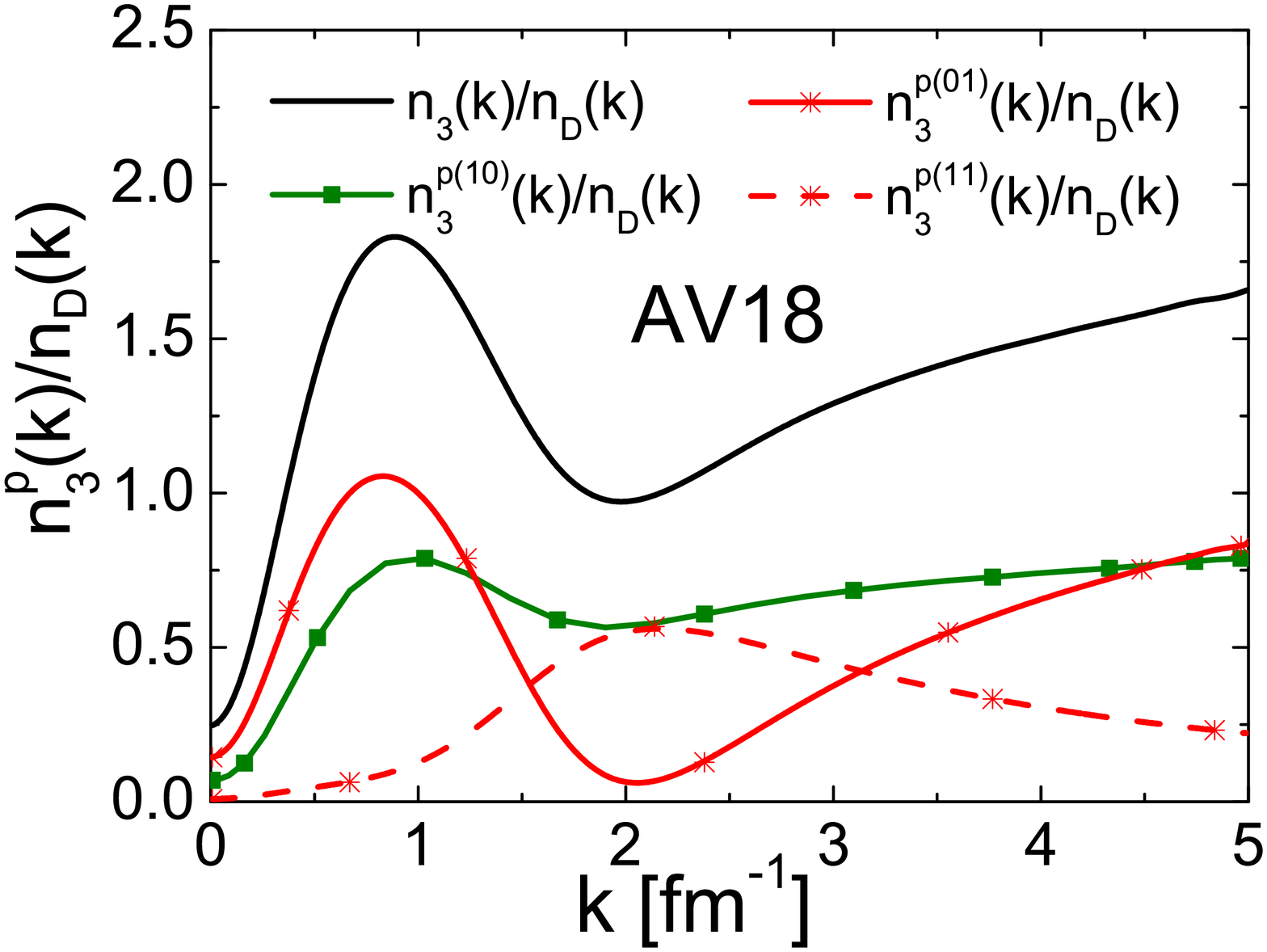}}
  \caption{(Color online) The various spin-isospin contributions to the ratio  of the
   proton  momentum distributions
    of \,$^{3}$He \, ~(Eq. (\ref{n3p1})) to the deuteron momentum distributions. Wave function from Ref.
    \cite{Kievsky:1992um}, AV18 interaction. }
  \label{Fig18}
\end{figure}

\begin{figure}[!htp]
\centerline{
\includegraphics[height=12.0cm]{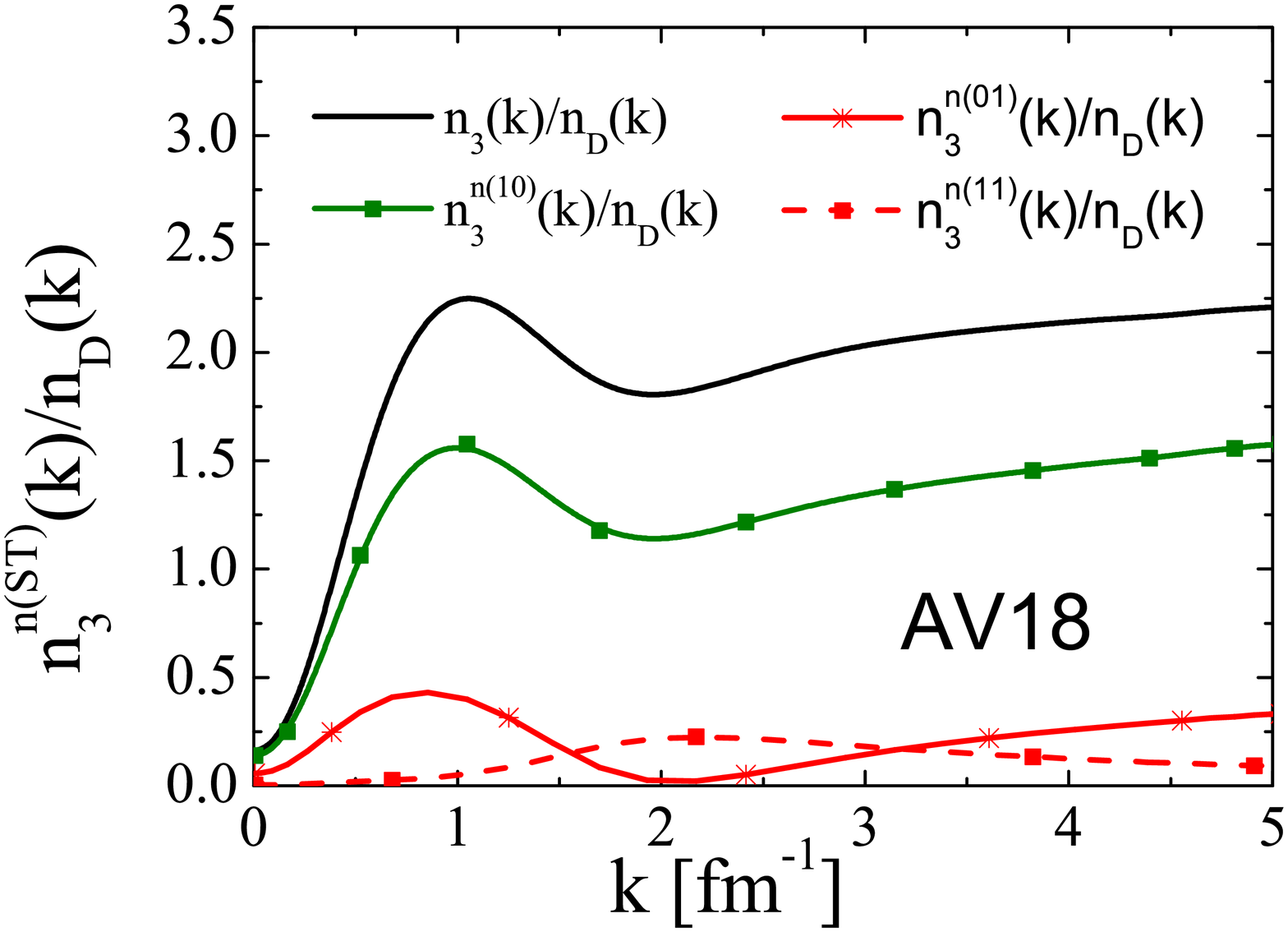}}
  \caption{(Color online) The same as in Fig. \ref{Fig18}, but for the neutron distribution.}
  \label{Fig19}
\end{figure}

\begin{figure}[!htp]
\centerline{
\includegraphics[height=12.0cm]{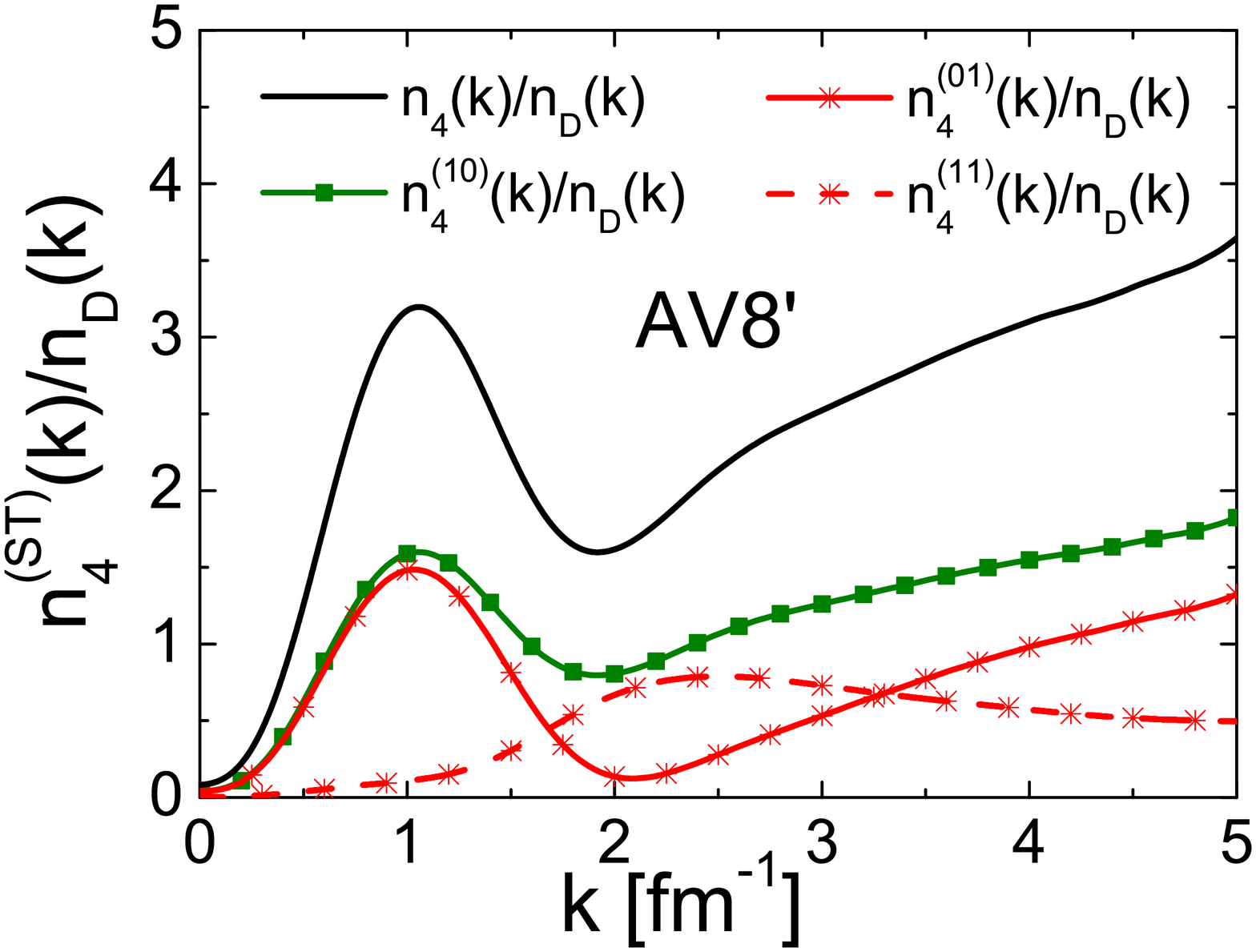}}
  \hspace{0.0cm}
  \caption{(Color online)  The various spin-isospin contributions to the ratio of the proton momentum distributions
    of $^{4}$He to the deuteron momentum distributions. Wave function from Ref.
    \cite{Kievsky:1992um}, AV8$^\prime$ interaction.}
  \label{Fig20}
\end{figure}

\begin{figure}[!htp]
\centerline{
\includegraphics[height=12.0cm]{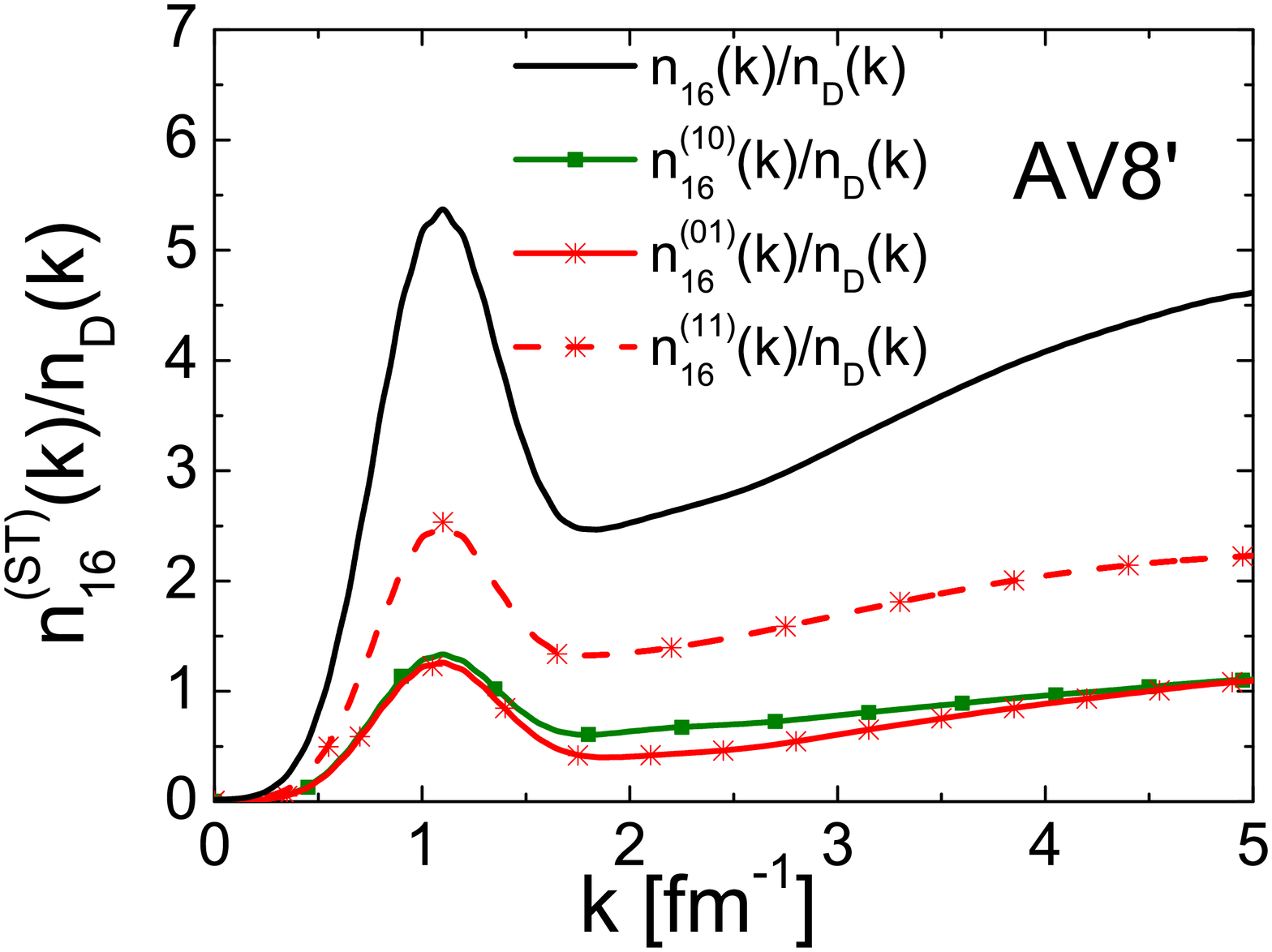}}
  \hspace{0.0cm}
  \caption{(Color online) The various spin-isospin contributions to  the  ratio of the  momentum distribution
    of $^{16}$O to the deuteron momentum distributions. Wave function from Ref. \cite{Alvioli:2005cz},
    AV8$^\prime$ interaction.}
  \label{Fig21}
\end{figure}

\begin{figure}[!htp]
\centerline{
\includegraphics[height=12.0cm]{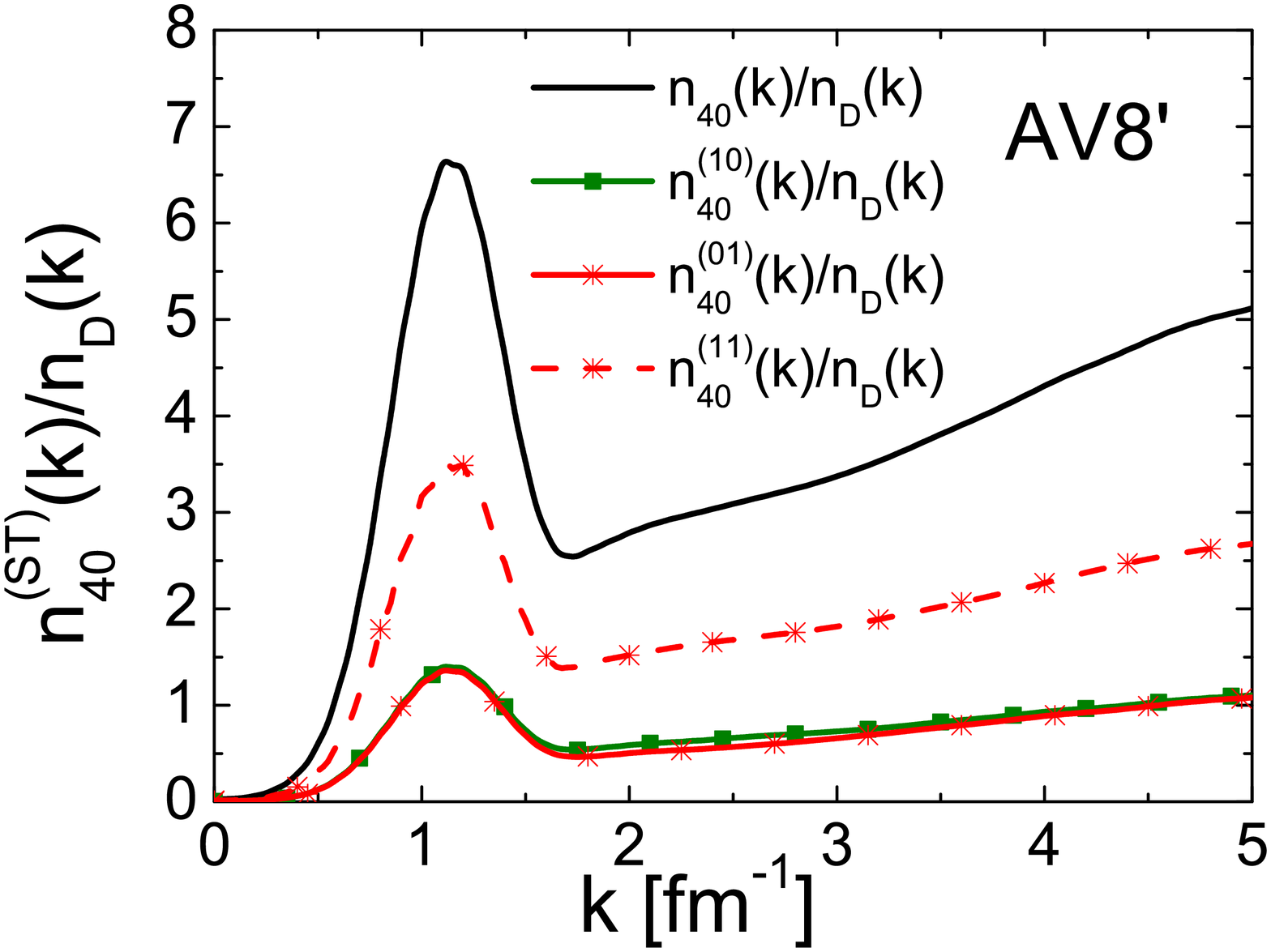}}
  \hspace{0.0cm}
  \caption{(Color online) The various spin-isospin contributions to  the  ratio of the  momentum distribution
    of $^{40}$Ca to the deuteron momentum distributions. Wave function from Ref. \cite{Alvioli:2005cz},
    AV8$^\prime$ interaction.}
  \label{Fig22}
\end{figure}

As it  clearly
appears in Fig. \ref{Fig6}, at $k\gtrsim 1.5-2\,fm^{-1}$ the
momentum distribution of  nuclei exhibits a trend similar to the
one of the deuteron\footnote{In the rest of the paper we frequently use the notation $^2H \equiv D$.}. However a quantitative analysis of the ratio
\beqy R_{A/D}(k)=\frac{n_A(k)}{n_D(k)} \label{rationanD} \eeqy
 is in order, because $n_A(k)$
 is usually  interpreted as the  scaled deuteron
momentum distribution, i.e.
$R_{A/D}(k)={n_A(k)}/{n_D(k)}\simeq const$. Such an
interpretation originated   long ago either from the use
of pioneering theoretical many-body calculations
\cite{Zabolitzky:1978cx,BenharCiofi,VanOrden:1979mt,Ji:1989nr} or
by assuming it as an input  for
the calculations of $n_A(k_1)$ at $k \geq k_F$ \cite{Schiavilla_old}
when variational Monte Carlo calculations were difficult to
perform at high values of the momentum, or
 by obtaining the momentum distributions from an average value of the $pn$ and $pp$
spectral functions \cite{CiofidegliAtti:1995qe}. Having nowadays at disposal
 advanced many-body calculations of the momentum distributions  performed with
realistic models of the two-nucleon interactions,
  a quantitative analysis of Eq. (\ref{rationanD}) is timing.
 To this end,   we show in Fig.~\ref{Fig16} the ratio $R_{A/D}(k)$  calculated with realistic
many-body wave functions. It clearly appears
that starting from $k \gtrsim 2\,fm^{-1}$, the ratio is not a constant but appreciably increases
 with $k$.
 Let us discuss the origin of such an increase. A first possible origin should be sought
 in the different role played by
 $pn$ and $pp$  correlations. As a matter of fact,
 the proton and neutron momentum distributions in $^3$He shown in Fig. \ref{Fig17},
 exhibit a different rate of increase, which  can qualitatively be
 understood in terms of SRCs as follows:
  in $^3$He the proton momentum distribution is affected by SRCs acting in one
 $pn$  and one $pp$ pairs, in the former pair the deuteronlike state $(10)$ is three
 times larger than the $(01)$ state,  whereas in the latter pair the
 deuteronlike state is totally missing; on the  contrary, the neutron distribution
  is affected by SRCs acting in two  proton-neutron pairs, with a pronounced dominance of the
  deuteronlike state $(10)$; therefore,
  one expects  that
  around $k \simeq 2\,fm^{-1}$,  where,
$np$ SRC  dominate over
 $pp$ SRC \cite{Schiavilla_old,Alvioli:2007zz},  $n_3^n/n_D \simeq 2$ and
  $n_3^p/n_D \simeq 1$, which  indeed seems to be the case. However, other effects
  of different origin can contribute to the deviation of  the ratio  $n_A(k)/n_D(k)$
 from a constant.
  These are attributable to the c.m. motion of a $pn$ pair in a nucleus,
to the different
  role played by
  the states $(01)$ and $(11)$ in different nuclei, and, particularly,  to the fact
  that being the one-nucleon momentum distribution
  the integral of the two-body distribution over $\bf{k}_2$, $n_{A}({\bf k}_1)=\int
   n_A({\bf k}_1,{\bf k}_2)d\,\Vec {k}_2$,   it may contain configurations different
   from
  the deuteron one (back-to-back nucleons).
  To better investigate these possibilities, let us  consider
  the spin-isospin ratio
\beqy
R_{A/D}^{(ST)}(k)=\frac{n_{A}^{(ST)}(k)}{n_D(k)}
\label{STratio}
\eeqy
  which is shown in  Figs.  \ref{Fig18}-\ref{Fig22} (note that, as stressed in the caption of the figures \ref{Fig18},
 the quantity $n_A^{(ST)}$ includes the proper coefficients which multiply
 the "elementary" quantities $n_{T}^{N_1N_2}$ ).
  It can be seen that
 the behavior  of the proton and neutron ratios for $^3$He clearly shows that
in the region $1.5 \lesssim k \lesssim 3$ fm$^{-1}$
 the former is  governed  by the
$(01)$ state in the $pp$ and $pn$ pairs; on the contrary, the neutron ratio
is fully dominated by the deuteronlike $(10)$ state in the two $pn$ pairs.
 The most interesting ratio is  $R_{A/D}^{(10)}(k)=n_{A}^{(10)}(k)/n_D(k)$,
because it  provides  information on the behavior of
 the deuteronlike pairs in  nuclei;
 it can
be seen that in the region of SRCs ($k\gtrsim 2$\,fm$^{-1}$)
  $R_{A/D}^{(10)}$
 increases with increasing value of $k$, with a different  rate
 of increase  for different nuclei: it is about $30 \%$ in the neutron momentum
 distribution of $^3He$,
and of the order of $100 \%$
in  other nuclei.  As already pointed out,
the increase of the ratio $R_{A/D}^{(ST)}(k)$ with $k$ could also  be attributable to the c.m.
motion of the pair in the nucleus. To take this into account is no easy task.
As a matter of fact, consider the simple case when the $(10)$
two-body momentum distribution factorizes in the following form \cite{CiofidegliAtti:1995qe}
\be
n^{pn}_{(10)}(\Vec{k}_{1},\Vec{k}_{2})=n^{pn}_{10}(\Vec{k}_{rel},\Vec{K}_{c.m.})
=n^{pn}_{10}({k}_{rel},{K}_{c.m.},\theta)\simeq
n_D(k_{rel})n_{c.m.}^A({K}_{c.m.}),
\label{2BMD_fact}
\ee
where $n_{c.m.}({K}_{c.m.})$, calculated from a many-body approach in
Ref. \cite{Alvioli:2012aa},  can be approximated by a $0S$ wave function.
In Refs. \cite{Alvioli:2012aa} and \cite{Scopetta},  Eq. (\ref{2BMD_fact})
 has indeed been
 shown to hold, but only in a restricted region of $k_{rel}$ and $K_{c.m.}$, namely
 \be
K_{c.m} \lesssim 1.0-2.0\,fm^{-1} \,\,\,\,\,\,\,\, k_{rel}
\gtrsim k_{rel}^{-}=f_A(K_{c.m.})
\label{Relation}
\ee
where the function $f_A$ depends upon    $K_{c.m.}$  and $A$  in  such a way that
the value of
$k_{rel}^{-}$ increases with increasing values of  $K_{CM}$. Thanks to momentum
conservation $\Vec{k}_{2}= -(\Vec{k}_{1}+ \Vec{K}_{c.m.})$, one can write
\be
n^{pn}_{(10)}({k}_{1})\simeq
\int n_D(|\Vec{k}_1-\frac{\Vec{K}_{c.m.}}{2}|)\,
 n_{c.m.}(|\Vec{K}_{c.m.}|)\,d\,\Vec{K}_{c.m.}
\label{Convolution}
\ee
which  shows that only in the case of  a $pn$ pair at rest, i.e.
$n_{c.m.}(\Vec{K}_{c.m.})=
 \delta(\Vec{K}_{c.m.})$,   one has  $n^{pn}_{10}({k}_{1})\simeq
 n_D({k}_1)$ $R_{A/D}^{(ST)}(k_1) \simeq const$. The convolution of the deuteron momentum distributions with the c.m. motion leads
 to an increase of $n^{pn}_{10}({k}_{1})$, whose magnitude and rate of increase depend
  upon
 the detailed forms of $n^{pn}_{10}({k}_{rel})$ and $n_{c.m.}({K}_{c.m.})$;
 moreover, because, as already stressed,  the one-body momentum distribution is the integral of the two-body momentum
 distribution,  configurations different from the factorized one
  (Eq. (\ref{2BMD_fact}))
 can contribute to the integral  (Eq. (\ref{Convolution})).

\subsection{On the short-range deuteronlike configurations in nuclei}
\begin{figure}[!htp]
\centerline{
\includegraphics[height=12.0cm]{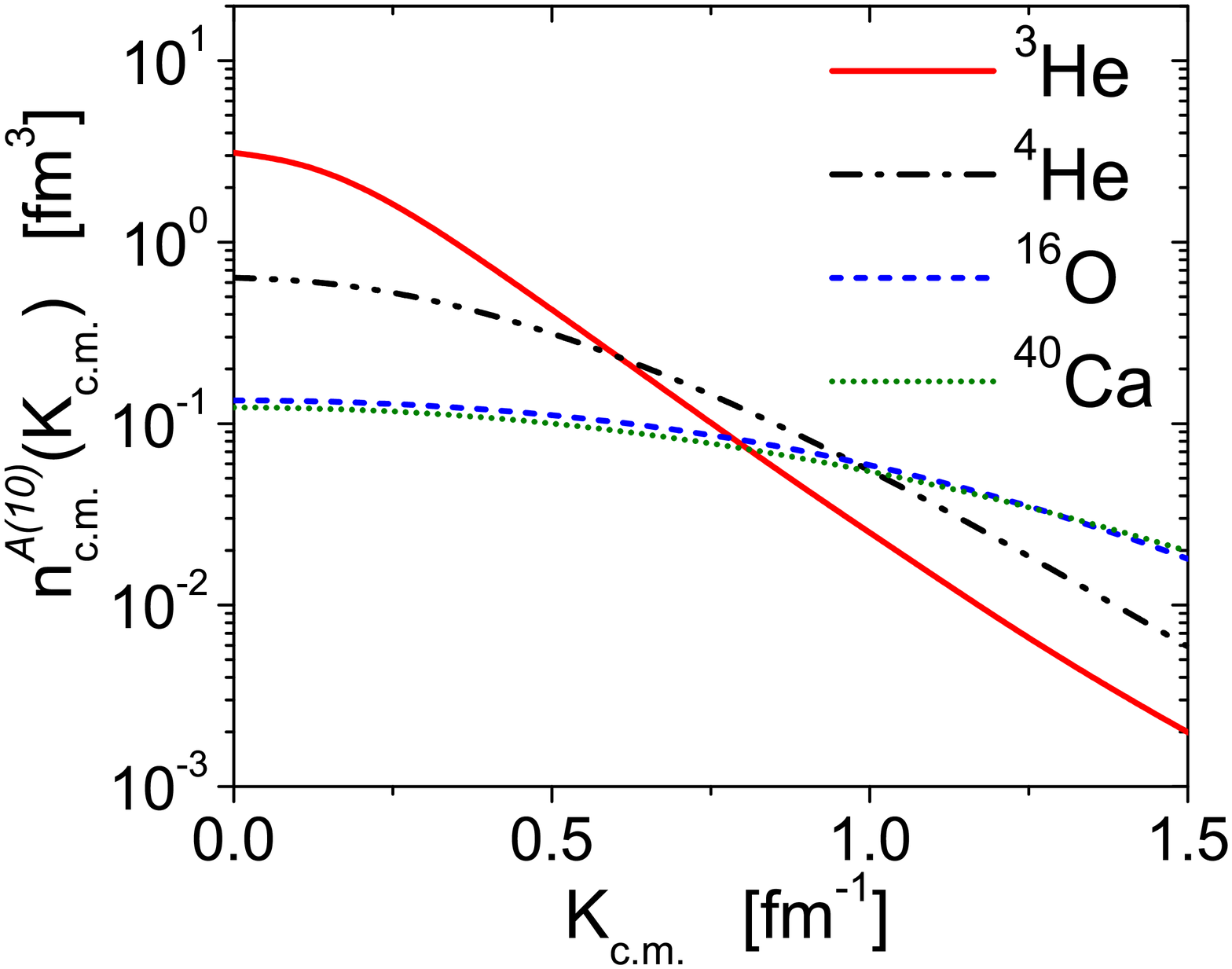}}
  \hspace{0.0cm}
  \caption{(Color online) The  c.m. momentum distribution in the state
  $(10)$ $n_{c.m.}^{(10)}({K}_{c.m.})$ in
   $^3$He, $^4$He, $^{16}$O and $^{40}$Ca.}
  \label{Fig23}
\end{figure}

 A particular useful quantity to understand SRC in nuclei is the one that
 is obtained by integrating
 the two-nucleon momentum
 distribution  of the state $(10)$ in a narrow range of the c.m. momentum
 ($K_{c.m.} \lesssim
 1-1.5\,fm^{-1}$), when the c.m. and
 relative motions are decoupled,  and Eq. (\ref{2BMD_fact}) is satisfied \cite{Alvioli:2012aa}, namely
\beqy
n_{D/A}^{pn}(k_{rel})=\int n^{pn}_{(10)}({k}_{rel},{K}_{c.m.},\theta)
\,d\,\Vec{K}_{C.M.}\simeq n_D(k_{rel})\,4\pi
\int_{0}^{K_{c.m.}^{+}}n_{c.m.}^A({K}_{c.m.}) \,K_{c.m.}^2d\,{K}_{c.m.}
\label{momdisD/A} \eeqy
 In Ref. \cite{Egiyan:2005hs} the $2N$ SRC probability in the deuteron
 has been defined as the integral of the deuteron momentum distribution in the range
 $k_{rel}\gtrsim 1.5$  fm$^{-1}$ (cf.  Table
 \ref{Table4}), therefore we can consider as the analog  in a nucleus the quantity
 \beqy
\mathcal{P}_{D/A}=4\pi\int_{1.5}^{\infty}\,n^{pn}_{D/A}(k_{rel}){k}_{rel}^2\,d\,{k}_{rel}
\label{nD/A}
 \eeqy
  where $n_{D/A}^{pn}$  is given by Eq. (\ref{momdisD/A}). We can also define
 the total number of quasi-deuteron short-range correlated pairs
 as follows
\beqy N_{D/A}=N_{(10)}^{A}\mathcal{P}_{D/A} \label{ND/A} \eeqy
 where the number
of $N_{(10)}^A$ pairs is listed in Table \ref{Table3}. The calculated  values of the partial
probability ${\mathcal P}^{N_1}=4\pi\int_{1.5}^{\infty}[n_0(k)+n_1(k)]\,k^2\,d{k}$
(Eq. (\ref{partialprob1}),
predicted by different
$NN$ interactions, is shown in Table \ref{Table4}, and the quantities
$\mathcal{P}_{D/A}$ and $N_{D/A}$ are given in Table \ref{Table5}. Because
${\mathcal P}^{N_1}$ includes all spin-isospin components and momentum configurations,
whereas only  deuteronlike configurations
[$(ST)=(10)$ and  ${\bf k}_1=-{\bf k}_1$]  are included in $\mathcal{P}_{D/A}$,
our result  $\mathcal{P}_{D/A} <  {\mathcal P}
^{N_1}$
is fully justified. Moreover,  the decreasing behavior
of $\mathcal{P}_{D/A}$ with $A$ can easily be understood as owing
to the increasing importance of higher  c.m. momentum components
 of the pair,  resulting in  flatter c.m.
 distributions in heavier nuclei (cf. Fig.
\ref{Fig23}), so that  only a smaller part of the distribution is included in the
integral over $K_{c.m.}$.
We have also considered  the quantity $a_{D/A}$,
 the per-nucleon probability of
deuteronlike  configurations in $A$ with respect to the probability of SRCs in the
deuteron ($\simeq 0.04$). Our values for $A <40$ are less than the values of $a_2$ extracted
from the $A(e,e')X$ experiments \cite{Egiyan:2005hs,Frankfurt:1993sp,
Fomin:2011ng}; however such a comparison is perhaps a premature one, because,
from one side,
 nondeuteronlike
configurations which occur outside the factorization region should be considered in the
theoretical calculation (e.g. the c.m. motion of the pair \cite{Fomin:2011ng,Vanhalst:2011es}), and, from the other side, a careful investigation of the effects
of FSI effects on the extraction of $a_2$ from the inclusive $A(e,e')X$
cross-section ratio should also be considered.  We should also mention, in this respect, that the values of
$a_2$ were also recently calculated in Ref. \cite{Vanhalst:2011es} within an approach
in which  only  $L=0$ pairs prone to SRCs were considered (cf. Table \ref{Table3}),
obtaining results that coincide with the ones obtained in the present
paper for $A=3,4$,
and which are
lower  for $A>4$.

\section{Summary and conclusions}\label{sec:sec6}

Recently,  several A-independent features of  SRCs in few-nucleon systems
($^2$H, $^3$H, $^3$He, and
 $^4$He) have been demonstrated  by calculating the dependence of two-body
 momentum distributions
 upon the relative momentum  $|{\bf k}_{rel}|\equiv k_{rel}$   of
 the correlated pair \cite{Feldmeier:2011qy}, as well
 as upon the c.m. momentum
 $|{\bf K}_{c.m.}|\equiv K_{c.m.}$   and the angle
 between ${\bf K}_{cm}$ and ${\bf k}_{rel}$ \cite{Alvioli:2012aa}. These calculations have
 been performed with exact wave functions resulting from the solution of the nonrelativistic Schr\"{o}dinger equation, using modern bare $NN$ interactions, featuring strong
 short-range repulsion and intermediate range tensor attraction, e.g.
   the Argonne-Urbana
 models.
In the present paper, using the same many-body approach and interactions,  we have addressed the  problem of the effects
of SRCs, and their spin-isospin components, on the one-nucleon
momentum distributions  $n_A(k)$ of few-nucleon systems and complex nuclei.
The momentum distribution, besides being
 {\it per se} a relevant quantity in nuclear theory,   plays a relevant
role in the interpretation of various experimental data, in particular in
inclusive experiments of lepton scattering off nuclei at medium and high energies.
 Using the proper diagonal and non diagonal one- and two-body
spin-  and isospin- dependent density matrices, we have derived
in Sec. \ref{sec:sec2}
the expression of the
 momentum
 distributions of a nucleon belonging to a $NN$ pair in a state with
total spin S and  isospin T.  In Sec. \ref{sec:sec3} we have presented some general
concepts concerning nucleon momentum distributions and a clear-cut way to separate
 them in
 mean-field and SRC contributions, and
 have analyzed the results of the most recent calculations of the momentum
distributions for nuclei with
A=2, 3, 4, 16 and 40,
performed within realistic many-body approaches and modern $NN$ interactions.
The aim was to ascertain  whether some general
 features
of the momentum distributions
could be established within the solution of the nuclear many-body problem, in terms of realistic
bare $NN$ interactions.
The results of our
  analysis have shown indeed that, even if quantitative differences are provided by
 different interactions and many-body approaches,  the following general
 features of
the momentum distributions can be singled out, namely: (i) at
 $ k \lesssim
1-1.5\, fm^{-1}$, the mean-field approach  dominates  the distributions, with a
resulting
sizeable $A$ dependence; (ii) at larger
values of $k$, of the order of $2\,fm^{-1}$, owing to the effects of SRCs,
the momentum distributions
 abruptly change their slope, and, apart from an A-dependent scaling factor, exhibit a
  $k$ dependence  which is very similar in different nuclei;
(iii) the correlated part of the momentum distribution is  by orders
of magnitude larger than the predictions of any mean-field approach, so that
 experiments providing  even  rough information on high-momentum
components would be able to rule out  mean-field predictions.
Similar conclusions,  reached in the past by phenomenological
calculations  (see e.g. Refs. \cite{CiofidegliAtti:1995qe} and \cite{Bohigas:1979kk}),
are therefore quantitatively confirmed by the present systematic  analysis.
 After having checked   that  the evaluation of the high-momentum part of $n_A(k)$
 is well
 under control, we turned  in Sec. \ref{sec:sec4} to the calculation
 of
the spin-isospin structure of the momentum distributions.
First of all we calculated the number of $NN$ pairs in various spin-isospin states
in different nuclei,
both within the independent particle models and in many-body approaches embodying
SRCs, finding agreement with calculations  performed by different groups,
confirming that SRCs have very small effects on the number
of isosinglet pairs in state (10), unlike what happens with isotriplet pairs
in state (01), whose number is decreased in favor of the pairs in (11) state.
We have calculated the contribution of the states
(ST)=(10),\,(00),\, (01) and (11)  to the momentum distributions, finding that all of them,
except the state (00), have comparable effects  in a wide range of
momentum. The contribution of the isosinglet state T=0 is almost entirely exhausted
by the (10) state, whereas both states (01) and (11) contribute to the isotriplet state
T=1. We found that at momentum values $k\gtrsim 2\,fm^{-1}$, the contribution of both
isosinglet and isotriplet states follow the same pattern, independently of A,
which represents further
 evidence
of the general scaling behavior of SRCs. A systematic and
quantitative
comparison of  $n_A(k)$, and its
spin-isospin components   $n_A^{ST}(k)$,
with the deuteron momentum distribution $n_D(k)$, has been presented in
 section \ref{sec:sec4}, by analyzing the ratios
$n_A^{ST}(k)/n_D(k)$. We found  that in the region of SRCs, $k\gtrsim 2\,fm^{-1}$, this
ratio does not stay constant but increases with increasing $k$, and  interpreted
such a behavior  as owing
to the presence in the momentum distribution  of  two-nucleon momentum
configurations  arising from the
c.m. motion of a pair  and   differing from the
back-to-back nucleons configuration.
  Our spin-isospin dependent approach allowed us to calculate also
(i) the
relative momentum
distribution of a proton-neutron pair moving with small c.m. momentum and
its integral in the range  $1.5 <k<\infty$, a quantity which is assumed to
represent the probability of two-nucleon  SRCs in a nucleus, finding similar values
($\simeq 0.04$) in a wide range of A, namely
$ 2 \leq A \leq 40 $; (ii) the total number of SRCs pairs in (10) state,
interpreting its A- dependence in terms of the A- dependence
of the c.m. momentum distribution; (iii)
 the per-nucleon probability of two-nucleon deuteronlike SRCs in nuclei, a quantity which is
under active experimental investigation. In closing this paper, we
would like to stress that the properties of SRCs we have found depend obviously upon
the wave function we have used to calculate the density matrices and momentum distributions.
In case of  A=2, 3, and 4 systems  the ground-state wave functions represent  the {\it ab initio} solution of the many-body nonrelativistic Schr\"{o}edinger
equation given in terms of modern bare $NN$ interactions, whereas, in the case of complex nuclei, they represent the variational
solution of the same equation. The high-momentum content of the ground-state wave function
will obviously depend upon the used $NN$ interaction.
In this respect it should be recalled that
phase shift data characterizing  elastic on-shell $NN$ scattering do not determine
uniquely the details of the short-range interaction;  moreover, in a many-body
bound nuclear
systems, two interacting nucleons that experience  interaction with surrounding
 nucleons are off shell; i.e. their energy is not related to their relative momentum,
 with the resulting complication that the off-shell behavior of the interaction cannot
 be determined uniquely
 from elastic phase shifts. As a result, a family of different phase-equivalent
 potentials
 can be derived (see, e.g. \cite{AV18,Machleidt,Epelbaum,Polyzou}) producing different
 high-momentum contents of the many-body nuclear wave functions. This fact points to the importance  of the
 investigation of the high-momentum part of the nucleon momentum distributions (see e.g. \cite{Vary}).
At the same time, it should also  be stressed that the interaction we have used
  (e.g. the AV18 or/and AV8$^\prime$ ones)
 are currently being used in that class of successful {\it  ab initio} many-body calculations (e.g.
  the Unitary Correlation Operator Method (UCOM)
 \cite{Roth:2010bm} and  the no-core shell-model approach \cite{Nocore_SM})  where  various
 renormalization groups (RG) methods \cite{RG} are used to soften the short-range and tensor interactions
  of the original
 bare interaction, so as to improve the convergence
 of the diagonalization of the many-body Hamiltonian. As a results  the
 finally  evolved ground-state wave function exhibits a low degree of  SRCs.

 It would appear that these methods
 are in conflict  with the traditional direct solution of the many-body Schr\"{o}dinger equation
  with bare  $NN$ interaction,
 producing  ground-state wave functions containing a large degree of SRCs, arising from the strong short-range
 repulsive  and the
 intermediate-range attractive tensor forces. This however is not the case, as discussed in two recent papers
  \cite{Anderson,Bogner} (see also Ref.
 \cite{Feldmeier:2011qy}), stressing the necessity to evolve, together with the $NN$ interaction,
 also the momentum distribution operator. Preliminary results for the two-body system \cite{Anderson},
 and Fermi and electron gases \cite{Bogner},
 show indeed that the  high-momentum content of the momentum distributions, and their scaling behavior stressed in
 the present and many other papers can also be predicted within low-momentum
  effective theories.


\section{Ackowledgments}

We thank the Pisa group for providing us with  the code for the calculation of deuteron
 momentum
distributions and the three-nucleon wave function.
 H. M. is grateful to INFN, Sezione di Perugia for warm
hospitality.
 We thank CASPUR for the grant SRCnuc3 - \textit{Short-Range Correlations in nuclei}, within the Standard
HPC grants 2012 programme.


\clearpage
\begin{table}[!ht]
\begin{center}
{\renewcommand\arraystretch{1.3}
\begin{tabular}{|c||c||c|c|} \hline
\multicolumn{4}{|c|}{MEAN FIELD AND CORRELATION PROBABILITIES} \\ \hline
$Nucleus$ &    $Potential$ &    \,\,\,$\mathcal{P}_{gr}$    & $\mathcal{P}_{ex}$  \\
\hline
\multirow{1}{*}{$^{3}He$ \cite{Kievsky:1992um}}        & AV18 \cite{AV18}   &  0.677 & 0.323  \\ \hline
\multirow{1}{*}{$^{4}He$ \cite{Akaishi,Schiavilla_old}} & RSC \cite{RSC} AV8$^\prime$\cite{AV8} & 0.85 & 0.15 \\
\hline
 $ Nucleus $  &  $ Potential$  &   $\mathcal{P}_0$   &   $\mathcal{P}_1$    \\ \hline
\multirow{1}{*}{$^{16}$O \cite{Pieper}} & V8' \cite{AV8} & 0.8  & 0.2   \\ \hline
\multirow{1}{*}{$^{40}$Ca \cite{Alvioli:2012aa}} & V8' \cite{AV8} & 0.8  & 0.2   \\
\hline
\hline
\end{tabular}}
\end{center}
\caption{The proton mean field and correlation probabilities $\mathcal{P}_{gr(0)}^p=\int  d\,{\bf k}_1 \, n_{gr(0)}^{p}({\Vec k}_1) $ [Eqs. (\ref{essegr}) and (\ref{essezero})]
  and $\mathcal{P}_{ex(1)}^p=\int  d\,{\bf k}_1  \, n_{ex(1)}^{p}({\Vec k}_1)$ [Eqs. (\ref{esseex}) and (\ref{esseuno})].}
 \label{Table1}
\end{table}
\begin{table} [!h]
\normalsize
\begin{center}
{\renewcommand\arraystretch{1.5}
\begin{tabular}{|c||c|c|c|c|c|c|c|c|c|c|}
\hline
 &$^2H$&{$^3$He(n)}&\multicolumn{2}{c|}{$^3$He(p)}&\multicolumn{2}{c|}{$^4$He}&\multicolumn{2}{c|}{$^{16}$O}&\multicolumn{2}{c|}{$^{40}$Ca}\\
\hline
 $k_1^-$ [fm$^{-1}$]&${\mathcal{P}}$& $\mathcal{{P}}_{1}$&$\mathcal{{P}}_{0}$ & $\mathcal{{P}}_{1}$& $\mathcal{{P}}_0$ & $\mathcal{{P}}_1$ & $\mathcal{{P}}_0$ & $\mathcal{{P}}_1$ & $\mathcal{{P}}_0$ & $\mathcal{{P}}_1$\\
\hline  
$0.00$ &$1.000$ &   $ 0.999 $     & $0.677$ &  $0.323$       & $0.84621$ & $0.15285$    & $0.79999$ & $0.20016$    & $0.80$ & $0.19321$\\
\hline
$0.50$ & $0.3078$ &  $0.568  $&  $0.277$  &   $0.201$   & $0.53643$ & $0.14032$    & $0.66972$ & $0.19635$          & $0.69997$ & $0.18301$\\
\hline
$1.00$ & $0.081$  &  $0.163 $&  $0.038  $ &  $0.0723 $  & $0.10479$ & $0.1045$   & $0.17588$ & $0.14794$    & $0.24706$ & $0.13771$\\
\hline
$1.50$ & $0.0366$  &  $0.067 $&  $0.0049 $ &  $0.036$    & $0.0079$ & $0.0791$   & $0.00792$ & $0.09417$      & $0.01022$ & $0.10143$ \\
\hline
$2.00$ & $0.0221$  & $0.041$ & $0.0015$ &  $0.024$  & $6.9512 \:10^{-4}$ & $0.06156$  & $5.9\:10^{-5}$ & $0.06344$  & $3.28\:10^{-4}$ & $0.07124$  \\
\hline
\hline
\end{tabular}}
\caption{The values of the partial probability, Eq. (\ref{partialprob1}),
for $^{3}$He, $^{4}$He, and $^{16}$O and $^{40}$Ca, calculated for different values of
the
momentum $k_1^-$ with  $k_1^+=\infty$.}
\label{Table2}
\end{center}
\end{table}
\begin{table}[h]
\begin{tabular}{|c||c||c|c|c|c|}
\hline
 \multicolumn{2}{|c||}{}&\multicolumn{4}{c|}{(ST)}\\
\cline{3-6}
 \multicolumn{2}{|c||}{Nucleus}&(10)& (01)& (00)& (11)\\
\hline
$^2$H& & 1 & - & - & - \\
\hline
\multirow{3}{*}{$^3$He}
& IPM & 1.50 & 1.50 & - & - \\
& SRC (Present work) & 1.488 & 1.360 &0.013 & 0.139 \\
& SRC \cite{Forest:1996kp}& 1.50 & 1.350 &0.01  & 0.14  \\
& SRC \cite{Feldmeier:2011qy} & 1.489 & 1.361 & 0.011 & 0.139 \\
\hline
\multirow{5}{*}{$^4$He} & IPM & 3 & 3 & - & - \\
& IPM($0s$ states) \cite{Vanhalst:2011es}& 3 & 3 & - & - \\
& SRC (Present work)& 2.99 & 2.57 & 0.01 & 0.43 \\
& SRC \cite{Forest:1996kp}&  3.02  & 2.5 & 0.01& 0.47\\
& SRC \cite{Feldmeier:2011qy}& 2.992 & 2.572 & 0.08 & 0.428 \\
\hline
\multirow{5}{*}{$^{16}$O} & IPM & 30 & 30 & 6 & 54\\
& IPM($0s$ states) \cite{Vanhalst:2011es}& 20 & 18 & - & - \\
& SRC(Present work) & 29.8 & 27.5 & 6.075&  56.7 \\
& SRC \cite{Forest:1996kp}& 30.05 & 28.4 &6.05  & 55.5\\
\hline
\multirow{3}{*}{$^{40}$Ca} & IPM & 165& 165& 45 & 405 \\
& IPM($0s$ states) \cite{Vanhalst:2011es}& 90 & 20 & - & - \\
& SRC(Present work) & 165.18 & 159.39 & 45.10 & 410.34 \\
\hline
\hline
\end{tabular}
\caption{The number of pairs $N_{(ST)}^A$, Eq. (\ref{Norm1new}), in various
  spin-isospin states in the independent particle model (IPM) and taking into account
  SRCs  within many-body theories with realistic interactions (in the approach of Ref. \cite{Vanhalst:2011es} pairs in relative $L=0$ motion were identified as those prone to SRCs).}
  \label{Table3}
\end{table}
\begin{table}[!ht]
\begin{center}
{\renewcommand\arraystretch{1.3}
\begin{tabular}{|c||c|c|c|c|c|c|} \hline
\multicolumn{7}{|c|}{ $2N$ SRC PROBABILITY} \\ \hline
$NN Interaction$ &  $^2$H &$^3$He(n) & $^3$He(p) & $^4$He & $^{16}$O & $^{40}$Ca\\
\hline
RSC & $0.04$ & - & - & $0.09$ & $0.12$ & -\\
\hline
AV14& $0.036$ & - & - & $0.11$ & $0.14$ & -\\
\hline
AV8$^\prime$& $0.036$ & - & - & $0.09$ & $0.10$ & $0.10$\\
\hline
AV18 & $0.037$ &  $0.067$  &  0.041  & $-$ & $-$ & -\\
\hline
CS & $0.033$  & 0.079 &$0.046$ & $0.09$ & $0.10$ & 0.14\\
\hline\hline
\end{tabular}}
\caption{The value  of the $2N$  SRC partial
probability (Eq. (\ref{partialprob1}))
in the  deuteron,
$4\pi\int_{1.5}^{\infty}n_D(k)\,k^2\,d{k}$,  and in complex nuclei
 $4\pi\int_{1.5}^{\infty}[n_0(k)+n_1(k)]\,k^2\,d{k}$ (cf. Table \ref{Table2})
obtained with momentum distribution resulting from many-body calculations
performed with  different
$NN$ interactions. The result (CS) of
the phenomenological model of Ref. \cite{CiofidegliAtti:1995qe} is also shown.}
\label{Table4}
\end{center}
\end{table}

\begin{table} [!h]
\vspace{-2.5cm}
\normalsize
\begin{center}
{\renewcommand\arraystretch{1.3}
\begin{tabular}{|c||c|c|c|c|c|c|c|c|c|c|c|c|c|c|c|}
\hline
 & \multicolumn{3}{c|}{$^2$H}
 & \multicolumn{3}{c|}{$^3$He}
 & \multicolumn{3}{c|}{$^4$He}
 & \multicolumn{3}{c|}{$^{16}$O}
 & \multicolumn{3}{c|}{$^{40}$Ca}\\
\hline
 $K_{c.m.}^{+}$ [fm$^{-1}$]& $\mathcal{P}_{D/A}$& $N_{D/A}$& $a_{D/A}$&
  $\mathcal{P}_{D/A}$ & $N_{D/A}$ & $a_2$& $\mathcal{P}_{D/A}$& $N_{D/A}$ & $a_{D/A}$ &
$\mathcal{P}_{D/A}$ & $N_{D/A}$ & $a_2$& $\mathcal{P}_{D/A}$& $N_{D/A}$& $a_{D/A}$\\
\hline
 1.5 & 0.04 & 0.04 & 1 &0.04 & 0.06 & 1 & 0.04 & 0.12 & 1.5 & 0.031 & 0.93 & 2.9 & 0.030 & 4.9 &6.1 \\
\hline
\hline
\end{tabular}}
\caption{The values of $\mathcal{P}_{D/A}$ (Eq. (\ref{nD/A})) and $N_{D/A}$
(Eq.(\ref{ND/A})) calculated in correspondence of  $K_{c.m.}^{+}=1.5\, fm^{-1}$. The quantity
$a_{D/A}=[(2/A)][N_{D/A}/N_{D/D}]$
is the per-nucleon probability of deuteronlike ((ST)=(10)) $2N$ SRC in $A$ with respect to
the deuteron.}
\label{Table5}
\end{center}
\end{table}
\clearpage
\clearpage
\clearpage
\clearpage
\clearpage
\clearpage
\clearpage
\clearpage
\clearpage
\clearpage
\clearpage
\clearpage
\clearpage
\clearpage
\clearpage
\clearpage
\clearpage
\clearpage
\clearpage
\clearpage
\clearpage
\clearpage
\clearpage
\clearpage


\begin{thebibliography}{99}
  %
\bibitem{Tang:2002ww}
  A.~Tang, J.~W.~Watson, J.~L.~S.~Aclander, J.~Alster, G.~Asryan,
   {\it et al.},
  Phys.\ Rev.\ Lett.\  {\bf 90}, 042301 (2003).
  %
\bibitem{Shneor:2007tu}
  R.~Shneor {\it et al.}  [Jefferson Lab Hall A Collaboration],
  Phys.\ Rev.\ Lett.\  {\bf 99}, 072501 (2007).
  %
\bibitem{Egiyan:2005hs}
  K.~S.~Egiyan {\it et al.}  [CLAS Collaboration],
  Phys.\ Rev.\ Lett.\  {\bf 96}, 082501 (2006).
  %
\bibitem{Frankfurt:1993sp}
  L.~L.~Frankfurt, M.~I.~Strikman, D.~B.~Day and M.~Sargsian,
  Phys.\ Rev.\ C {\bf 48}, 2451 (1993).
  %
\bibitem{Fomin:2011ng}
  N.~Fomin, J.~Arrington, R.~Asaturyan, F.~Benmokhtar, W.~Boeglin {\it et al.},
  Phys.\ Rev.\ Lett.\  {\bf 108}, 092502 (2012).
  %
\bibitem{Subedi:2008zz}
  R.~Subedi, R.~Shneor, P.~Monaghan, B.~D.~Anderson, K.~Aniol {\it et al.},
  Science {\bf 320}, 1476 (2008).
  %
\bibitem{Frankfurt:1981mk}
  L.~L.~Frankfurt and M.~I.~Strikman,
  Phys.\ Rept.\  {\bf 76}, 215 (1981).
  %
\bibitem{Sargsian:2005ru}
  M.~M.~Sargsian, T.~V.~Abrahamyan, M.~I.~Strikman and L.~L.~Frankfurt,
  Phys.\ Rev.\ C {\bf 71}, 044615 (2005).
  %
\bibitem{Schiavilla:2006xx}
  R.~Schiavilla, R.~B.~Wiringa, S.~C.~Pieper and J.~Carlson,
  Phys.\ Rev.\ Lett.\  {\bf 98}, 132501 (2007).
  %
\bibitem{Alvioli:2007zz}
  M.~Alvioli, C.~Ciofi degli Atti and H.~Morita,
  Phys.\ Rev.\ Lett.\  {\bf 100}, 162503 (2008).
  %
  \bibitem{Alvioli:2012aa}
  M.~Alvioli, C.~Ciofi degli Atti, L.~P.~Kaptari, C.~B.~Mezzetti, H.~Morita and S.~Scopetta,
  Phys.\ Rev.\ C {\bf 85}, 021001 (2012).
  %
\bibitem{CiofidegliAtti:1995qe}
  C.~Ciofi degli Atti and S.~Simula,
  Phys.\ Rev.\ C {\bf 53}, 1689 (1996).
  %
\bibitem{Review_1} J. Arrington, D. W. Higinbotham, G. Rosner, and M. Sargsian,
  Prog. Part. Nucl. Phys. {\bf 67}, 898 (2012)
  %
\bibitem{Review_2} L. Frankfurt, M. Sargsian, M. Strikman, Int. J. Mod. Phys.  {\bf A23}, 2991 (2008).
  %
\bibitem{Bohigas:1979kk} M. Dal Ri, O.~Bohigas and S.~Stringari,
  Nucl. Phys. {\bf A376}, 81 (1982).
  %
\bibitem{CiofidegliAtti:1990rw} 
  C.~Ciofi degli Atti, E.~Pace and G.~Salme,
  Phys.\ Rev.\ C {\bf 43}, 1155 (1991).
  %
\bibitem{Zabolitzky:1978cx} 
  J.~G.~Zabolitzky and W.~Ey,
  Phys.\ Lett.\  {\bf 76}B, 527 (1978).
  %
\bibitem{VanOrden:1979mt} 
  J.~W.~Van Orden, W.~Truex and M.~K.~Banerjee,
  Phys.\ Rev.\ C {\bf 21}, 2628 (1980).
  %
\bibitem{BenharCiofi} O. Benhar, C. Ciofi degli Atti, S. Liuti, G. Salm\`e,
  Phys. Lett.{\bf B177}, 135 (1986).
  %
\bibitem{Ji:1989nr}
  X.~-D.~Ji and J.~Engel,
  Phys.\ Rev.\ C {\bf 40}, 497 (1989).
  %
\bibitem{Schiavilla_old} R. Schiavilla, V. R. Pandharipande, R. B. Wiringa,
Nucl.\ Phys.\ A {\bf 449}, 219 (1986).
  %
  \bibitem{Pieper}
  S.~C.~Pieper, R.~B.~Wiringa and V.~R.~Pandharipande,
  Phys.\ Rev.\ C {\bf 46}, 1741 (1992).
  %
\bibitem{Feldmeier:2011qy}
  H.~Feldmeier, W.~Horiuchi, T.~Neff and Y.~Suzuki,
Phys.\ Rev.\ C {\bf 84}, 054003 (2011).
  %
\bibitem{Akaishi}
  H. Morita, Y. Akaishi, H. Tanaka,
  Prog.\ Theor.\ Phys. {\bf 79}, 863 (1988).
  %
  \bibitem{Hiko} H. Morita {\it et al}, to be published
%
\bibitem{Kievsky:1992um}
  A.~Kievsky, S.~Rosati and M.~Viviani,
  Nucl.\ Phys.\ A {\bf 551}, 241 (1993).
  %
\bibitem{Gloeckle:1995jg}
  W.~Gloeckle, H.~Witala, D.~Huber, H.~Kamada and J.~Golak,
  Phys.\ Rept.\  {\bf 274}, 107 (1996).
  %
\bibitem{Arias de Saavedra:2007qg}
  F.~Arias de Saavedra, C.~Bisconti, G.~Co' and A.~Fabrocini,
  Phys.\ Rept.\  {\bf 450}, 1 (2007).
  %
\bibitem{Wiringa_review}  S. C. Pieper, and R. B. Wiringa, Annu. Rev. Nucl. Part. Sci.
  {\bf51} 53 (2001).
  %
\bibitem{Roth:2010bm}
  R.~Roth, T.~Neff and H.~Feldmeier,
  Prog.\ Part.\ Nucl.\ Phys.\  {\bf 65}, 50 (2010).
  %
\bibitem{Alvioli:2005cz}
  M.~Alvioli, C.~Ciofi degli Atti and H.~Morita,
  Phys.\ Rev.\ C {\bf 72}, 054310 (2005).
  %
\bibitem{Varga:1995dm}
  K.~Varga and Y.~Suzuki,
  Phys.\ Rev.\ C {\bf 52}, 2885 (1995).
  %
\bibitem{Suzuki:2008fg}
  Y.~Suzuki, W.~Horiuchi, M.~Orabi and K.~Arai,
  Few Body Syst.\  {\bf 42}, 33 (2008).
%
  \bibitem{Suzuki_1} Y. Suzuki and  W. Horiuchi, Nucl. Phys. {\bf A818}, 188 (2009).
  %
\bibitem{RSC} R.V. Reid, Ann. Phys. ~N.Y. {\bf50}, 411 (1968).
%
\bibitem{Paris}  M. Lacombe et al., Phys. Rev. {\bf C 21}, 861 (1980).
%
\bibitem{AV8} B. S. Pudliner, V. R. Pandharipande, J. Carlson, S. C. Pieper,
and, R. B. Wiringa, Phys. Rev. {\bf C56} 1720(1997).
%
\bibitem{AV14}
  R. B. Wiringa, R. A. Smith, and T. A. Ainsworth, Phys.
Rev. {\bf C29}, 1207B (1984).
  %
\bibitem{AV18}
  R.~B.~Wiringa, V.~G.~J.~Stoks and R.~Schiavilla,
  Phys.\ Rev.\ C {\bf 51}, 38 (1995).
  %
\bibitem{CPS_momdistr} C. Ciofi degli Atti, E. Pace, and
G. Salm\`{e}, Phys. Lett. {\bf 141B} 14 (1984).
  %
\bibitem{CiofiLiuti} C. Ciofi degli Atti, S. Liuti, Phys. Lett. {\bf B225}, 215 (1989).
  %
\bibitem{eeprime_exp1}
  F.~Benmokhtar {\it et al.}  [Jefferson Lab Hall A Collaboration],
  Phys.\ Rev.\ Lett.\  {\bf 94}, 082305 (2005).
  %
\bibitem{eeprime_exp2} J.F.J. Van Den Brandt {\it et al.}, Phys. Rev. Lett. {\bf 60}, 2006 ~(1988);
A. Magnon et al., Phys. Lett. {\bf B222} 352 ~(1989).
  %
\bibitem{Forest:1996kp}
  J.~L.~Forest, V.~R.~Pandharipande, S.~C.~Pieper, R.~B.~Wiringa, R.~Schiavilla and A.~Arriaga,
  Phys.\ Rev.\ C {\bf 54}, 646 (1996).
  %
\bibitem{Wiringa:2006ih}
  R.~B.~Wiringa,
  Phys.\ Rev.\ C {\bf 73}, 034317 (2006).
  %
\bibitem{Vanhalst:2011es}
  M.~Vanhalst, W.~Cosyn and J.~Ryckebusch,
  Phys.\ Rev.\ C {\bf 84}, 031302 (2011).
  %
 \bibitem{Vanhalst:2012} M.~Vanhalst, W.~Cosyn and J.~Ryckebusch, Phys. Rev. C {\bf86},
 044619 (2012).
  %
  \bibitem{Scopetta} C. Ciofi degli Atti, L. P. Kaptari, H. Morita,  and S. Scopetta, Few
  Body Sys. {\bf 50} 243 (2011).
  %
  \bibitem{Machleidt} R. Machleidt, Phys. Rev. C{\bf63},
 024001 (2001).
  %
\bibitem{Epelbaum} E. Epelbaum, W. Gl\"{o}ckle, U. -G Meissner, Nucl. Phys. ,
 A{\bf747},362 (2005).
  %
  \bibitem{Polyzou} W. N. Polizou, W. Gl\"{o}ckle, Few-Body Systems {\bf9}, 97 (1990).
  %
  \bibitem{Vary} J. P. Vary, Phys. Rev. C{\bf7}, 521 (1973).
   %
   \bibitem{Nocore_SM} P. Navratil, S. Quaglioni, I. Stetcu, B. Barrett, J. Phys. G {\bf 36}
   083101(2009).\\
   P. Maris, J. P. Vary, A. M. Shirokov, Phys. Rev. C{\bf79} 014308 (2009).
   %
    \bibitem{RG} S. K. Bogner, R. J. Furnstahl, and A. Schwenk, Prog. Part. Nucl.
Phys. {\bf 65}, 94 (2010).
%
   %
\bibitem{Anderson} E. R. Anderson, S. K. Bogner, R. J. Furnstahl, and R. J. Perry,
Phys. Rev. C {\bf 82}, 054001 (2010).
%
\bibitem{Bogner}
S. K. Bogner, D. Roscher, Phys. Rev C {\bf 86}, 064304 (2012)
\end{thebibliography}
\end{document}